\documentclass[%
 reprint,onecolumn,
superscriptaddress, 12pt, 
nofootinbib,
 amsmath,amssymb,
floatfix,
]{revtex4-2}

\usepackage{graphicx}
\usepackage{dcolumn}
\usepackage{bm}

\usepackage{cancel}
\usepackage{amsfonts}
\usepackage{xcolor}
\usepackage{tcolorbox}

\newcommand{\beqa}{\begin{eqnarray}}
\newcommand{\eeqa}{\end{eqnarray}}
\newcommand{\nn}{\nonumber}
\usepackage{subcaption} 

\begin{document}
\title{  Vortex Dynamics in Tubular Fluid Membranes }
\author{Udaya Maurya}
\affiliation {Institute for Plasma Research, Bhat, Gandhinagar, India}
\author{Surya Teja Gavva}
\affiliation {Department of Computer Science,
Rutgers University, Piscataway, NJ 08854-8019, USA}
\author{Arpan Saha}
\affiliation{IMECC UNICAMP, Campinas, São Paulo, Brazil}
\author{Rickmoy Samanta \footnote{corresponding author, Email : rickmoysamanta@gmail.com}}
\affiliation{ Birla Institute of Technology and Science Pilani, Hyderabad 500078, India}
\date{\today}

\begin{abstract}
Abstract: Thin cylindrical membranes arise in a wide variety of biological systems ranging from tubular structures on and within cell membranes to in-vitro experiments on artificial vesicles. Motor proteins embedded in such fluidic membranes often induce vortex-like flows. In this work, we construct a class of two-dimensional (2D) vortex flow in a thin tubular membrane, coupled to three-dimensional (3D) external embedding fluids. The cylinder topology enforces the creation of an additional saddle in the flow field, such that a global flow constraint emerging from topological considerations is satisfied (Poincar\'e Index Theorem). In this setup, the incompressibility of the membrane fluid can be utilized to cast the dynamics of a multi-vortex system in the form of a Hamiltonian, This Hamiltonian also incorporates the specific couplings of the 2D membrane flow with the 3D external fluids. The cylinder geometry breaks the in-plane rotational symmetry of the membrane and leads to several interesting features in the multi-vortex dynamics, such as orbit pinching,  For a two-vortex system of same circulation, we observe closed orbits with the inter-vortex separation oscillating in time, unlike flat and spherical fluid membranes, where the separation remains constant.  Vortex pairs (vortices with opposite circulation) move together along helical geodesics in accordance with a conjecture by Kimura Ref.~\cite{km}, \textit{Proceedings of the Royal Society A.455245–259 (1999)}, now extended to tubular geometries. We also explore relative equilibria of multi-vortex systems in this setup and demonstrate vortex leapfrogging via numerical simulations.  Our results will be interesting in the context of microfluidic flows arising in nature as well as experimental studies in membrane tubes similar to Ref.~\cite{bss11}, \textit{PNAS 108 (31) 12605-12610 (2011)}.
\end{abstract}

\maketitle
\section{Introduction} The collective dynamics of biological nanomachines at fluid interfaces and membranes is currently an active arena for studying fascinating aspects of dynamical systems. These self-driven motors are capable of converting chemical or other forms of energy into mechanical motion. Moreover, mutual hydrodynamic interactions can often drive the system into an organized state. Biological membranes are particularly interesting in this context because of their fluidic nature \cite{sn72}, which is vital for many living processes. Such a membrane can be well approximated as a thin 2D sheet of viscous fluid with associated flows governed by low Reynolds hydrodynamics \cite{purcell}. Moreover, the membrane fluid can exchange momentum with the external embedding fluids in the ambient space, thus they are essentially quasi-2D in nature.  The early works of Saffman and Delbr\"{u}ck Ref.~\cite{saff1,saff2} explored the low Reynolds hydrodynamics of inclusions embedded in such quasi-2D, flat membranes, followed by several advancements Ref.~ \cite{hughes,evans,lg96,staj,fischer,nlp07,oppdiamant1,oppdiamant2,oppdiamant3}. The quasi-2D nature of the membrane introduces an additional length scale (Saffman length) which is the ratio of the membrane viscosity ($\eta_{2D}$) and the viscosity of the external fluid ($\eta$). Beyond the Saffman length, the in-plane flows are governed by the traction stress from the external embedding fluids. In addition to this interesting feature, membranes found in nature typically have a confined geometry with non-zero curvature. Moreover, they host a large number of inclusions. The impact of membrane topology and curvature on dynamics of such inclusions has been studied in many recent works, Ref.~\cite{dt07, henlev2008,henlev2010, bss11, wg2012, wg2013, atzbergershape, dan2016, atz2016, atz2018,atz2019, sosm20, rs21, atz22, sarthak22, jain23, smn22} and also being probed in recent experiments, Ref.~\cite{henlev2008,bss11, wg2013}. In particular, good agreement with Saffman Delbr\"{u}ck theory has been demonstrated in membrane tubes Ref.~\cite{bss11}.\\\\
The physics of rotating motors in fluid interfaces has recently been the subject of many theoretical and experimental investigations, Ref.~\cite{rt0,rt1,rt2,rt3,rt4,rt5,rt6,rt7,rt8,rt9,  lushiv2015,ylv2015,mzc}. In this paper, we focus on the hydrodynamic interactions of rotating inclusions embedded in tubular fluid membranes. The inclusions considered in this work may be considered as driven-to-rotate entities which induce vortex-like flows in membranes. Such vortices are known to self-organize into crystallized states under suitable conditions and exhibit hyperuniform order \cite{lenz2003,lenz2004, nmsh19, nmsh22} in flat membranes. The dynamics of such rotating units in spherical fluid membranes is being explored as well, see Ref.~\cite{rs21}. However, analogous studies in membranes of tubular geometries have been relatively less explored. The formation of membrane tubes is an active area of research, Ref.~\cite{Derenyi2002,Bozic2001,Bukman1996,Heinrich1999,Zhang1999,Smith2004,Powers2002,Raucher1999,Fygenson1997,Daniels2005,Brochard-Wyart2006}, arising in a wide variety of biological systems due to polymerization of actin fibers Ref.~\cite{Boal2001} and microtubules Ref.~\cite{Alberts2002}. and hosting motor proteins Ref.~\cite{Berk1992,Roux2005}. Such tubular
structures exist within and on cell membranes, often bridging two nearby membranes. Moreover, forces exerted on vesicles also lead to tubule formation, as observed in vitro Ref.~\cite{Fygenson1997,Roux2005,Roux2002,Koster2003,Reiner2006,Rustom2004,Schnur1993,Evans1996,Roopa2003}. For example, such tubular membrane geometries
of varying radii can be created by pulling a
membrane-tethered bead from the surface of a cell or vesicle using a laser trap.
Biological membranes in the Endoplasmic Reticulum
 and the Golgi apparatus can also form tubular structures Ref.~\cite{Upadhyaya2004,Shemesh2003,Sciaky1997,Watson2005,Ladinsky1999,Cole1996,Weiss2003,Sbalzarini2005,Weiss2004}, playing vital role in inter-membrane transport Ref.~\cite{Demontis2004,Ellenberg1997}. Such membranes can host a large class of proteins, some of which may be rotating, such as ATP synthase arising in mitochondrial cristae and thylakoids in chloroplasts, Ref.~\cite{Hahn2018,Abrahams1994,Stock2000}. \\

The tubular membrane geometry that we explore in this paper is distinct from flat and spherical membranes, since the in-plane rotational symmetry is broken. We find that this breakdown of rotational symmetry, along with the confined cylinder topology (in the angular direction) leads to several interesting features in the dynamics of microvortex assemblies in membrane tubes. First,we show that the cylinder topology enforces the creation of an additional saddle in the flow field, such that a global flow constraint emerging from topological considerations is satisfied (Poincar\'e Index Theorem). Next, utilizing the incompressibility condition of  the membrane fluid flow, we cast the dynamics of the multi-vortex system in terms of a Hamiltonian on the cylinder, similar to the approach followed in flat membranes \cite{nmsh19,nmsh22}. The Hamiltonian  depends on the system parameters, the Saffman length $\lambda$, the tube radius R and the circulation strength of the vortices. This offers experimentally accessible tuning parameters to control the dynamics of the multi-vortex system. Focusing on a simple system of two vortices of same circulation strength $\tau$, we show that the breaking of rotational symmetry leads to non-conservation of the inter-vortex separation, unlike flat and spherical fluid membranes. The vortices generically move in closed orbits, which can be severely distorted by tuning the initial inter-vortex separation as well as the parameter $\frac{\lambda}{R}$. For example, upon reducing the value of $\frac{\lambda}{R}$ we find that the single orbit breaks into two smaller orbits. Also, when they are initially sufficiently separated along the longitudinal (z-axis) direction of the tube, we observe that the vortices prefer two distinct orbits that cover the entire circumference of the tube.  Two vortices with opposite circulation generically move together along helical trajectories, in accordance with a conjecture by Kimura \cite{km}. We also explore relative equilibria and  relative periodic orbits including vortex leapfrogging.  A mathematical understanding of confined versus unconfined dynamics of vortices as well as the pinching of vortex orbits emerges from the associated conservation laws of the multivortex system.\\ \\Besides being of experimental relevance Ref.~\cite{bss11}, we believe our results will be interesting for understanding the collective dynamics of vortices in tubular fluid interfaces and membranes. Moreover, many of our results will be applicable to more setups involving rotating motors (both living and non-living) in tubular fluid interfaces. \\\\
It's worth commenting on how some of our results compare with spherical (Ref.~\cite{rs21}) and flat membranes. The in-plane  fluid flows in both spherical and tubular membranes depart from flat membranes in the regime $\frac{\lambda}{R} \gg 1$. For the spherical membrane, modifications to fluid flow arise from the intrinsic Gaussian curvature as well as the non-local interactions mediated via the external 3D fluids ( related to extrinsic geometry), leading to a  global rotation ( see Ref.~\cite{henlev2010} for details) of the entire membrane fluid (along with internal fluids). However, the in-plane rotational symmetry ensures that many features of fluid flow and vortex dynamics in flat membranes continue to hold in spherical membranes, see Ref.\cite{rs21}, even in the high curvature regime. On the other hand, for cylindrical membranes, although the intrinsic Gaussian curvature is zero, the extrinsic geometry of the cylinder comes into play via the couplings to the external 3D fluids, leading to distinct features in flows along the transverse and longitudinal directions. The cylindrical geometry breaks the in-plane rotational symmetry and leads to new phenomenon, such as non-conservation of inter-vortex separation and orbit pinching of same sign vortices, helical trajectories of vortex dipoles,  leapfrogging in multi-vortex systems, which are the main results of the paper.\\\\
The paper is organized as follows: In Sec.~\ref{mhd} we provide a brief review of hydrodynamics of generic curved membranes, following \cite{henlev2010} and \cite{rs21,sarthak22, jain23}. Specializing to  membrane tubes, we construct the point vortex solution, explore topological aspects of the flow and provide the multi-vortex Hamiltonian in Sec.~\ref{cylvortex}. The dynamics governed by the Hamiltonian is next explored in Sec.~\ref{cylsim}, initially focusing on two vortices and then extending to multi-vortex systems, illustrating the conserved quantities. Investigations on vortex equilibria, relative periodic orbits and vortex leapfrogging in the tubular geometry are also provided.   We finally conclude with a future outlook in Sec.~\ref{cncl}.
\begin{figure}[htbp!]
\includegraphics[width=9cm]{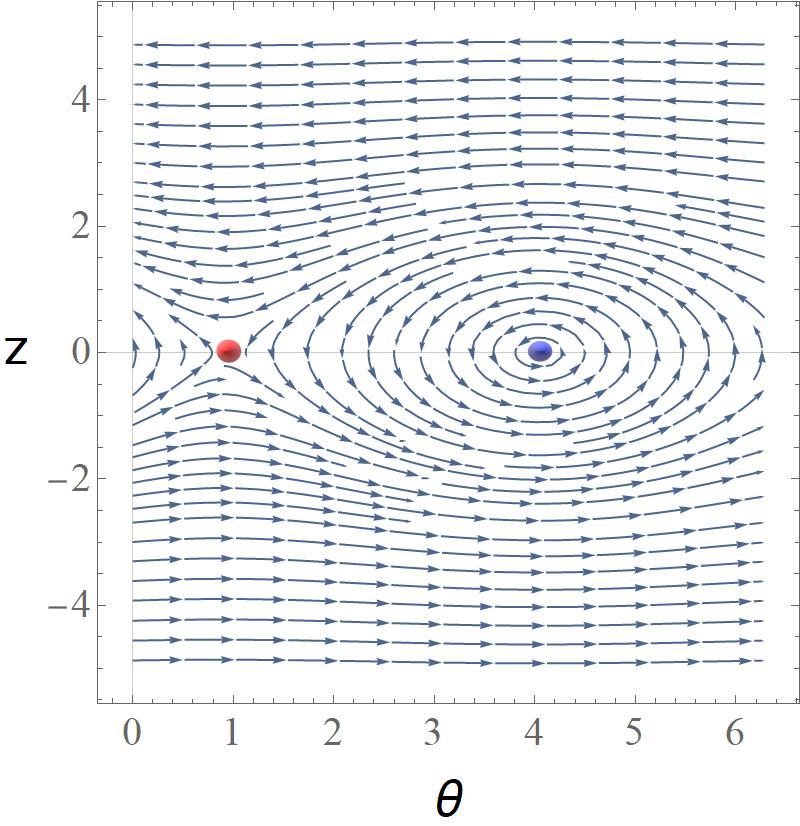}
   \caption{ Plot of the vortex solution Eq.~(\ref{vhc}) with the blue and red dot representing the vortex core and saddle respectively. }
     \label{kernel}
\end{figure}
\section{Membrane Hydrodynamics}
\label{mhd}
In this section, we present a brief review of viscous hydrodynamics in membranes of fixed geometries, closely following Ref.~\cite{henlev2010} and Ref.\cite{rs21,sarthak22}. We will describe the setup for arbitrarily curved but fixed membrane geometry and later specialize to cylindrical membranes. In the fluid approximation, the membrane is considered as a two-dimensional, curved, incompressible monolayer of viscosity $\eta_{2d}$  surrounded by external embedding fluids of different viscosities, both outside and inside the membrane. In order to simplify the analysis, we consider situations where the membrane geometry is fixed in time and only consider in-plane shear flows with no normal component. The membrane fluid is coupled to the external embedding fluids, and hence the membrane is essentially quasi 2D. The key hydrodynamic equations appropriate for this setup are as follows:
\beqa
&D^\alpha v_\alpha =0
\label{membeq1}\\
&\sigma^{ext}_{\alpha} = -\eta_{2D} \left( K(\vec{x})~ v_\alpha  + D^\mu D_\mu v_\alpha \right) + D_\alpha p+T_\alpha
\label{membeq2}\\
&~ \nabla \cdot {\bf v}_{\pm} = 0, ~~~\eta _{\pm}\nabla^2 {\bf v}_{\pm} = \nabla_{\pm} p^{\pm}
\label{membeq3}\\
& T_\alpha= \sigma_{\alpha r}^{-}|_ {r=R} -\sigma_{\alpha r}^+ |_ {r=R}, ~~ \sigma_{ij}^\pm =\eta_\pm \left(D_i v_j^\pm +D_j v_i^\pm\right)- g_{ij} p_{\pm}
\label{membeq4}\\
&{\bf v}_\pm | _{r=R}=v.
\label{membeq5}
\eeqa

We refer to Appendix A of Ref.~\cite{rs21} by one of the authors for a detailed derivation of the equations. We use Greek indices to represent the  in-plane 2D membrane coordinates. Latin indices are used for 3D coordinates to describe flows in the ambient 3D embedding fluids surrounding the 2D membrane.  Eq.~(\ref{membeq1}) is the incompressibility condition for the 2D membrane fluid velocity field $v_\alpha$. $D$ is  the metric compatible covariant derivative. Note that on a curved manifold, incompressibility implies that the divergence is taken with respect to the covariant derivative.  Eq.~(\ref{membeq2}) is the boundary condition for tangential stress balance at the membrane surface.  $\sigma^{ext}_{\alpha} $ represents the local stress exerted by the rotating inclusions embedded in the membrane (this will be a point torque for the following discussions, see Ref.\cite{rs21} for details ) and is balanced by the in-plane membrane stress and the tangential traction stress from the external embedding fluids (denoted by $T_\alpha$)  in the equation. Generic membrane surface coordinates are denoted by ``x". Notably, in the low Reynolds regime, inertia terms are absent and fluid flows are dictated by Stokes equations, with viscosity $\eta_{2D}$ being the viscosity of the membrane fluid. $K(\vec{x})$ is the local Gaussian curvature  and $p$ represents the 2D membrane pressure.  The membrane fluid is surrounded by external 3D embedding fluids with viscosities $\eta_\pm$  and pressure  $p_\pm$  where  $``\pm"$ denotes the outer/inner fluid respectively. Eq.~(\ref{membeq3}) are the relevant Stokes equations for the external incompressible embedding fluids in 3D. Eq.~(\ref{membeq4}) gives the expression for the traction vector $T_\alpha$.  The coordinate $r$ represents a generic coordinate along the membrane normal.  Finally, Eq.~(\ref{membeq5}) is the no-slip boundary condition. We also define two Saffman length scales
\beqa
\lambda_+ =\frac{ \eta_{2d}}{\eta_+},~~~\lambda_- =\frac{ \eta_{2d}}{\eta_-}.\nn
\eeqa
We refer to Appendix A and B of Ref.~\cite{rs21}  for more details on  the appearance of  Gaussian curvature and external torque in the stress balance condition Eq.(\ref{membeq2}).  In general, the viscosities $\eta_+$ and $\eta_-$ of the external fluids give rise to two Saffman lengths $\lambda_+$ and $\lambda-$. However, the geometric asymmetry between the two embedding fluids for the cylindrical case is not much,  since both the internal and external fluids are unbounded, unlike spherical membranes where the internal fluid is bounded. As a result, the effects of asymmetry in the viscosities are much weaker for cylindrical membranes \cite{henlev2010}.
In what follows, we will choose $\eta \equiv \eta_+=\eta_-$ and thus restrict to a unique Saffman length $\lambda \equiv \frac{\eta_{2D}}{\eta}$, with full expressions for asymmetric viscosities in Appendix \ref{avcons}. Higher values of $\frac{\lambda}{R}$ indicate the high curvature regime (thin membrane tube). Typical values of $\lambda$ range from 0.1 to 10 micrometers while  typical membrane tethers have radius of the order of few nanometers, justifying the thin tube approximation i.e. the high curvature regime  ($\frac{\lambda}{R} \gg 1$)  of this work. Let us note that both these assumptions can be relaxed if one uses the full vortex solution constructed in Appendix \ref{avcons}.\\
We summarize the notations in a table:
\begin{center}
\begin{tabular}{|l|l|}
\hline
$x$ & Generic surface coordinates \\
$v_{\alpha}$ & 2D velocity of the membrane fluid \\
$\alpha$ & Surface coordinate index for the 2D membrane \\ 
$\sigma^{ext}_{\alpha}$ & Local stress exerted by the inclusions embedded in the membrane \\
$\eta_{2D}$ & Viscosity of the 2D membrane fluid \\
$D$ & Covariant derivative compatible with the metric \\
$K(\vec{x})$ & The local Gaussian curvature of the membrane \\
$p$ & Local 2D membrane pressure \\ 
$\sigma^{3D}$ & Stress tensor of the external fluids  \\
$z$ & A generalized coordinate normal to the membrane surface \\
$\lambda$ & Saffman length \\
$\tau$ & Vortex circulation strength \\
\hline
\end{tabular}
\end{center}

\section{Vortex Solution and Multi-vortex Hamiltonian}
\label{cylvortex}
\begin{figure}
\centering
\begin{subfigure}{0.9\textwidth}
\begin{tabular}{lcccccccc}
\includegraphics[height=0.13\textheight]{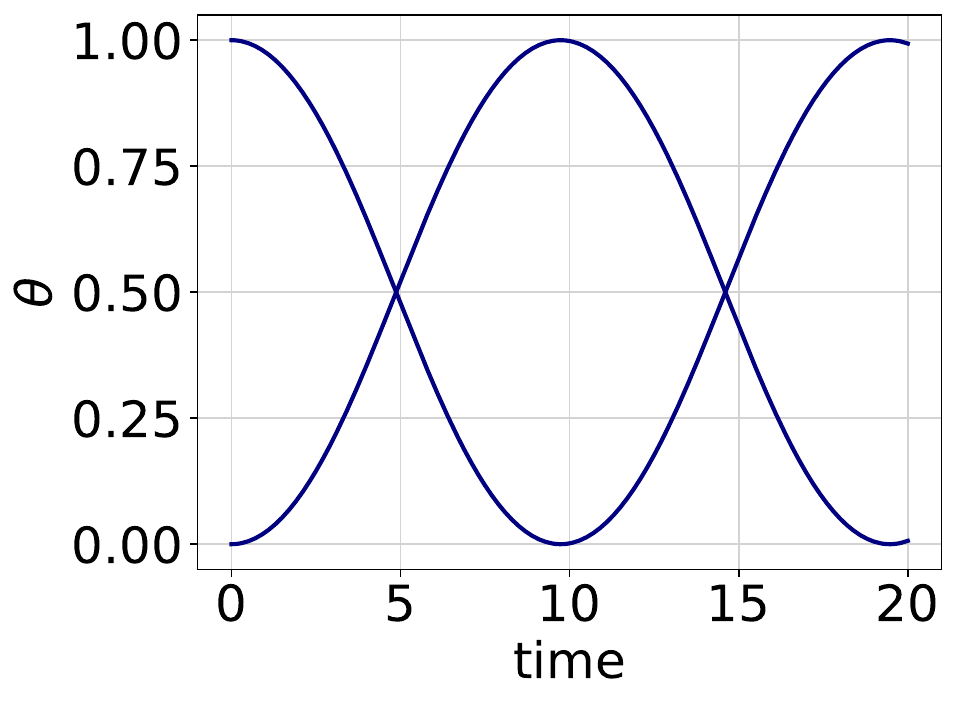}&&
\includegraphics[height=0.13\textheight]{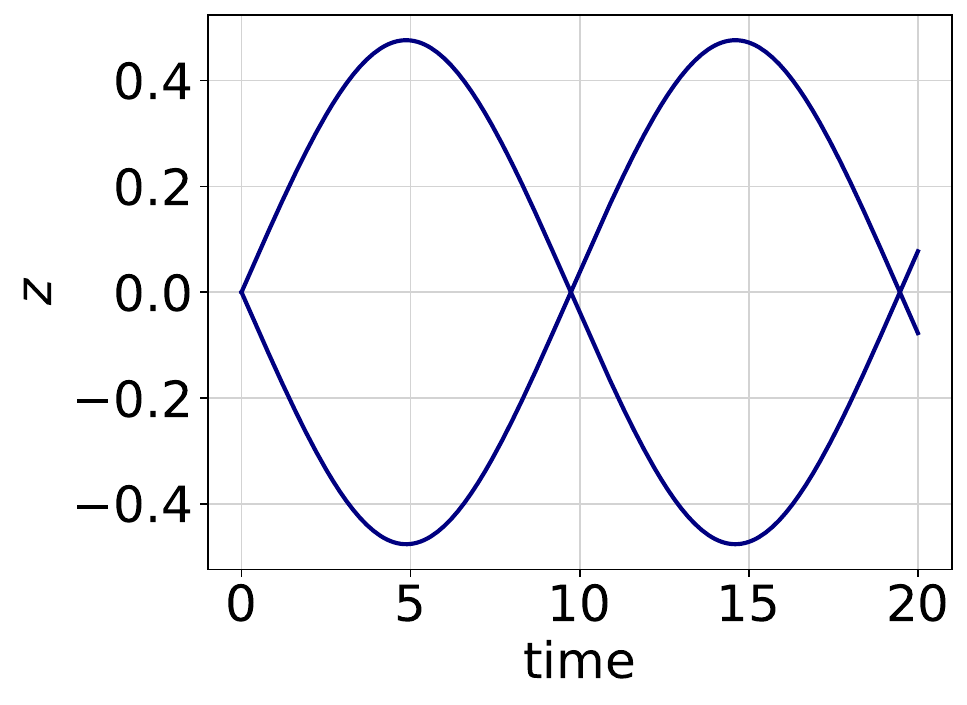}&&
\includegraphics[height=0.13\textheight]{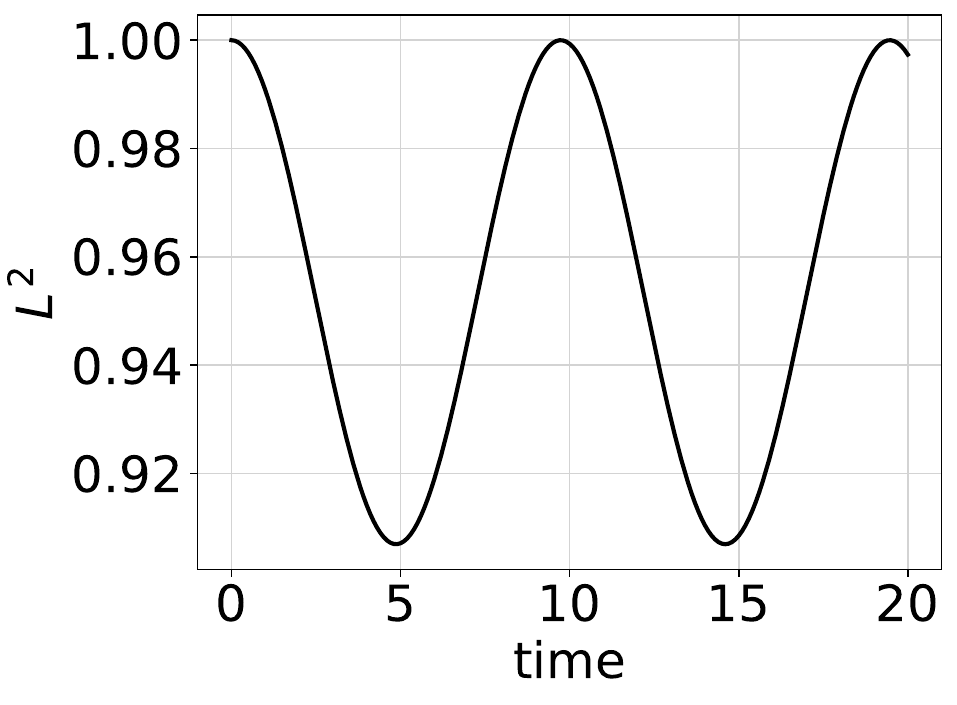}&&\hspace{0.02\textwidth}
\includegraphics[height=0.13\textheight]{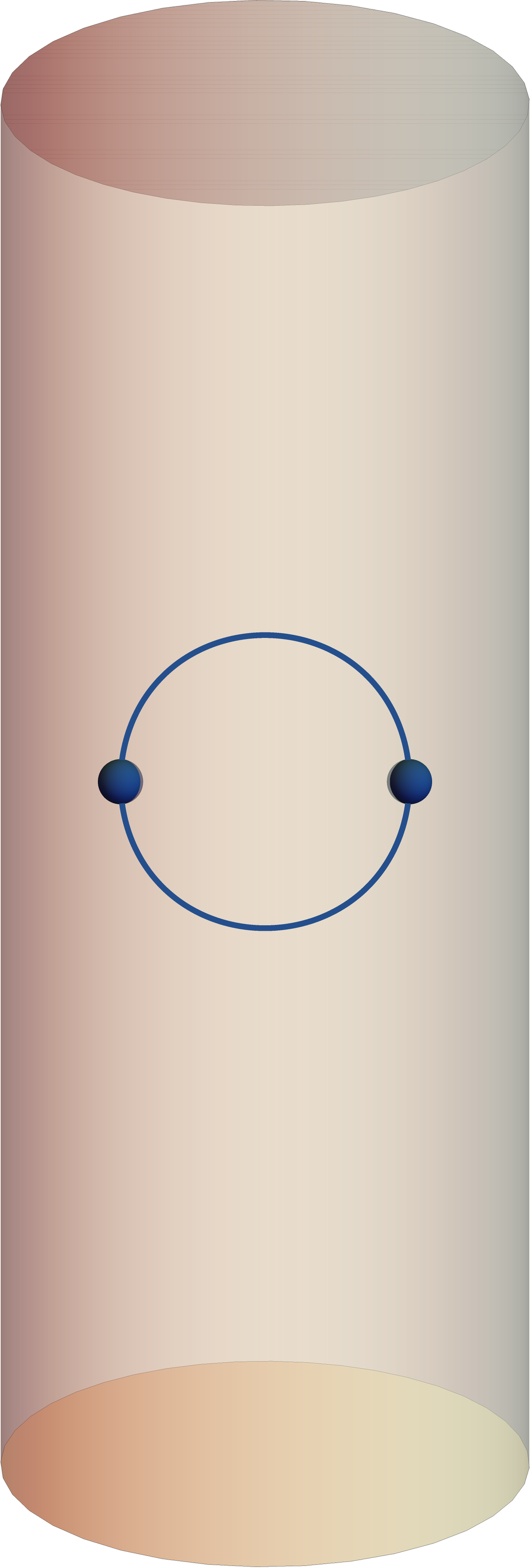}
\end{tabular}
\caption{Case A }
\label{ca}
\end{subfigure}

\begin{subfigure}{0.9\textwidth}
\begin{tabular}{lcccccccc}
\includegraphics[height=0.13\textheight]{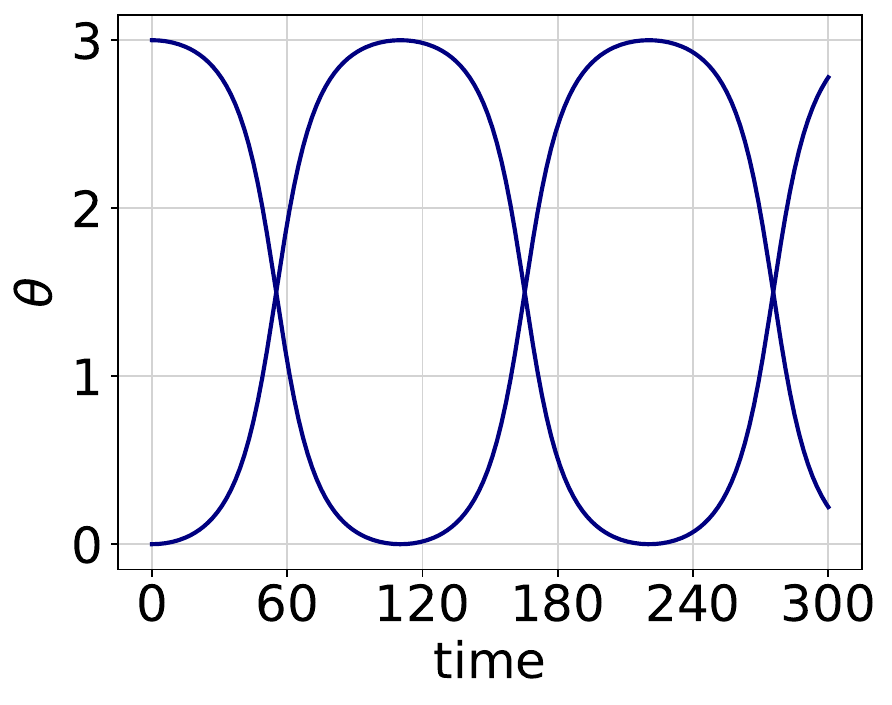}&&
\includegraphics[height=0.13\textheight]{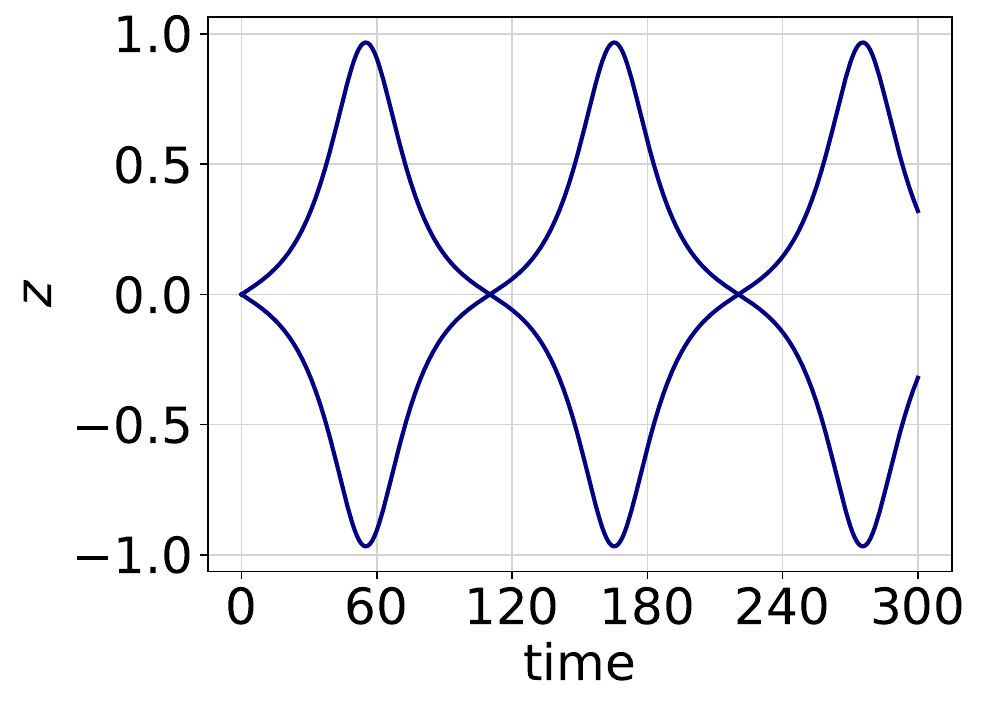}&&
\includegraphics[height=0.13\textheight]{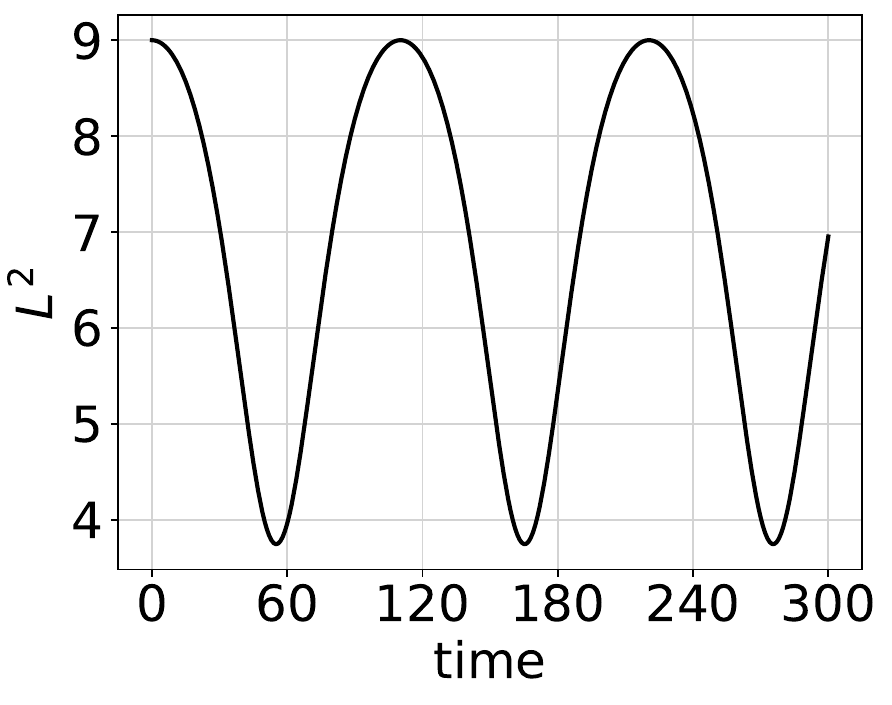}&&\hspace{0.02\textwidth}
\includegraphics[height=0.13\textheight]{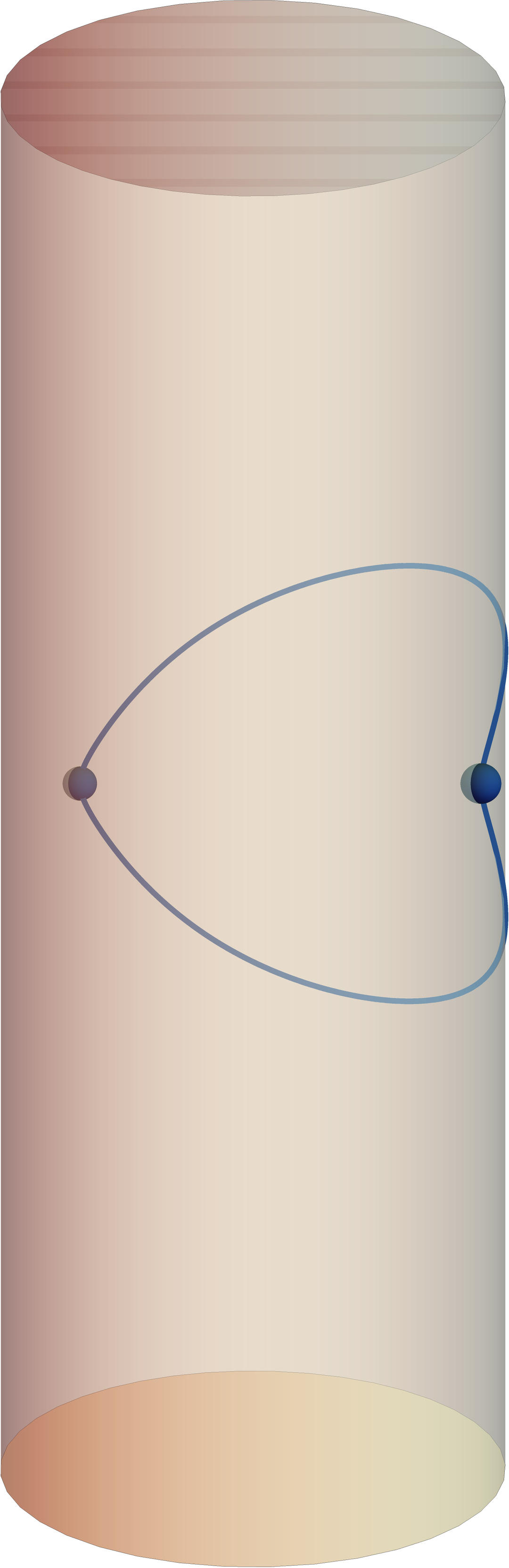}
\end{tabular}
\caption{Case B }
\label{cb}
\end{subfigure}
\begin{subfigure}{0.9\textwidth}
\begin{tabular}{lcccccccc}
\includegraphics[height=0.13\textheight]{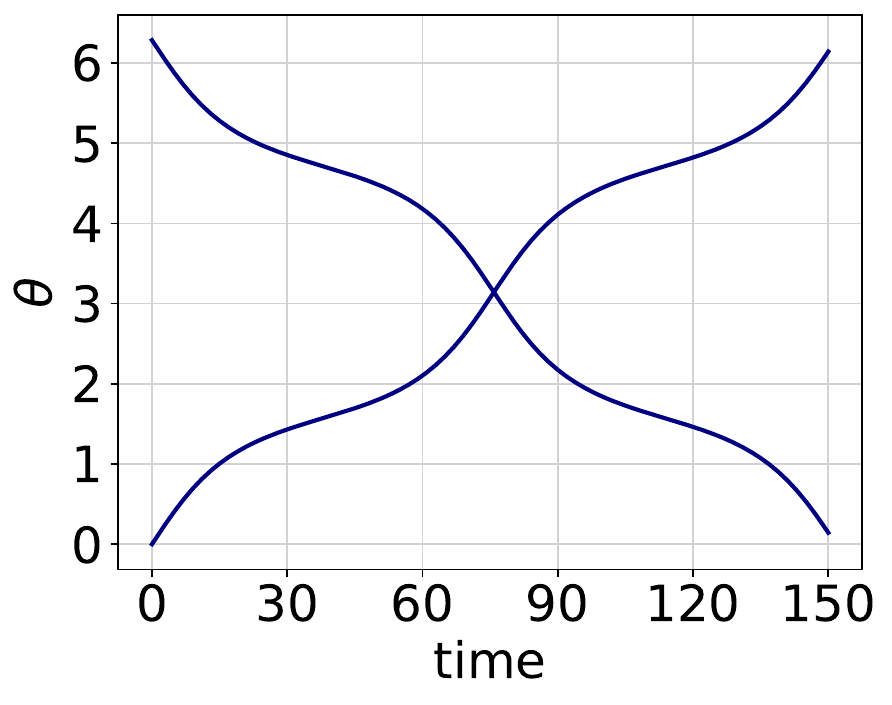}&&
\includegraphics[height=0.13\textheight]{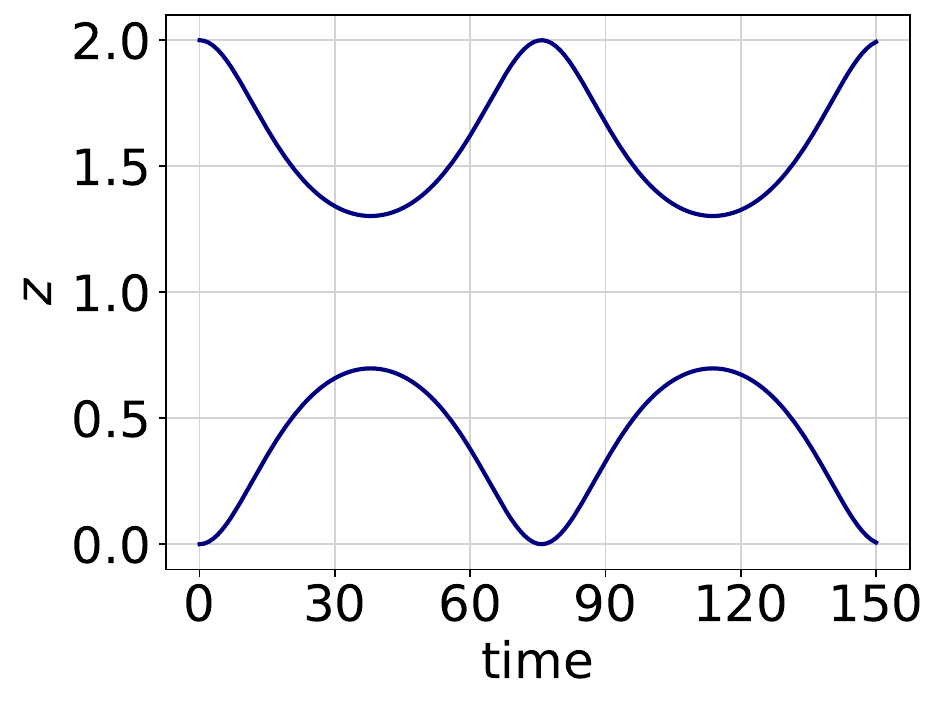}&&
\includegraphics[height=0.13\textheight]{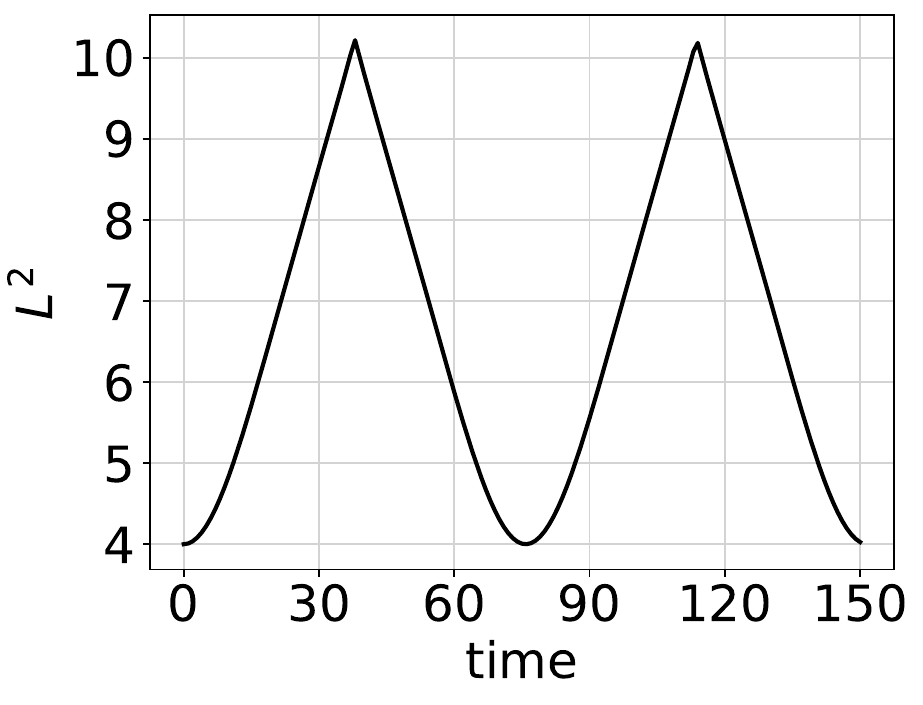}&&\hspace{0.02\textwidth}
\includegraphics[height=0.13\textheight]{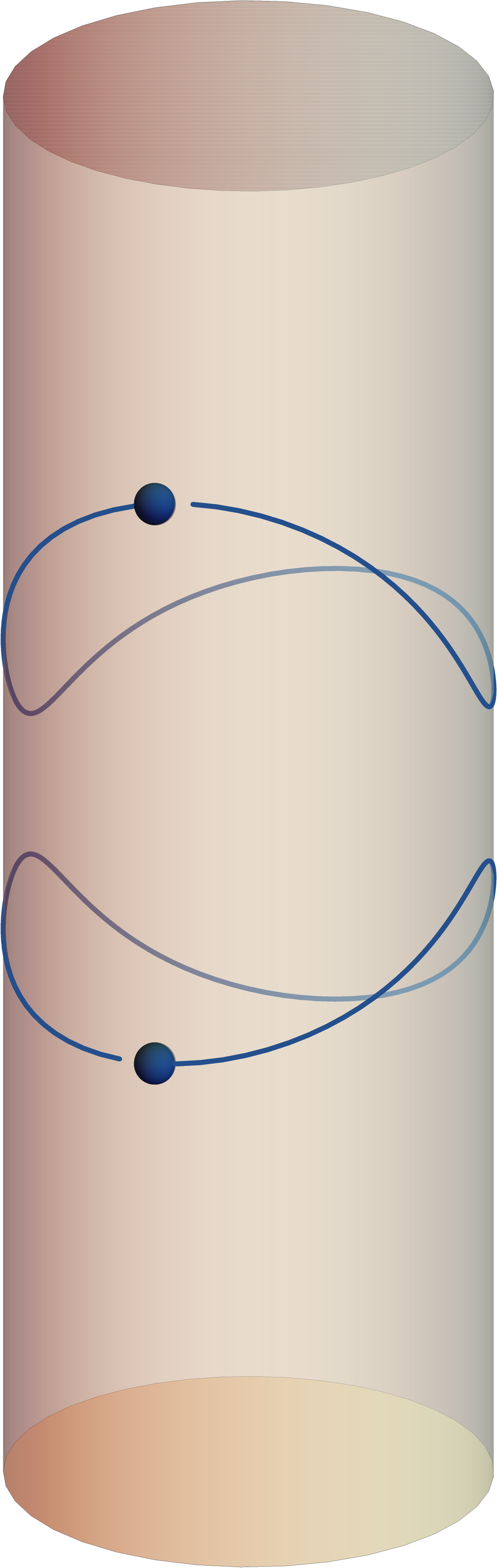}
\end{tabular}
\caption{Case C }
\label{cc}
\end{subfigure}
\begin{subfigure}{0.9\textwidth}
\begin{tabular}{lcccccccc}
\includegraphics[height=0.13\textheight]{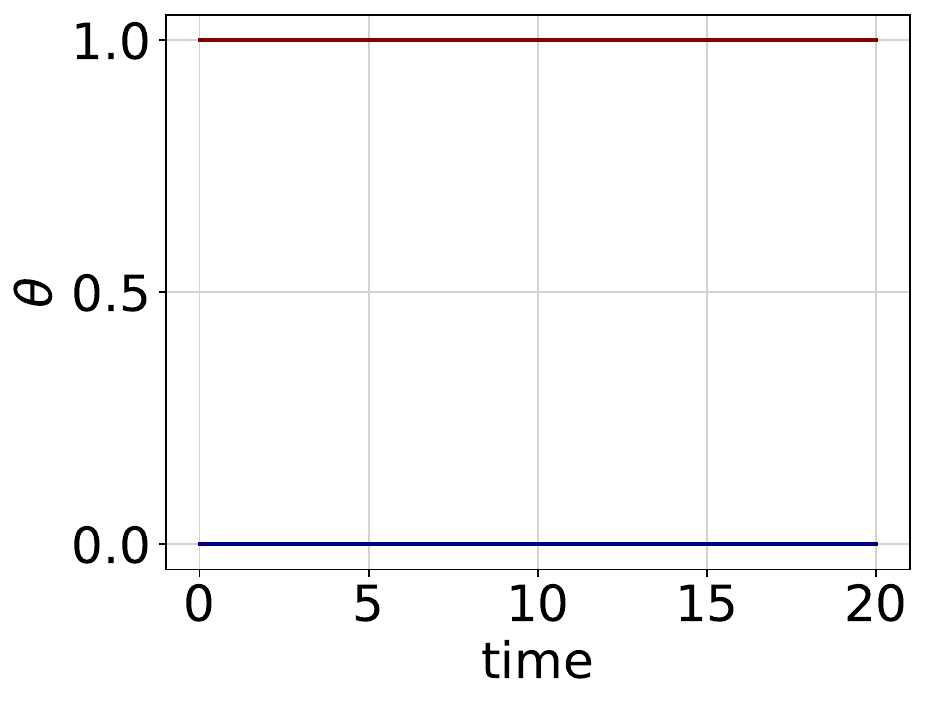}&&
\includegraphics[height=0.13\textheight]{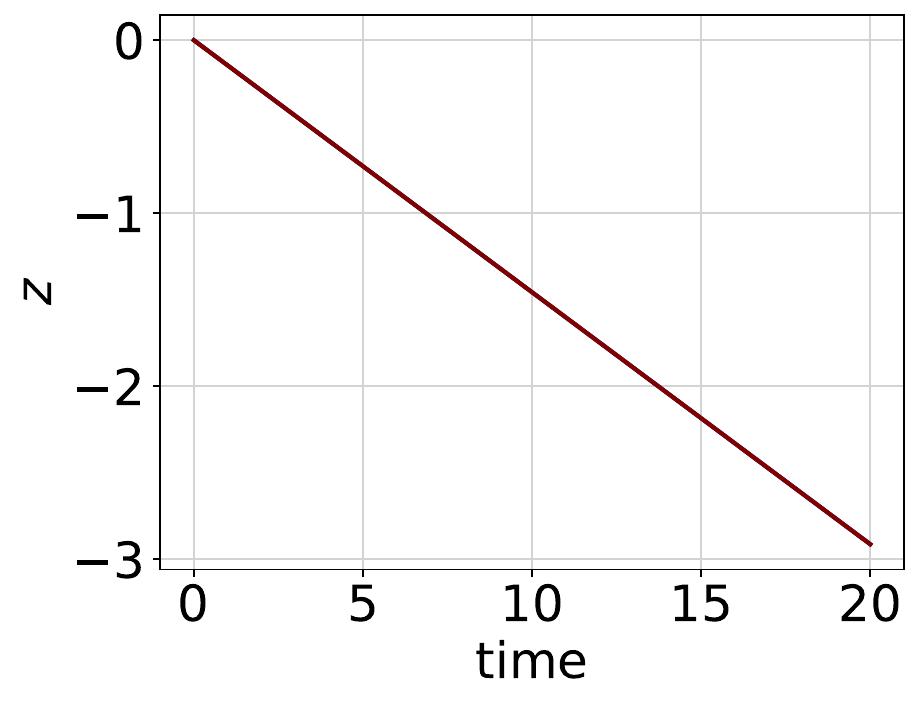}&&
\includegraphics[height=0.13\textheight]{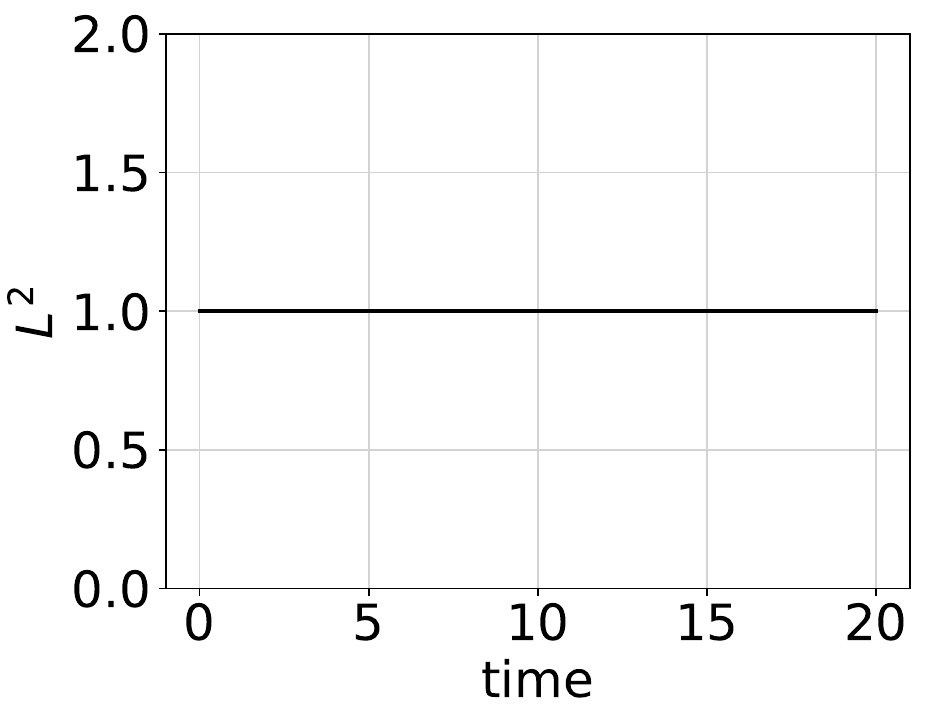}&&\hspace{0.02\textwidth}
\includegraphics[height=0.13\textheight]{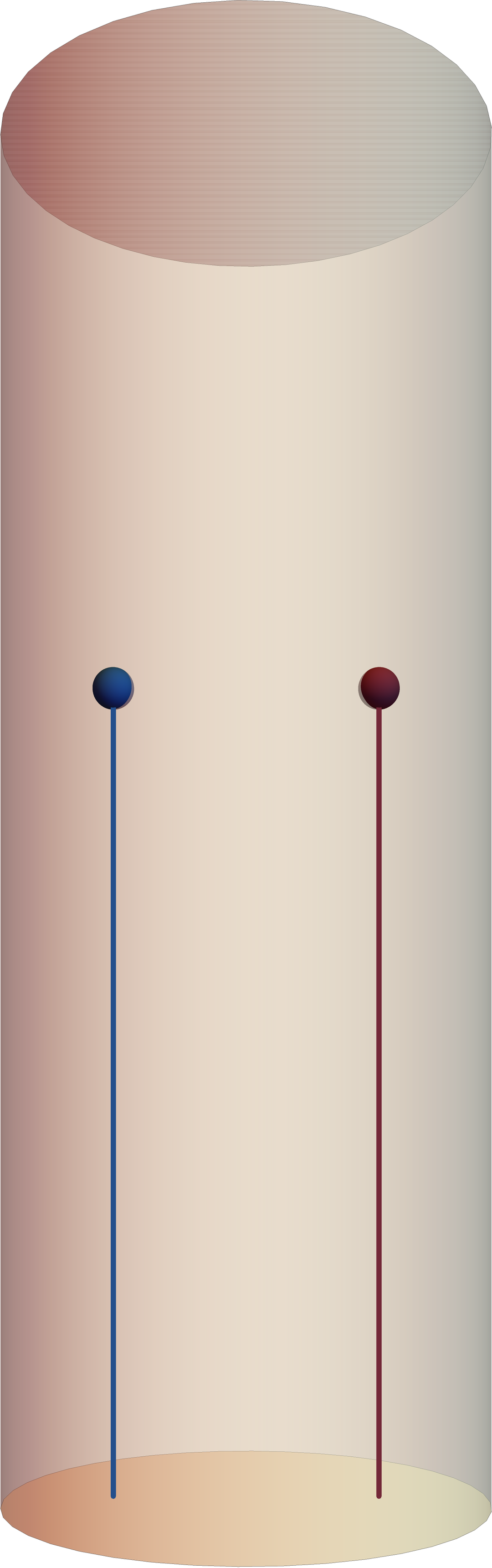}
\end{tabular}
\caption{Case D }
\label{cd}
\end{subfigure}
\begin{subfigure}{0.9\textwidth}
\begin{tabular}{lcccccccc}
\includegraphics[height=0.13\textheight]{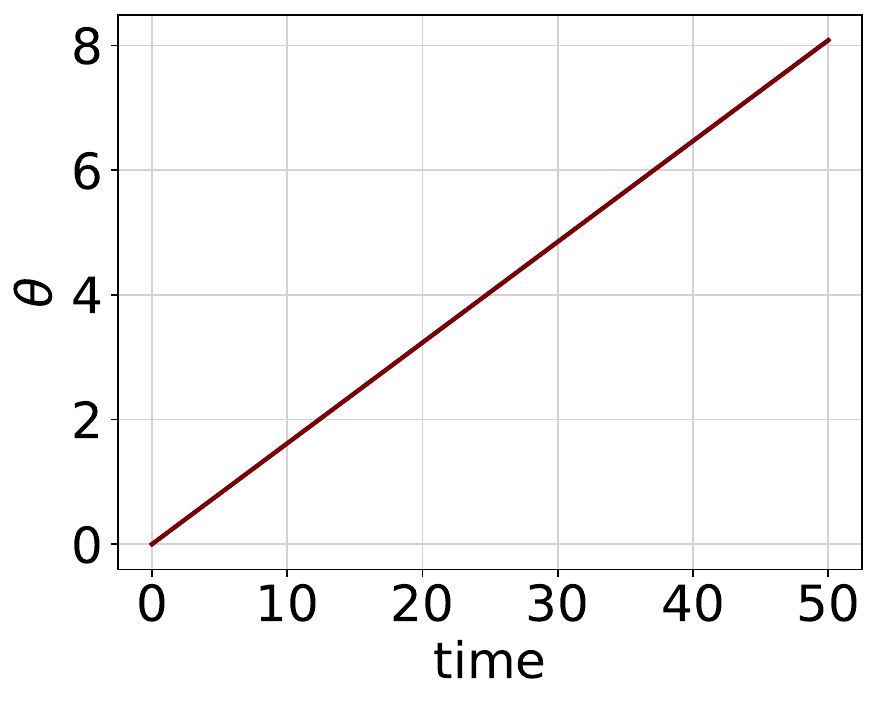}&&
\includegraphics[height=0.13\textheight]{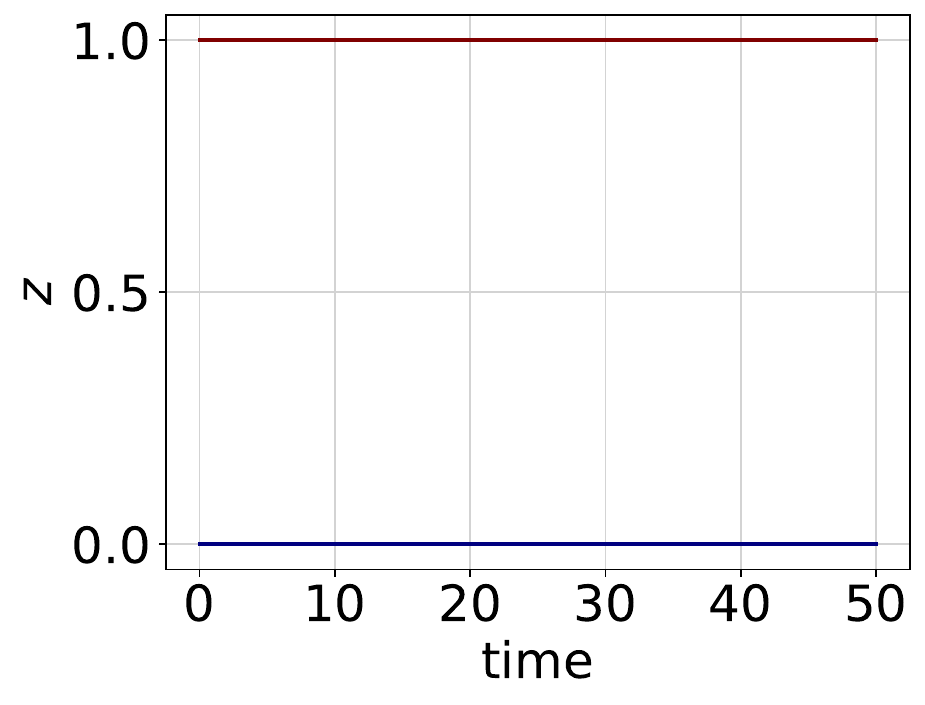}&&
\includegraphics[height=0.13\textheight]{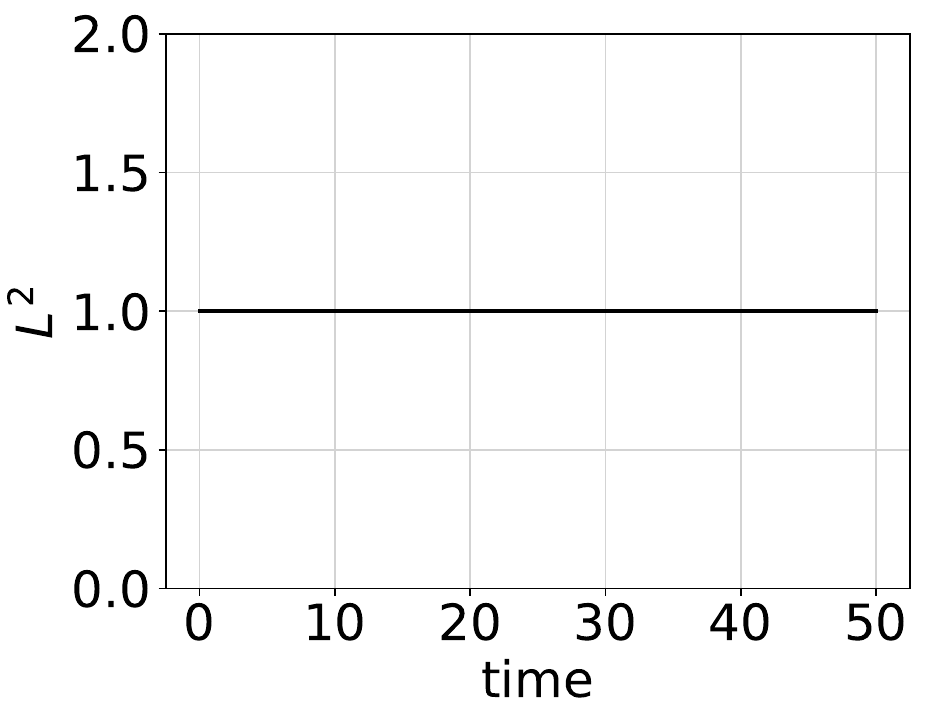}&&\hspace{0.02\textwidth}
\includegraphics[height=0.13\textheight]{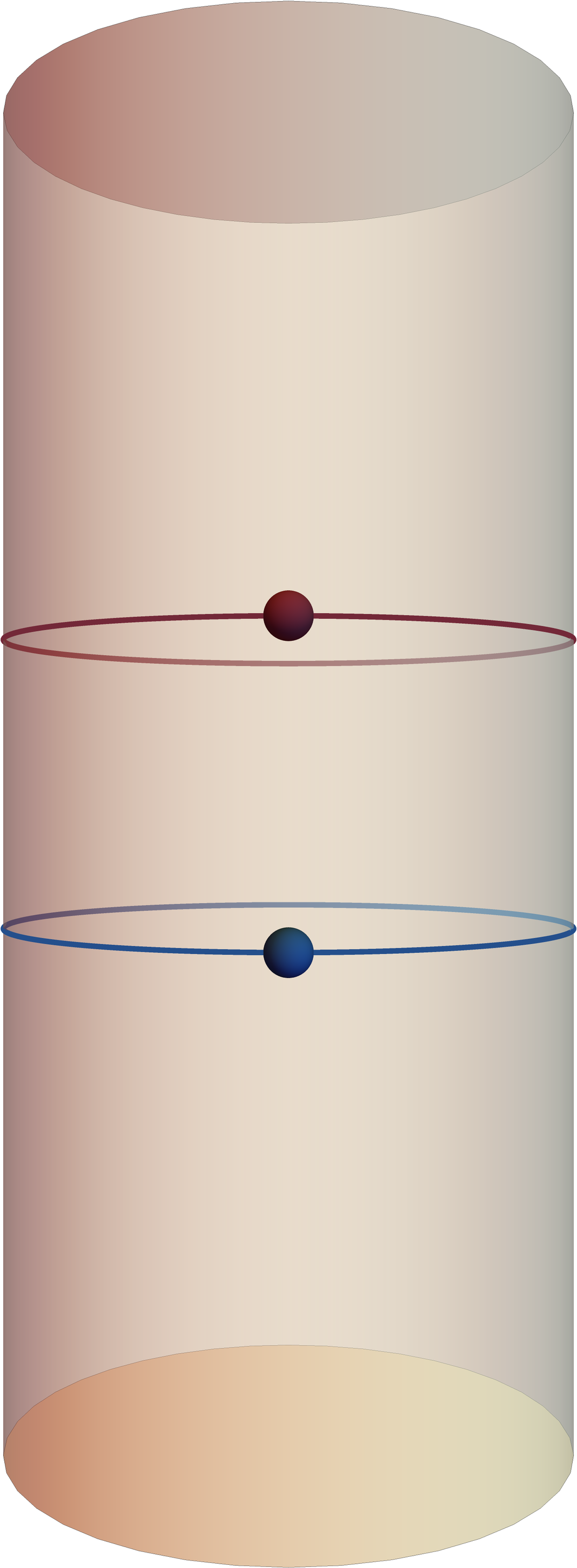}
\end{tabular}
\caption{Case E }
\label{ce}
\end{subfigure}
\end{figure}
\begin{figure}
\ContinuedFloat
\begin{subfigure}{0.9\textwidth}
\begin{tabular}{lcccccccc}
\includegraphics[height=0.13\textheight]{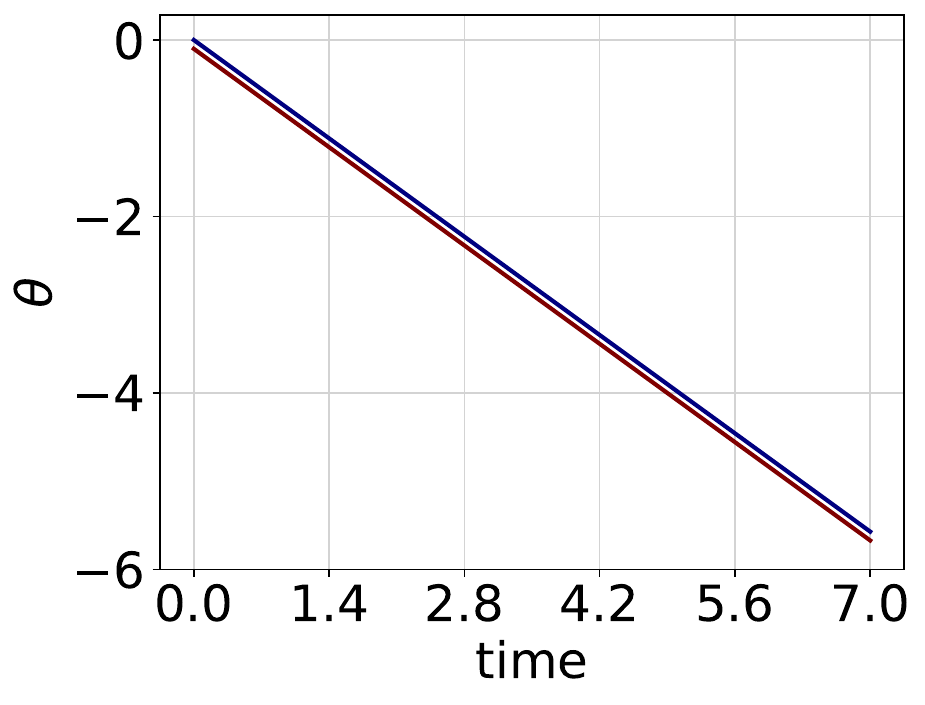}&&
\includegraphics[height=0.13\textheight]{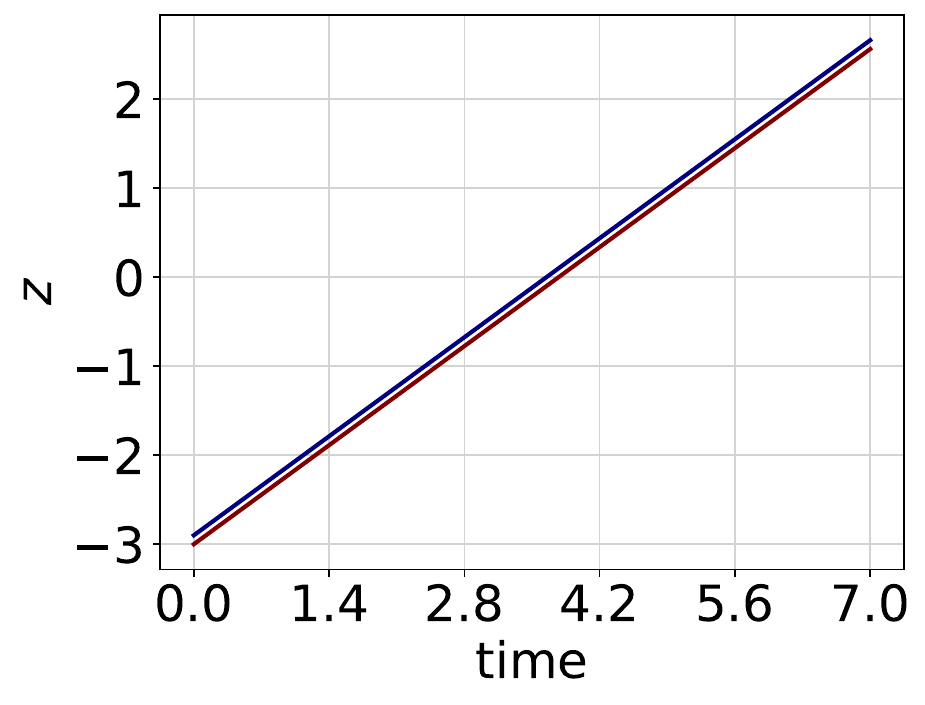}&&
\includegraphics[height=0.13\textheight]{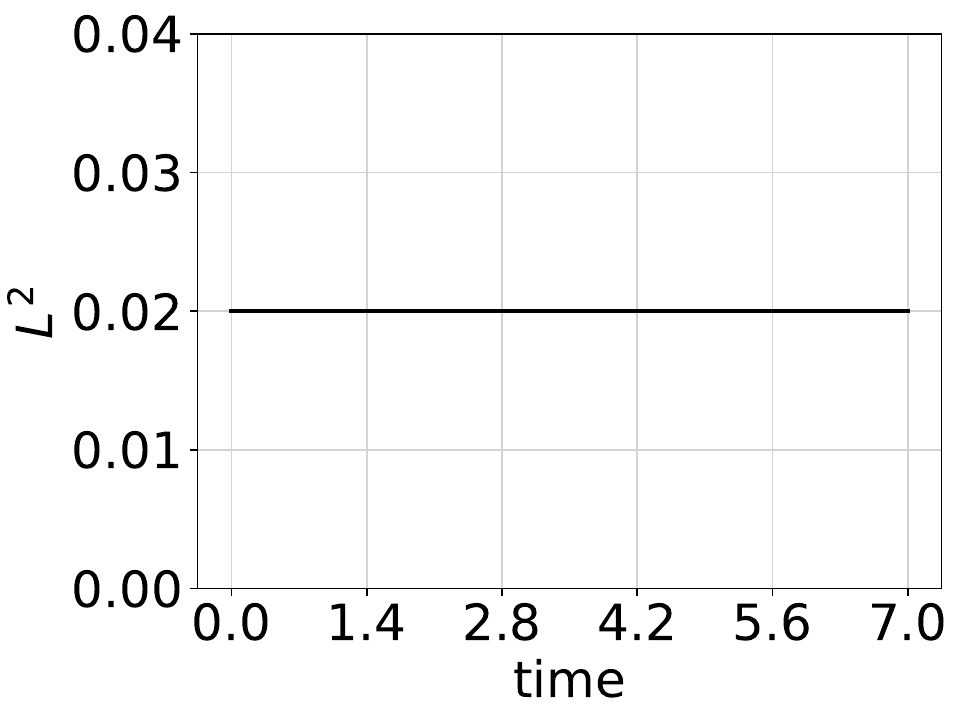}&&\hspace{0.02\textwidth}
\includegraphics[height=0.13\textheight]{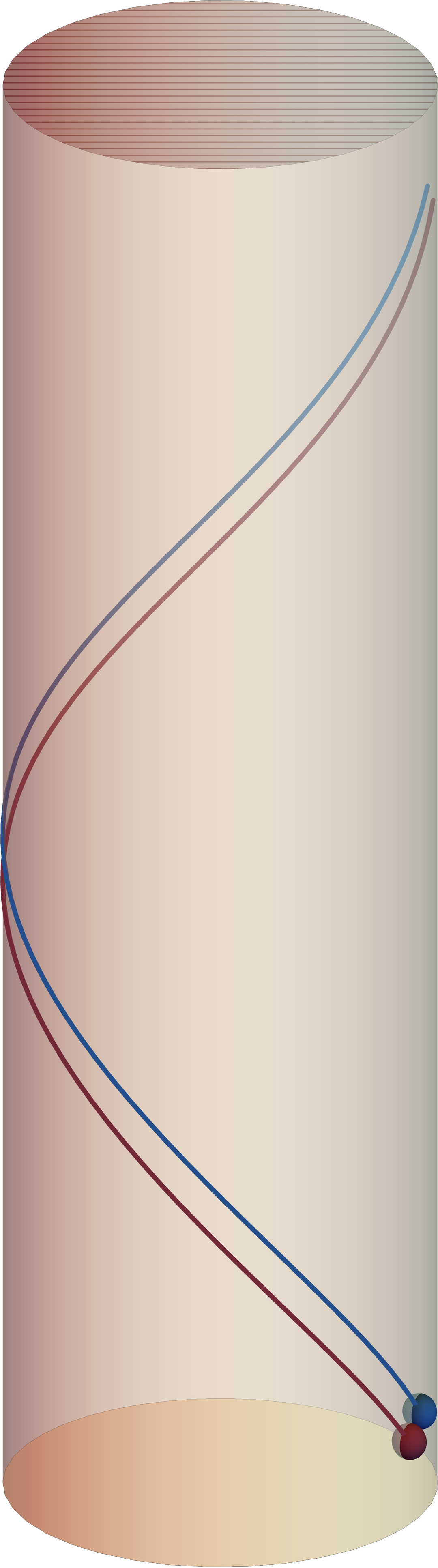}
\end{tabular}
\caption{Case F }
\label{cf}
\end{subfigure}
\caption{Two vortex dynamics in the membrane tube with $\frac{\lambda}{R}=100$. We show variations in vortex locations $(\theta,z)$ and the evolution of $L^2$ defined in Eq.~(\ref{Ldef}) with respect to time. For the 3D cylinder plot appearing on the extreme right, blue dots indicate initial location of vortices with positive circulation, while red dots indicate vortices with negative circulation. The same color code is used for the vortex trajectories.}
\end{figure}
\begin{figure}
\centering
\begin{subfigure}{0.9\textwidth}
\begin{tabular}{lcccccccc}
\includegraphics[height=0.13\textheight]{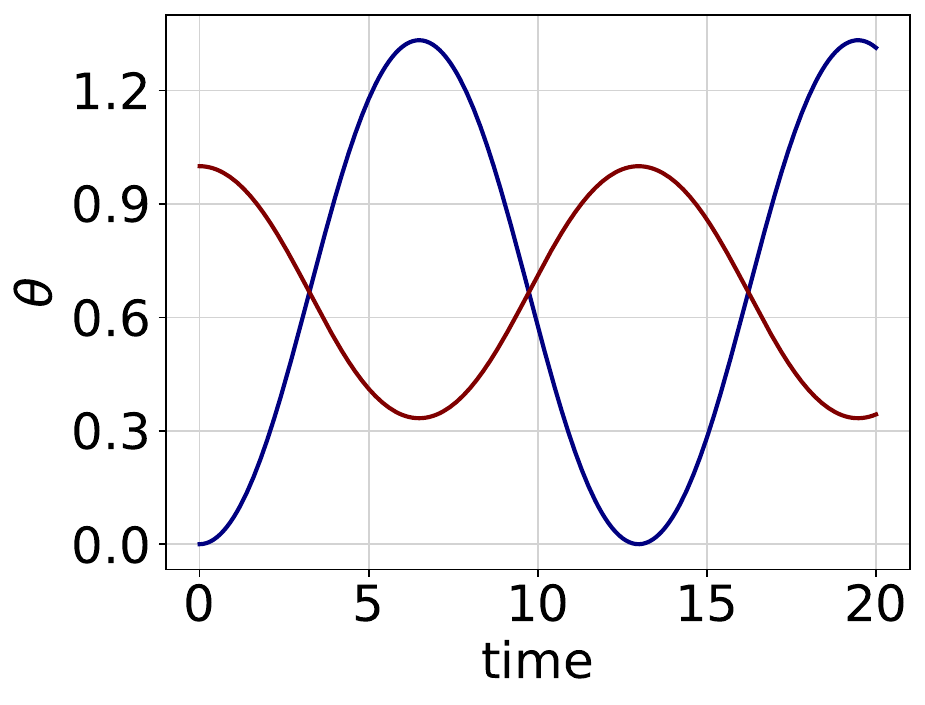}&&
\includegraphics[height=0.13\textheight]{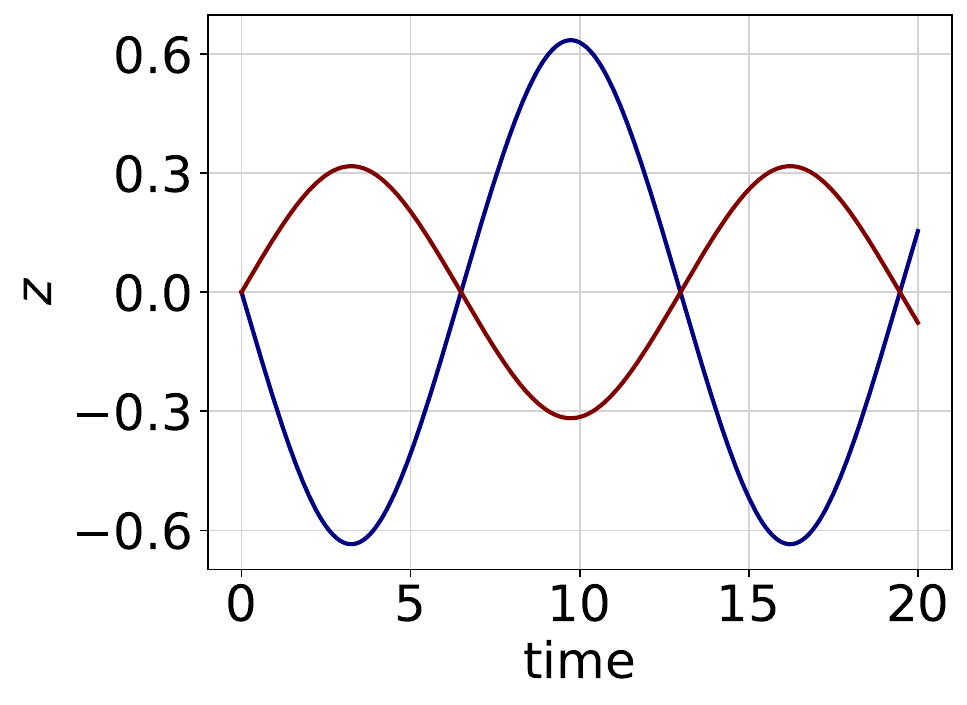}&&
\includegraphics[height=0.13\textheight]{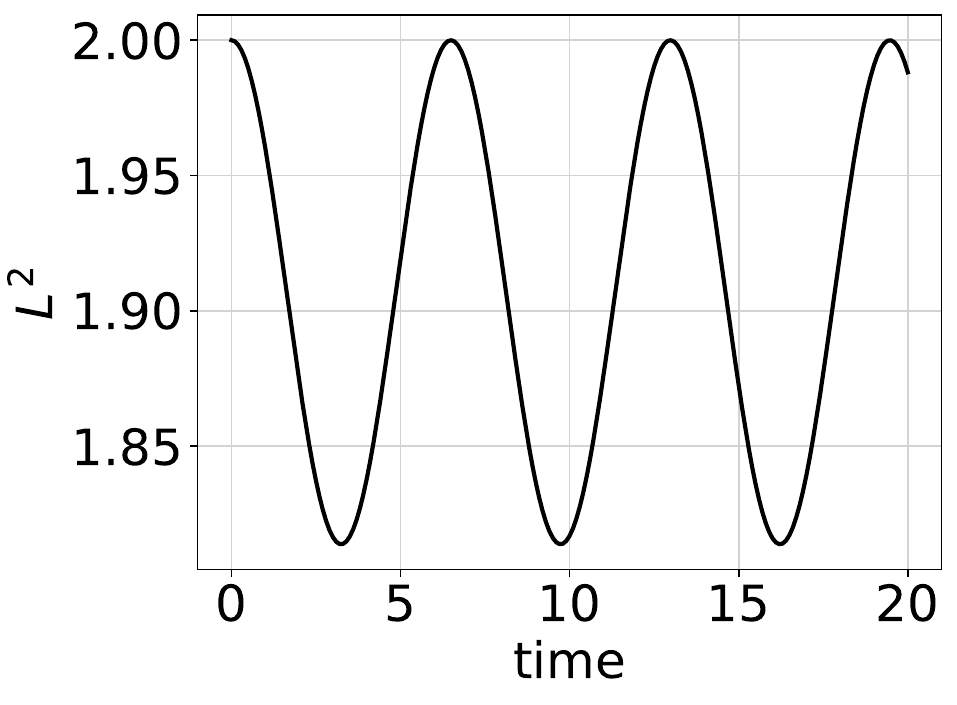}&&\hspace{0.02\textwidth}
\includegraphics[height=0.13\textheight]{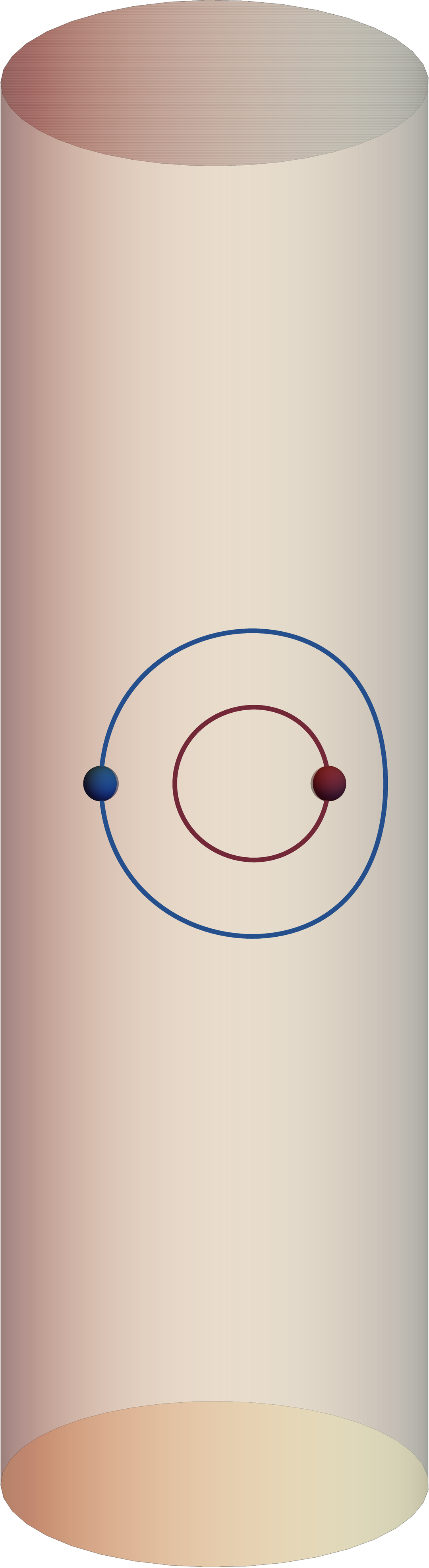}
\end{tabular}
\caption{Same configuration as Fig.~(\ref{ca}) but with unequal vortex strengths of same sign.}
\label{figuneq}
\end{subfigure}\vskip1ex
\begin{subfigure}{0.9\textwidth}
\begin{tabular}{lcccccccc}
\includegraphics[height=0.13\textheight]{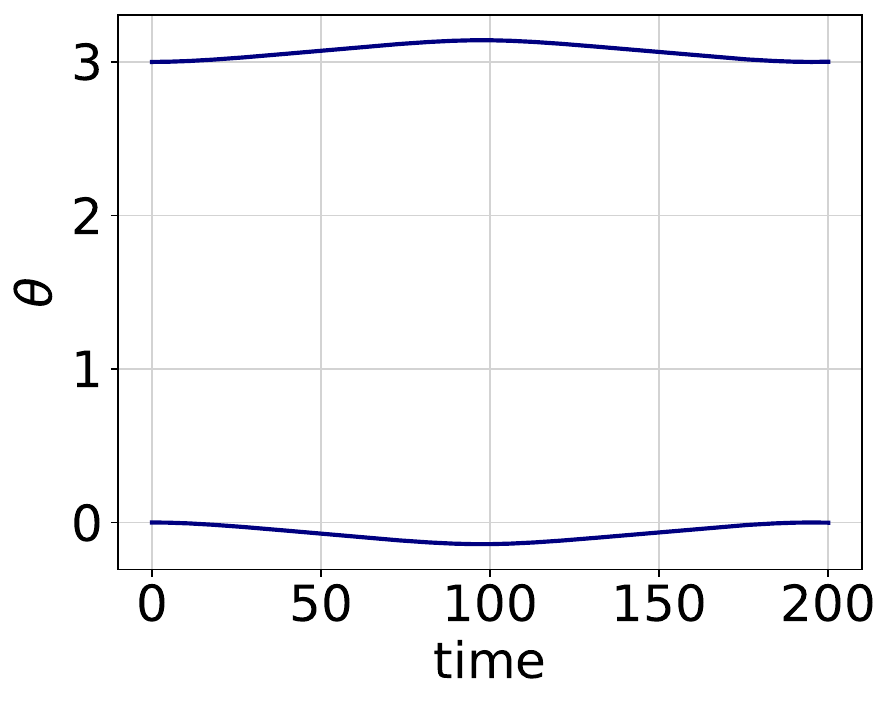}&&
\includegraphics[height=0.13\textheight]{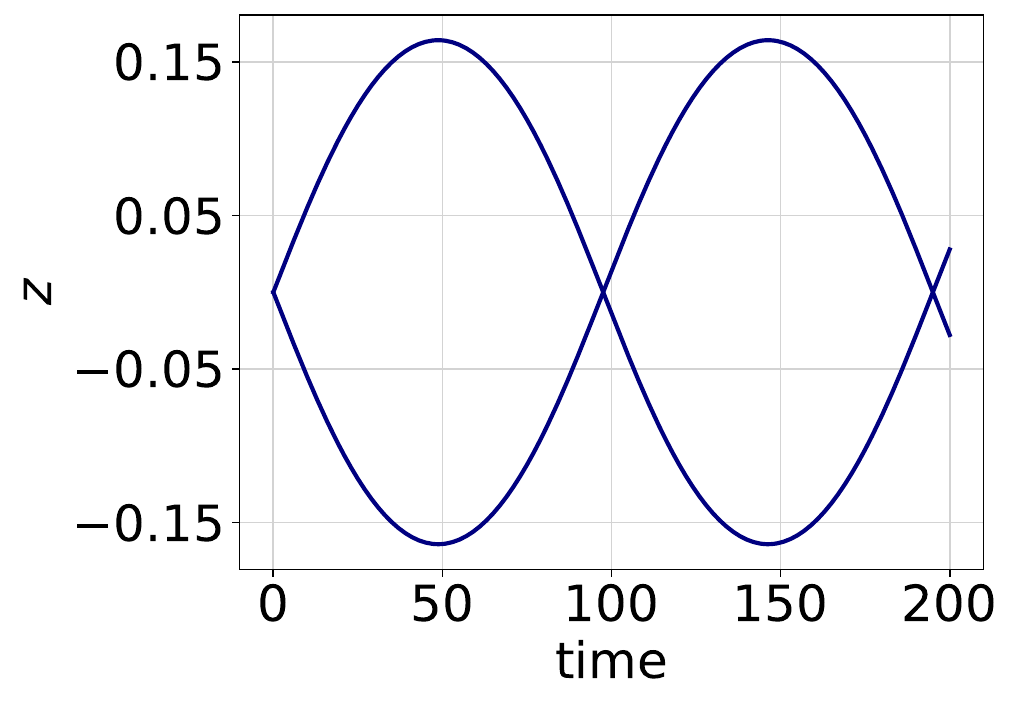}&&
\includegraphics[height=0.13\textheight]{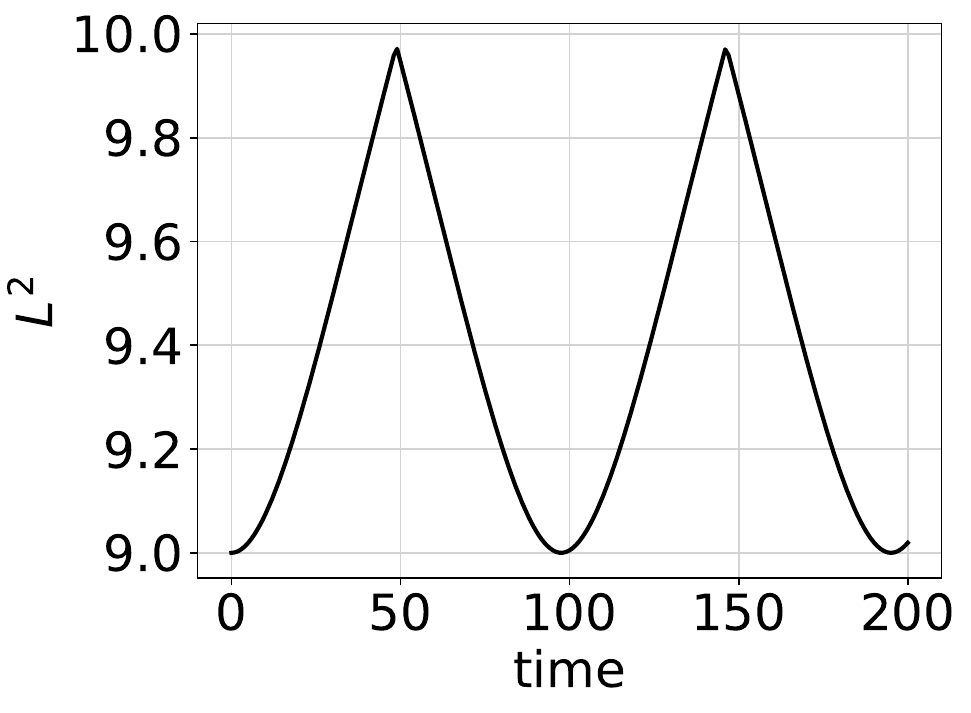}&&\hspace{0.02\textwidth}
\includegraphics[height=0.13\textheight]{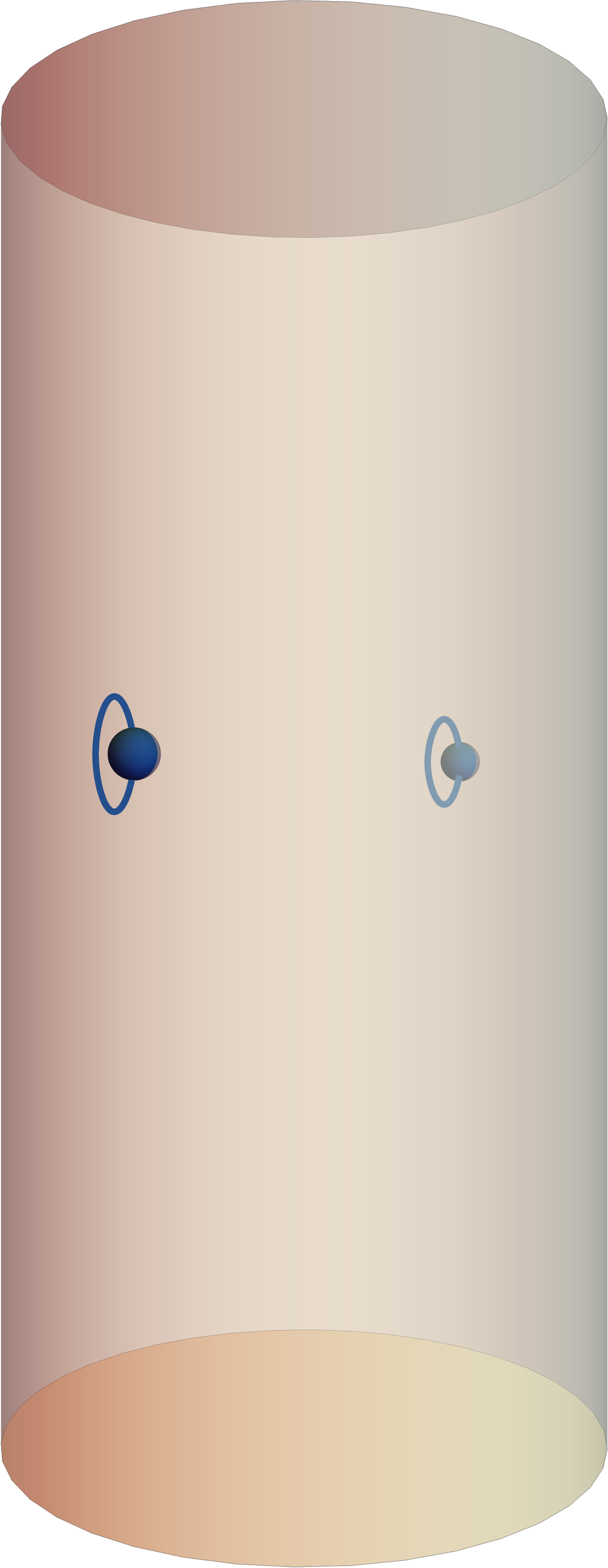}
\end{tabular}
 \caption{ Same configuration as Fig.~(\ref{cb}) but with $\frac{\lambda}{R} =5$ }
 \label{figlc}
 \end{subfigure} 
 \caption{ Two vortex dynamics with tuning of parameters: In each row, we show the vortex locations $(\theta,z)$ and the evolution of $L^2$ defined in Eq.~(\ref{Ldef}) with respect to time. The 3D cylinder plot appearing on the extreme right,shows the initial vortex locations (blue dots) as well as the vortex trajectories.}
\end{figure}

\begin{figure}
\centering
\begin{subfigure}{0.9\textwidth}
\begin{tabular}{lcccccccc}
\includegraphics[height=0.13\textheight]{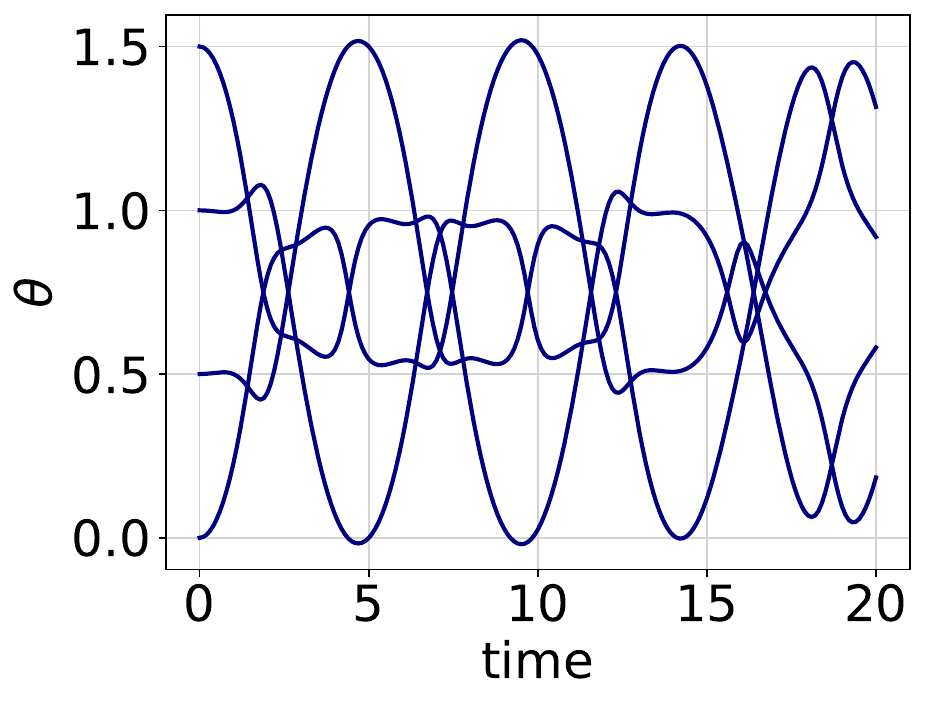}&&
\includegraphics[height=0.13\textheight]{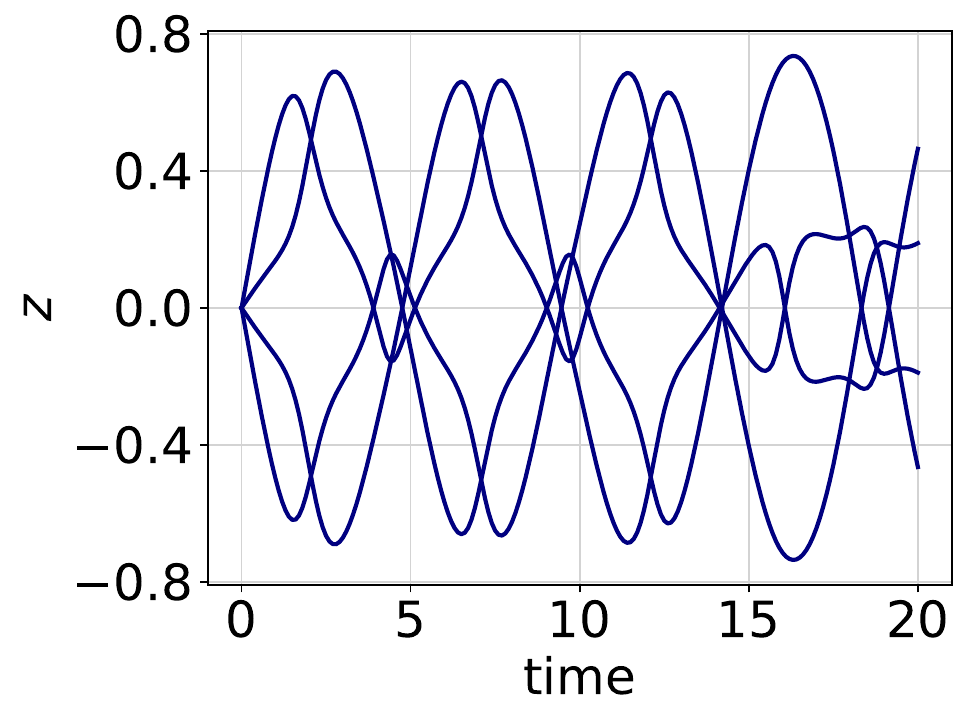}&&
\includegraphics[height=0.13\textheight]{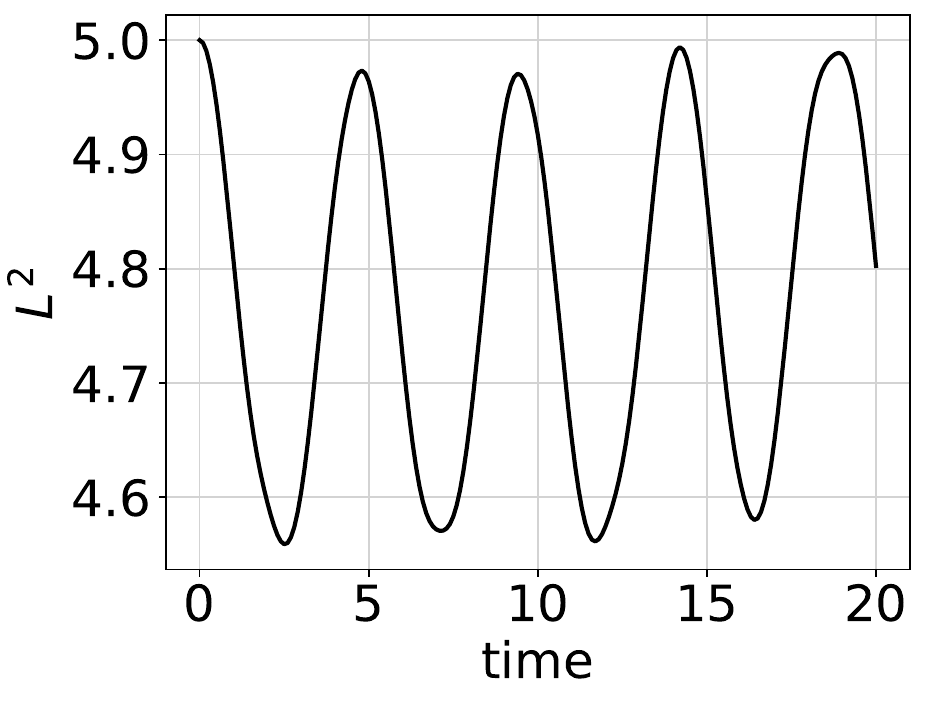}&&\hspace{0.02\textwidth}
\includegraphics[height=0.13\textheight]{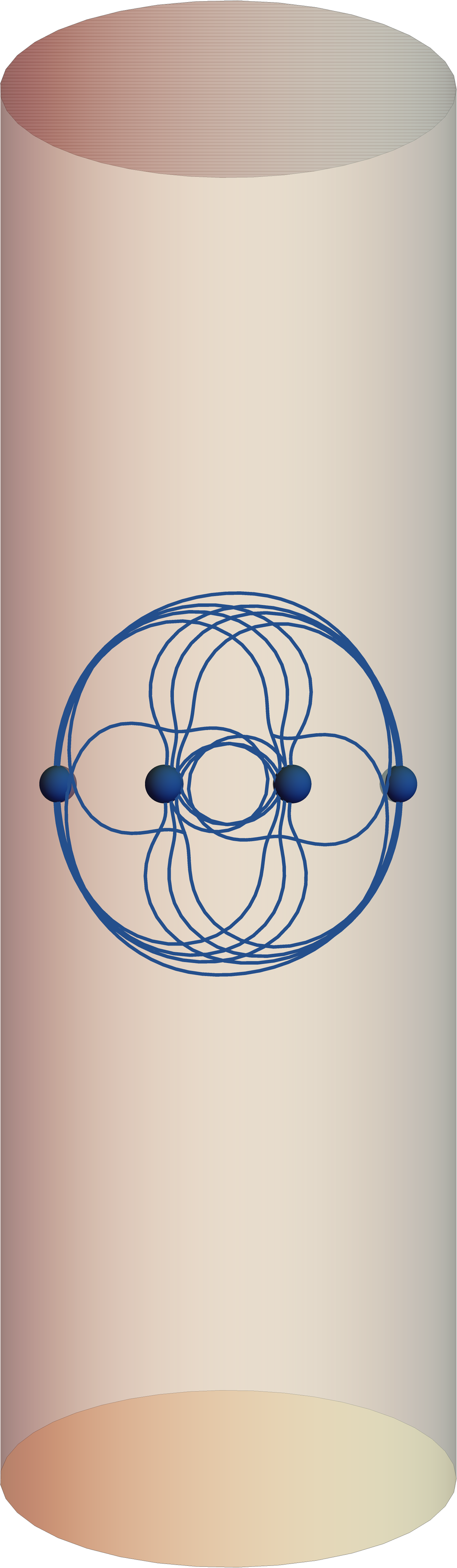}
\end{tabular}
\caption{Case A }
\label{mvca}
\end{subfigure}

\begin{subfigure}{0.9\textwidth}
\begin{tabular}{lcccccccc}
\includegraphics[height=0.13\textheight]{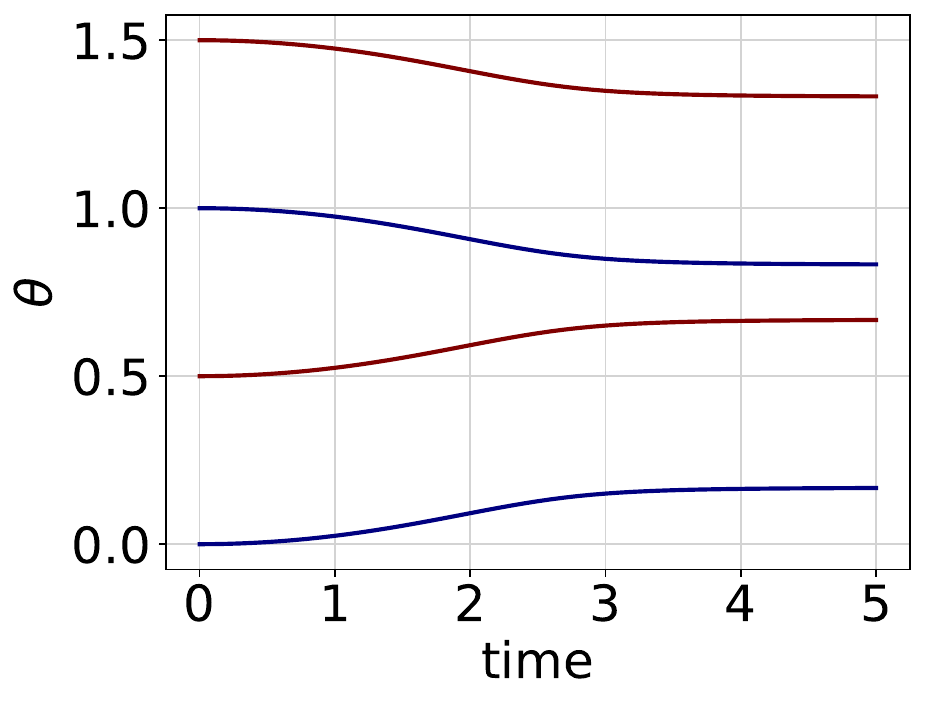}&&
\includegraphics[height=0.13\textheight]{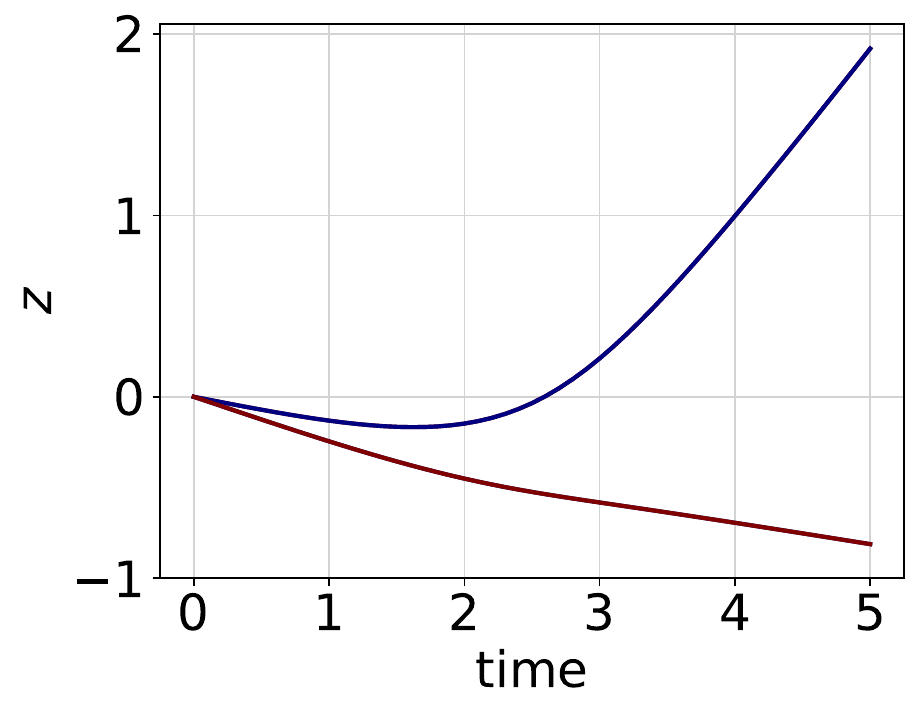}&&
\includegraphics[height=0.13\textheight]{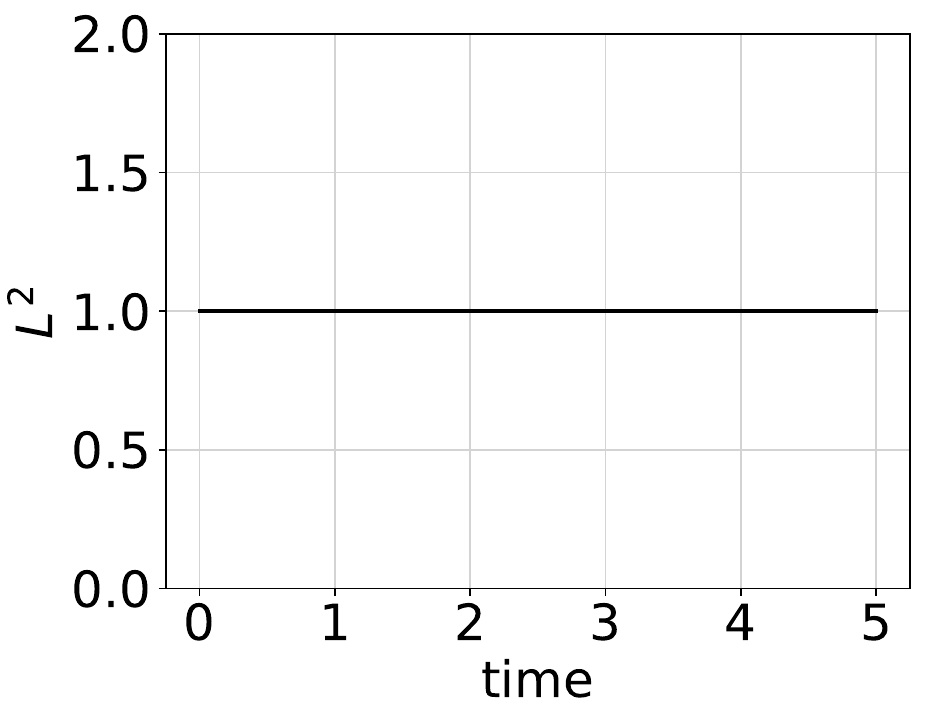}&&\hspace{0.02\textwidth}
\includegraphics[height=0.13\textheight]{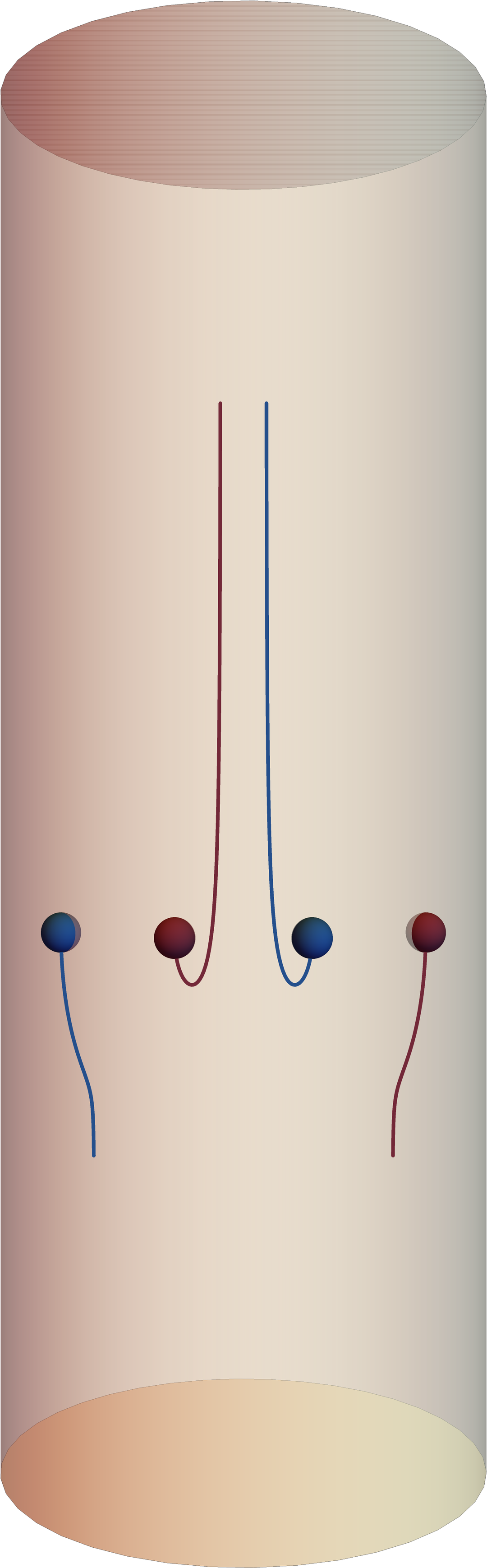}
\end{tabular}
\caption{Case B }
\label{mvcb}
\end{subfigure}

\begin{subfigure}{0.9\textwidth}
\begin{tabular}{lcccccccc}
\includegraphics[height=0.13\textheight]{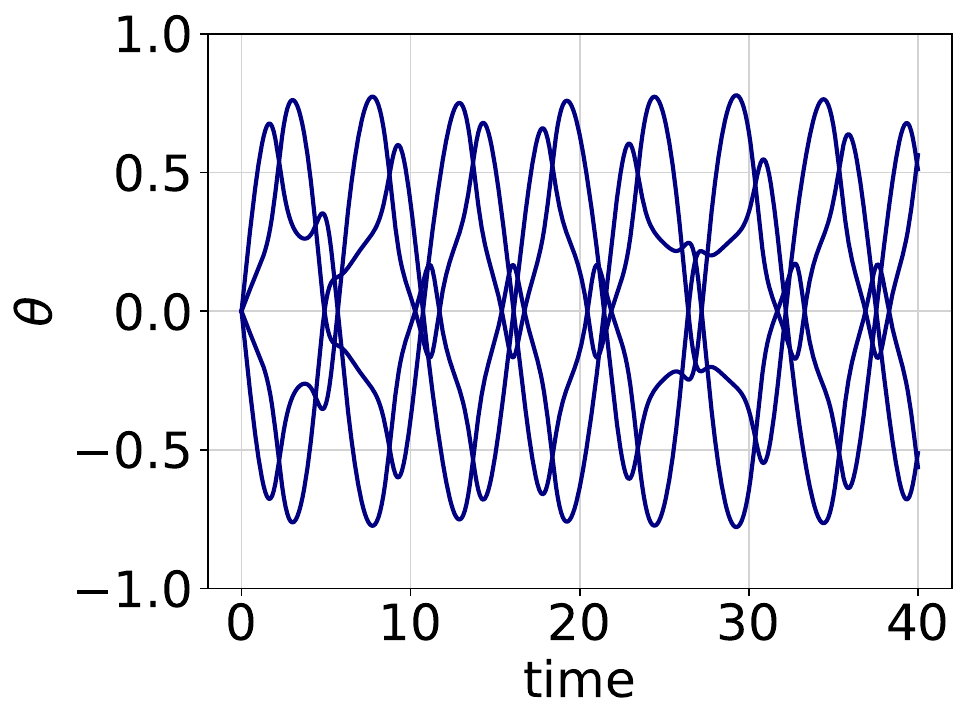}&&
\includegraphics[height=0.13\textheight]{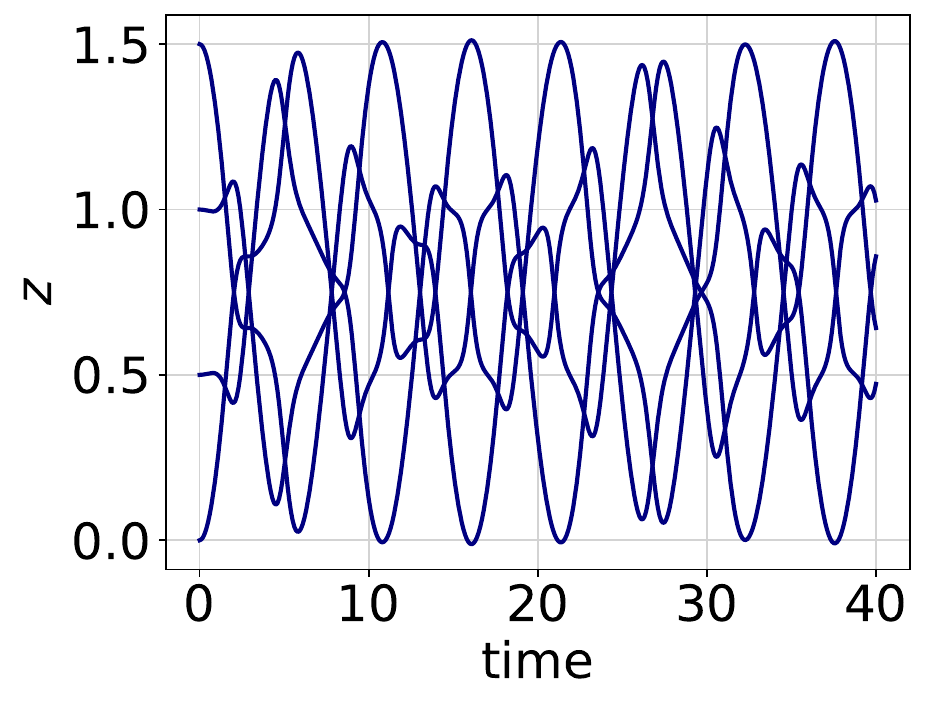}&&
\includegraphics[height=0.13\textheight]{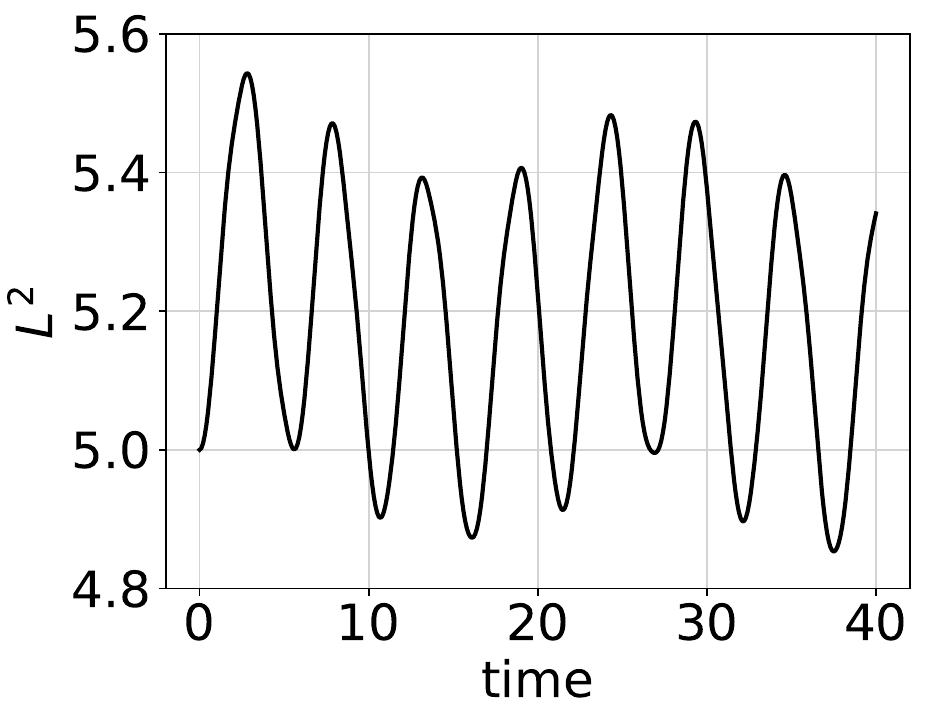}&&\hspace{0.02\textwidth}
\includegraphics[height=0.13\textheight]{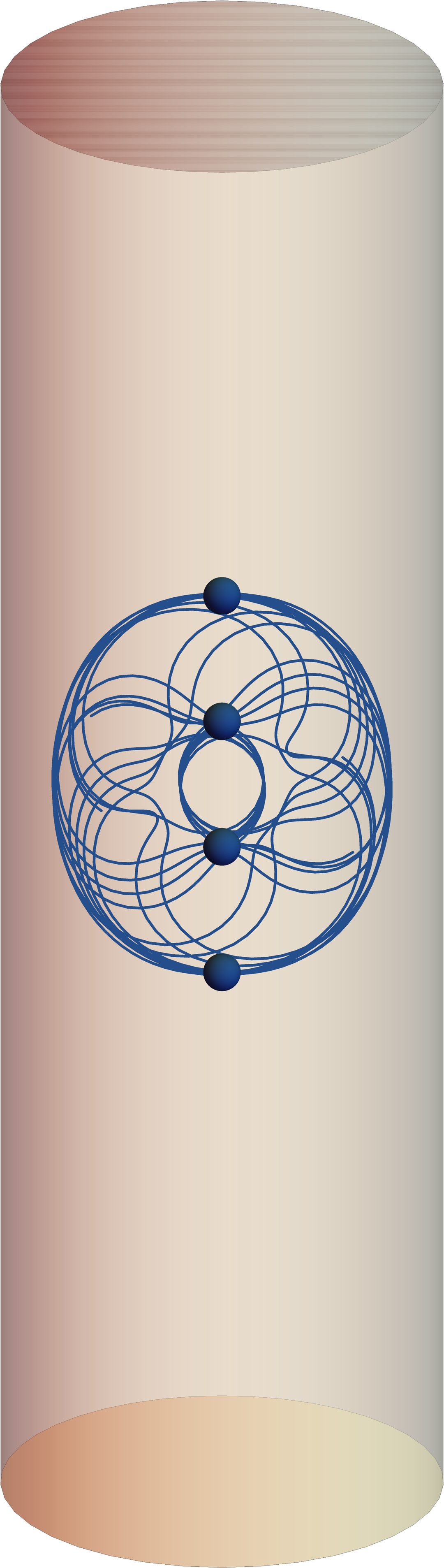}
\end{tabular}
\caption{Case C }
\label{mvcc}
\end{subfigure}

\begin{subfigure}{0.9\textwidth}
\begin{tabular}{lcccccccc}
\includegraphics[height=0.13\textheight]{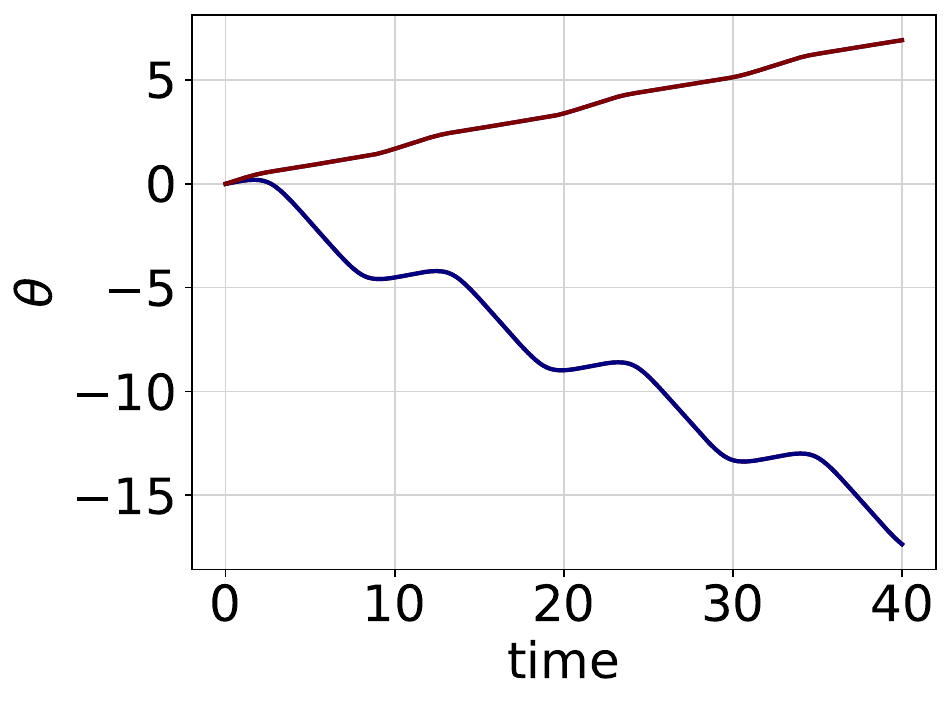}&&
\includegraphics[height=0.13\textheight]{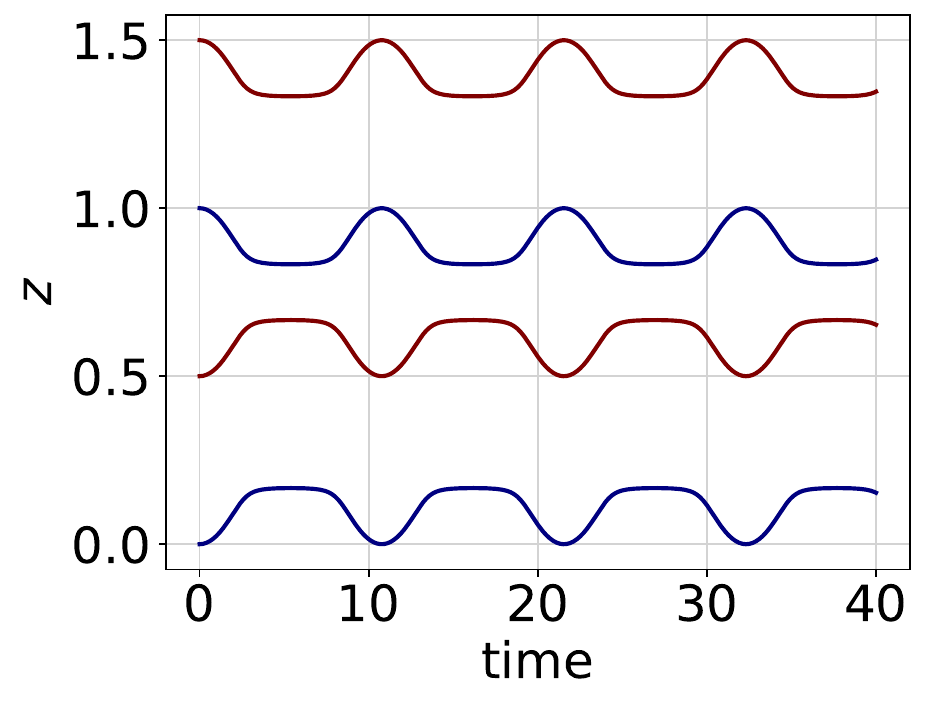}&&
\includegraphics[height=0.13\textheight]{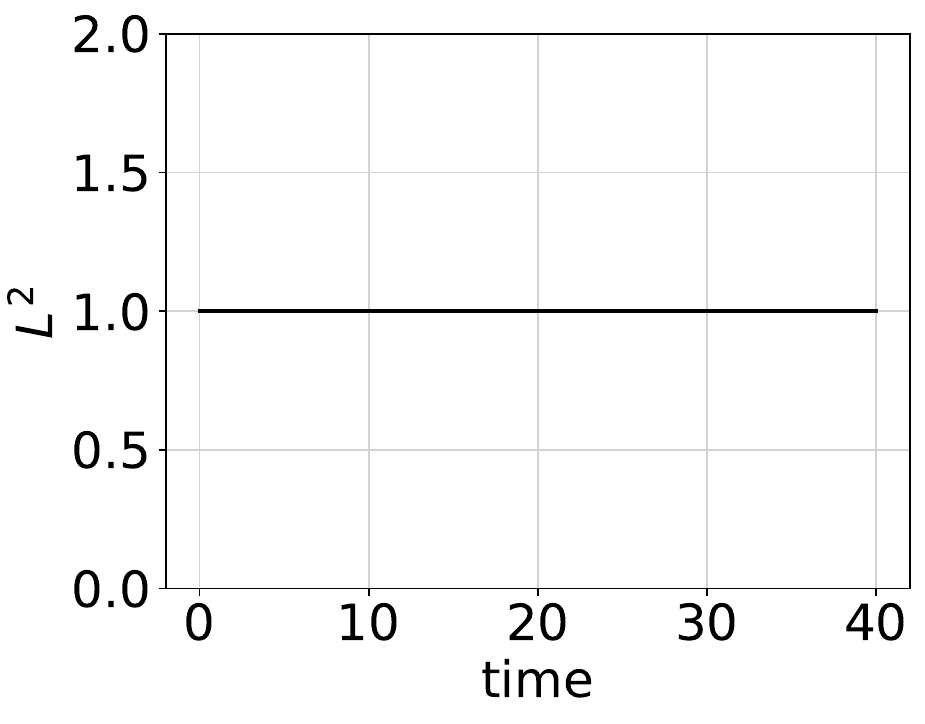}&&\hspace{0.02\textwidth}
\includegraphics[height=0.13\textheight]{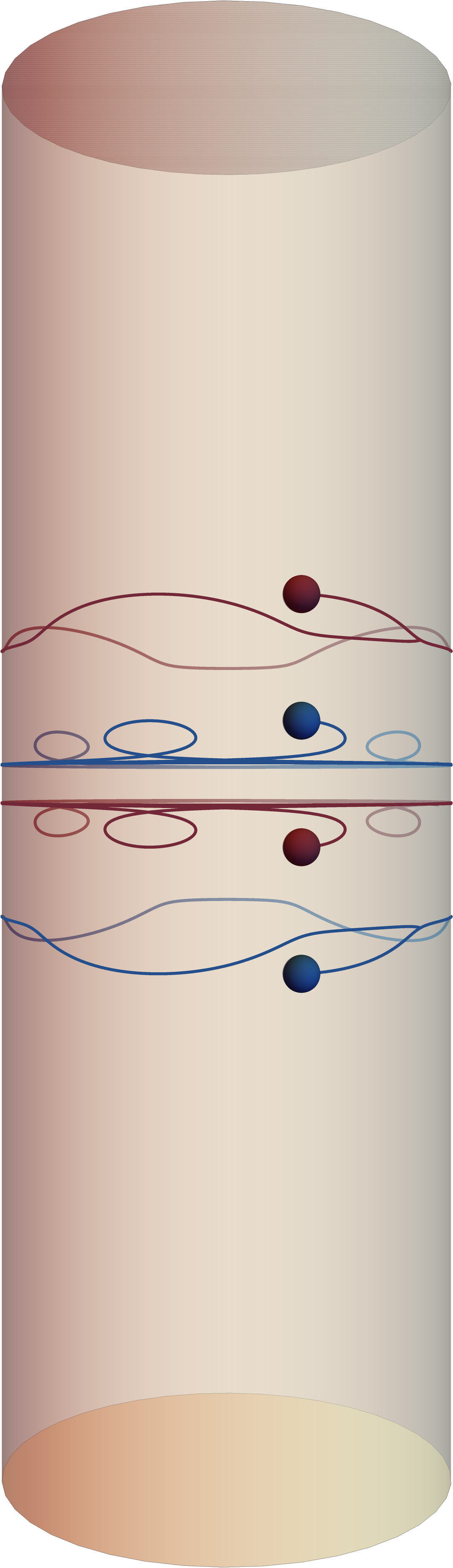}
\end{tabular}
\caption{Case D }
\label{mvcd}
\end{subfigure}

\caption{Multi-vortex dynamics in the membrane tube :  In each row, we show the locations $(\theta,z)$ of all 4 vortices and the evolution of $L^2$ defined in Eq.~(\ref{Ldef}) with respect to time. The 3D cylinder plot appearing on the extreme right, shows the initial vortex locations as well as the vortex trajectories. Blue dots indicate initial location of vortices with positive circulation, while red dots indicate vortices with negative circulation. The same color code is used for the vortex trajectories.}
\end{figure}

\begin{figure}[htbp!]
\begin{tabular}{lcccccccc}
\includegraphics[height=0.13\textheight]{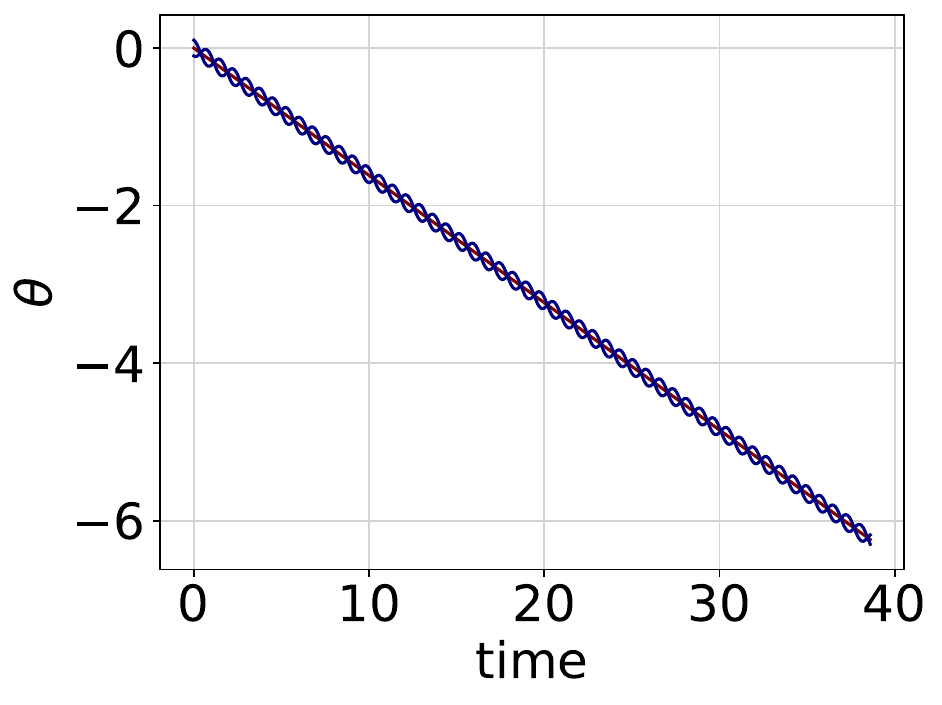}&&
\includegraphics[height=0.13\textheight]{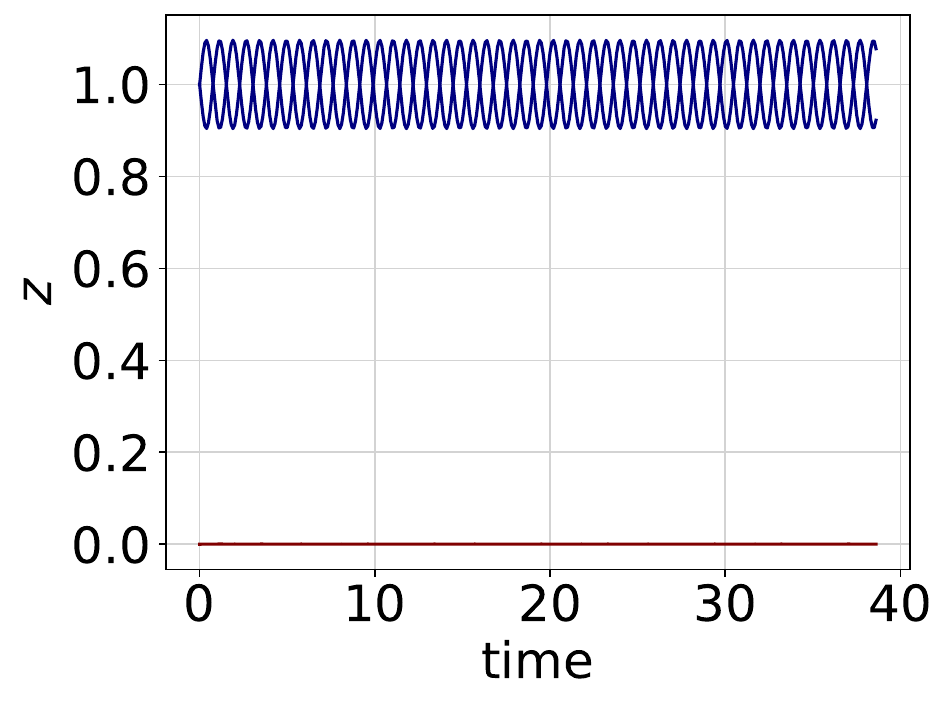}&&
\includegraphics[height=0.13\textheight]{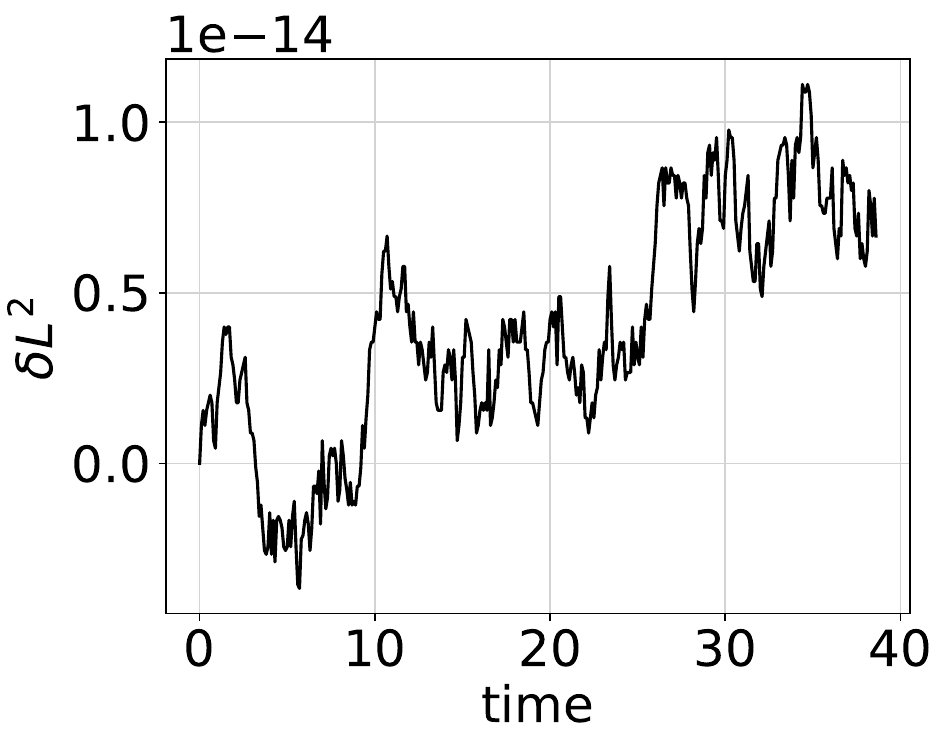}&&\hspace{0.02\textwidth}
\includegraphics[height=0.13\textheight]{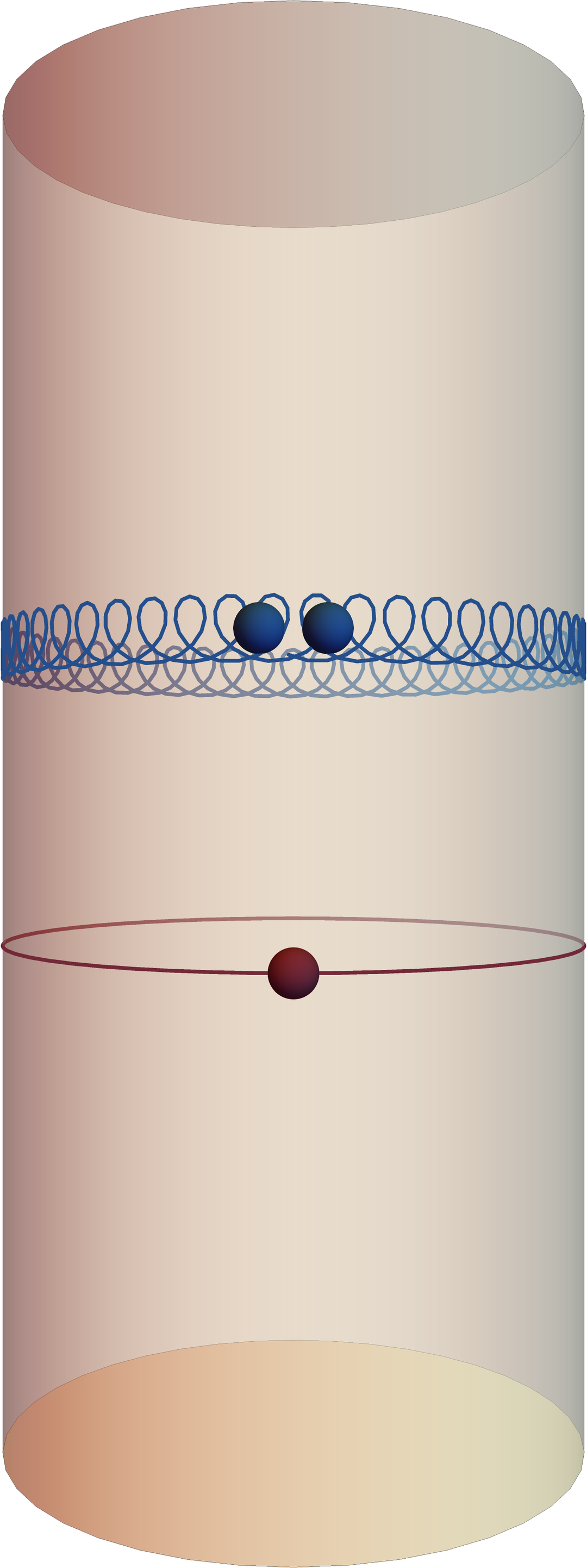}\\
\includegraphics[height=0.13\textheight]{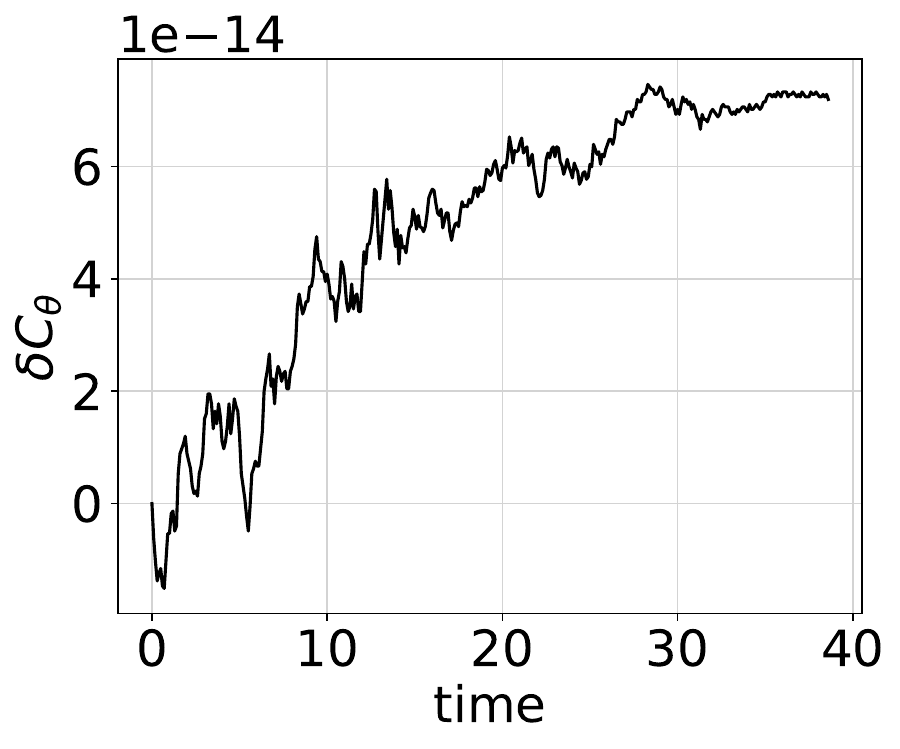}&&
\includegraphics[height=0.13\textheight]{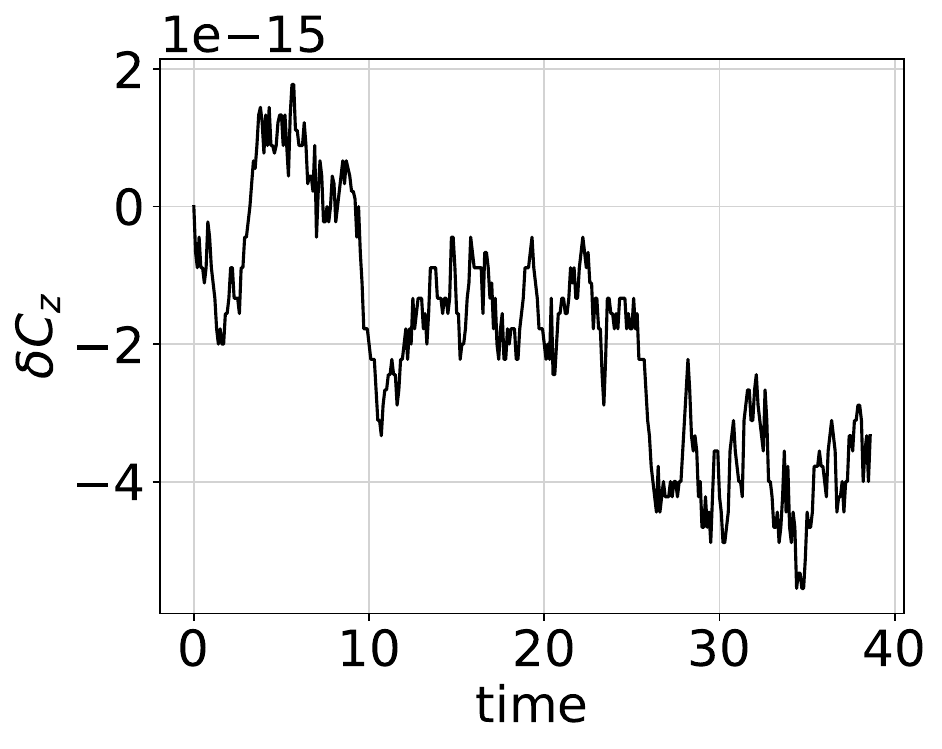}&&
\end{tabular}
 \caption{ Vortex leapfrogging for $\frac{\lambda}{R} =100$ in a 3 vortex system. The vortex with positive circulation $\tau=+1$ is marked red,  while  two vortices of circulation $\tau=-\frac{1}{2}$ each is shown in blue, with the same color code for trajectories. Numerical errors in $L^2, C_\theta, C_z$ are also shown. }
 \label{lpf0} 
\end{figure}

\begin{figure}[htbp!]
\begin{tabular}{lcccccccc}
\includegraphics[height=0.13\textheight]{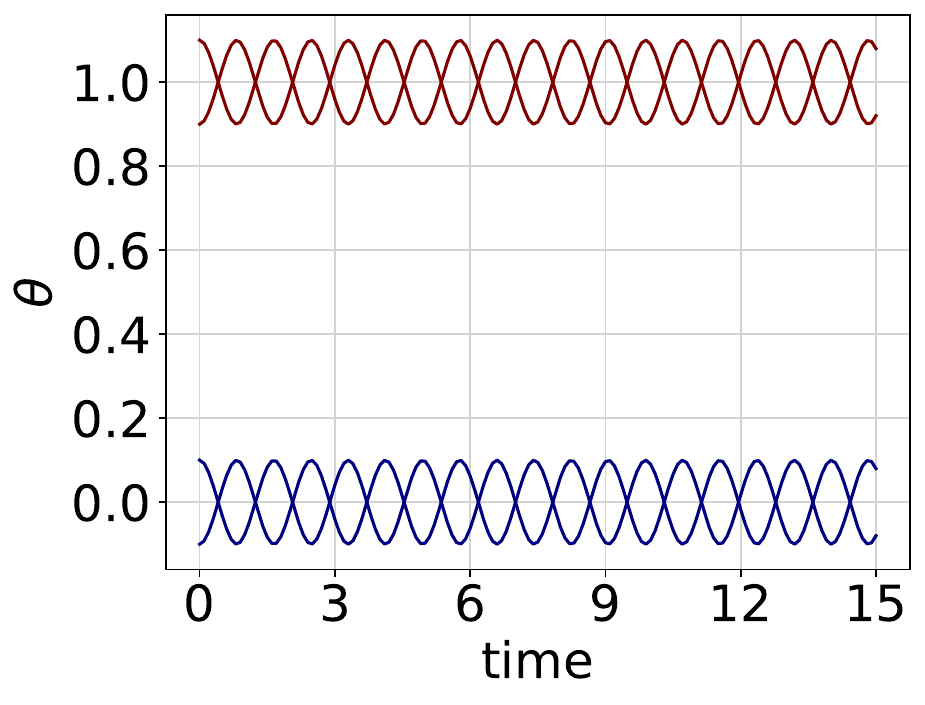}&&
\includegraphics[height=0.13\textheight]{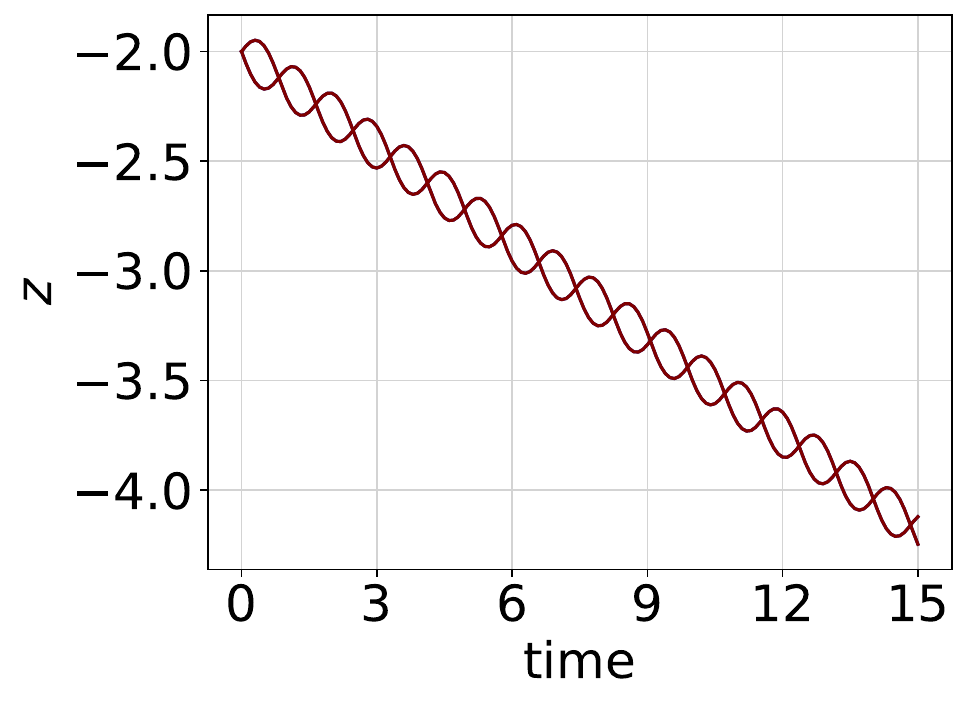}&&
\includegraphics[height=0.13\textheight]{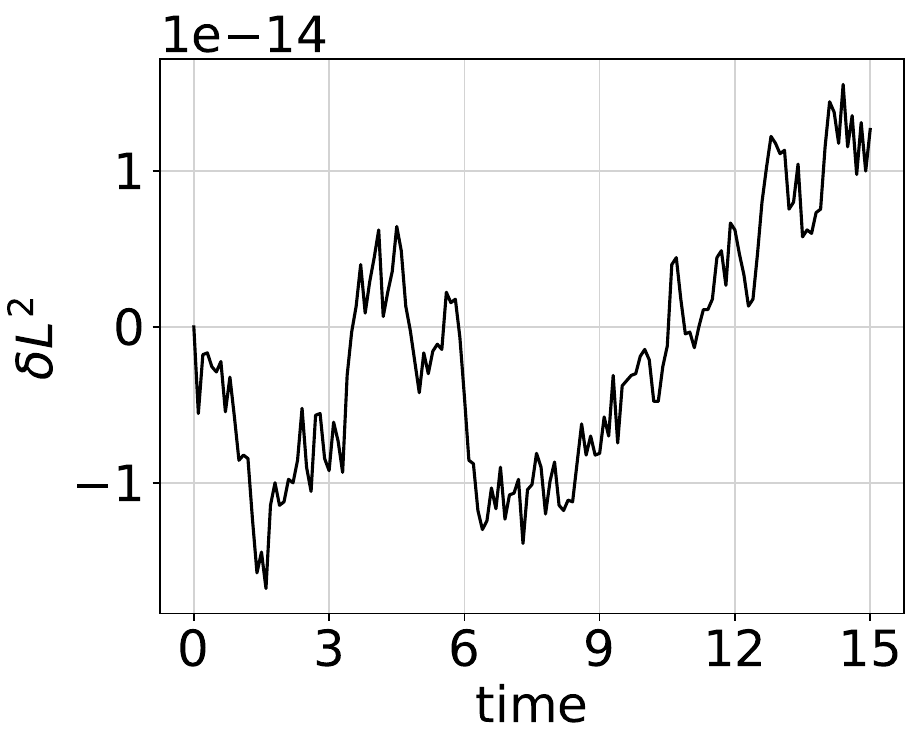}&&\hspace{0.02\textwidth}
\includegraphics[height=0.13\textheight]{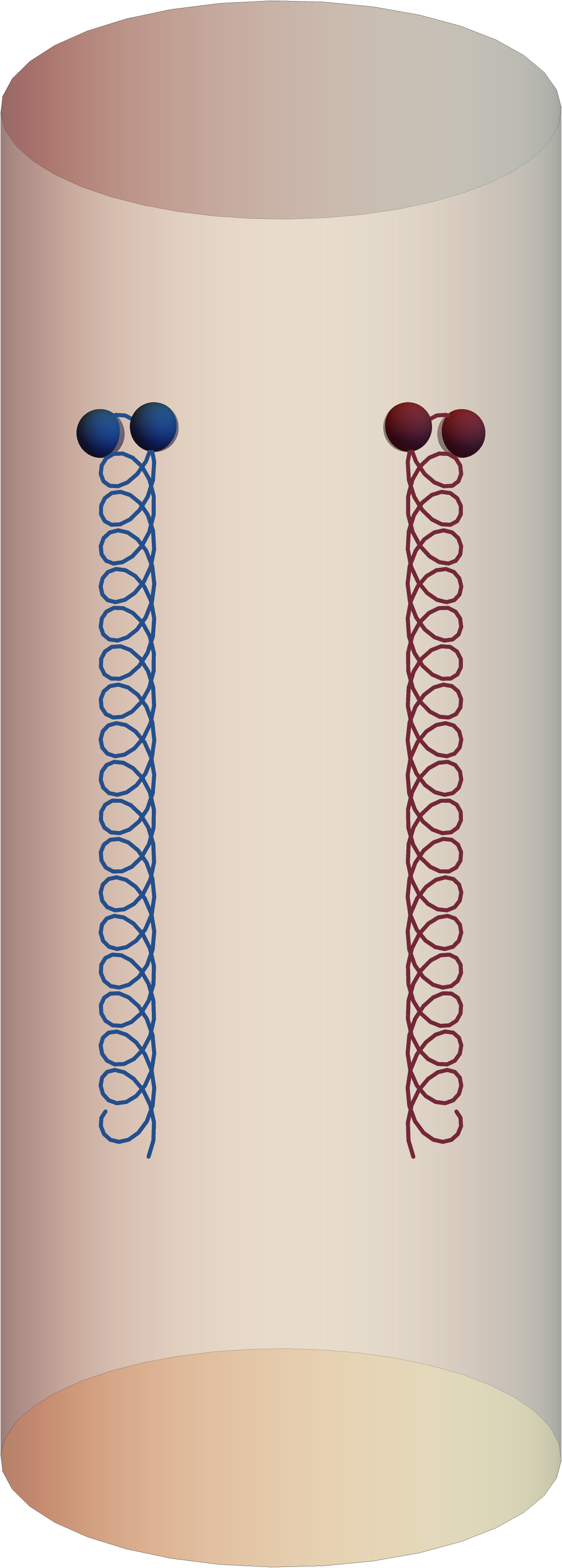}\\
\includegraphics[height=0.13\textheight]{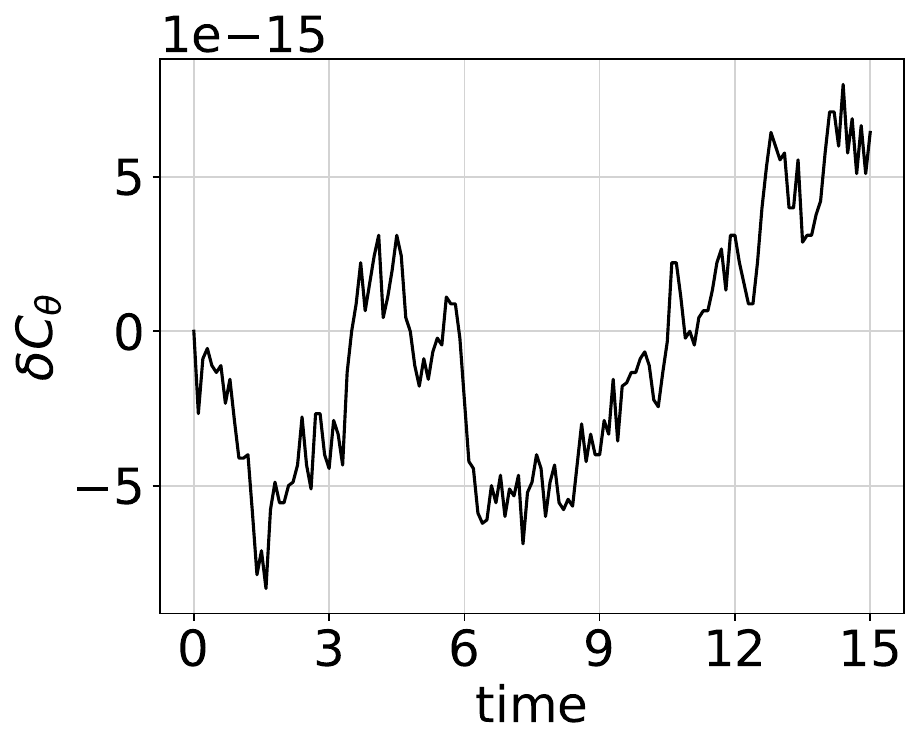}&&
\includegraphics[height=0.13\textheight]{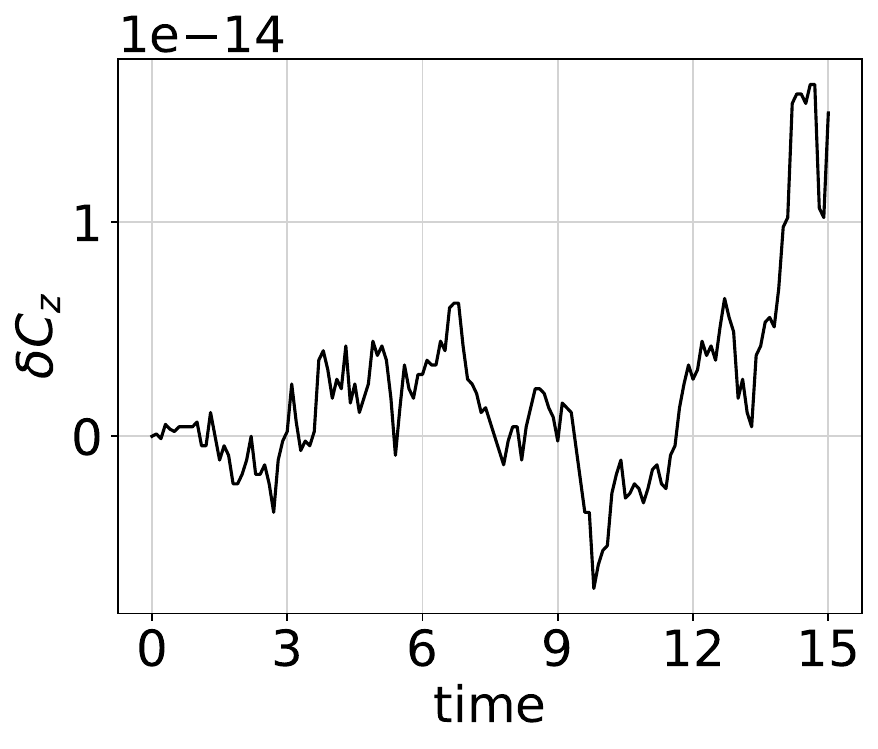}&&
\end{tabular}
 \caption{ Vortex leapfrogging for $\frac{\lambda}{R} =100$ in a 4 vortex system. Two vortices with positive circulation $\tau=+1/2$ each are marked red,  while  two vortices of circulation $\tau=-\frac{1}{2}$ each are shown in blue, with the same color code for trajectories. Numerical errors in $L^2, C_\theta, C_z$ are also shown. }
 \label{lpf1} 
\end{figure}

\begin{figure}[htbp!]
\begin{tabular}{lcccccccc}
\includegraphics[height=0.13\textheight]{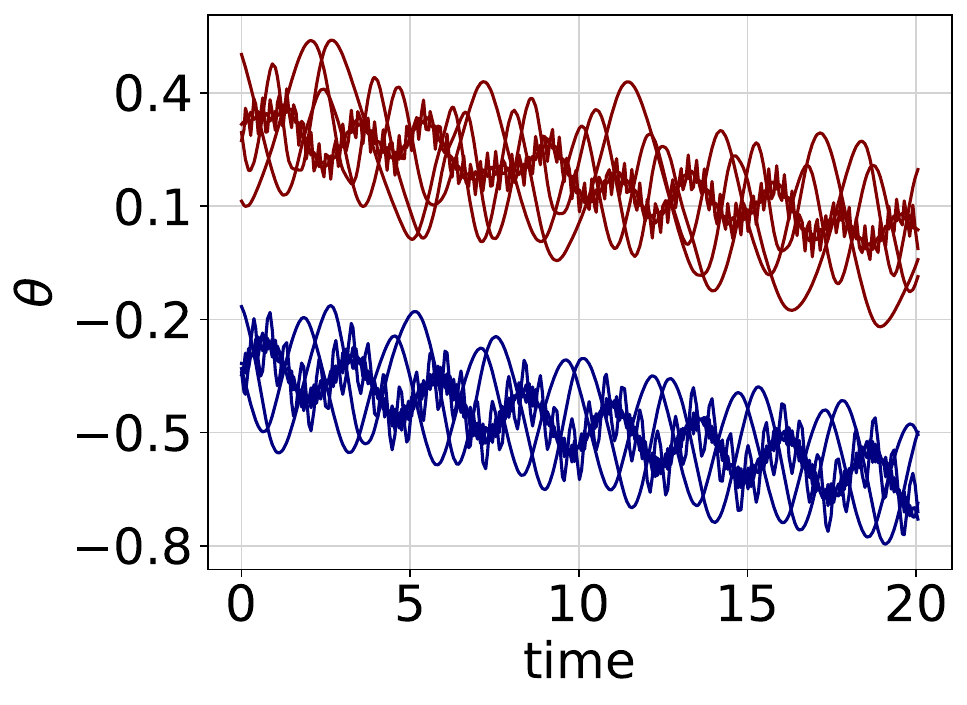}&&
\includegraphics[height=0.13\textheight]{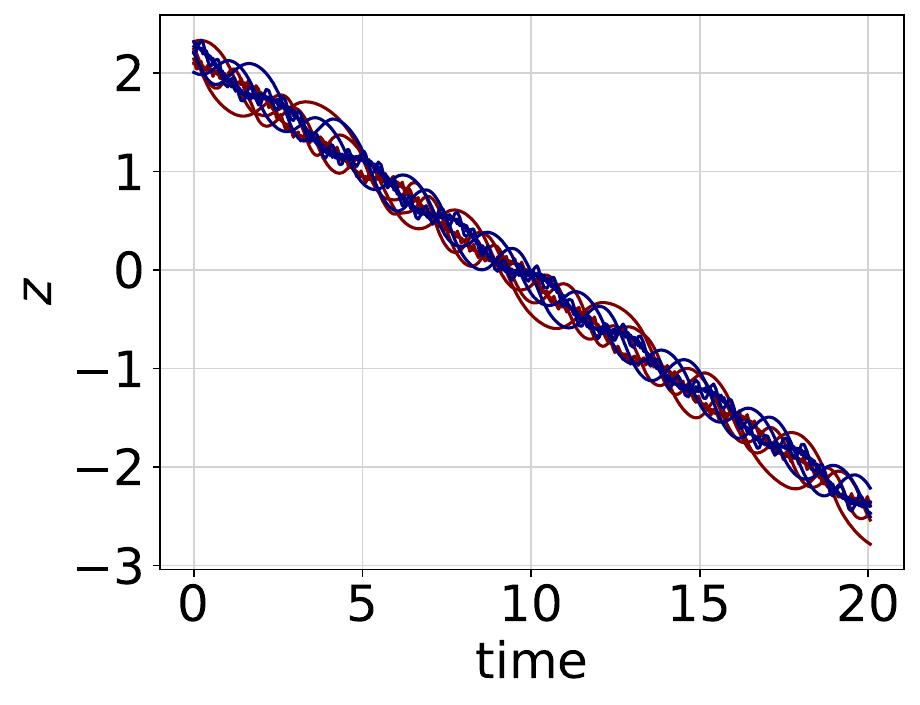}&&
\includegraphics[height=0.13\textheight]{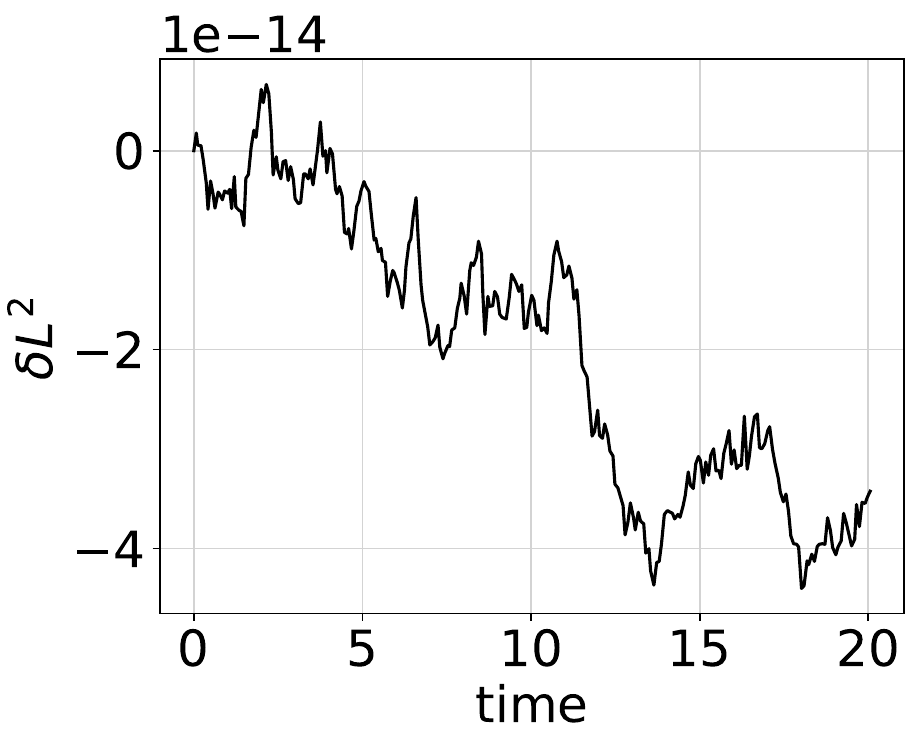}&&\hspace{0.02\textwidth}
\includegraphics[height=0.13\textheight]{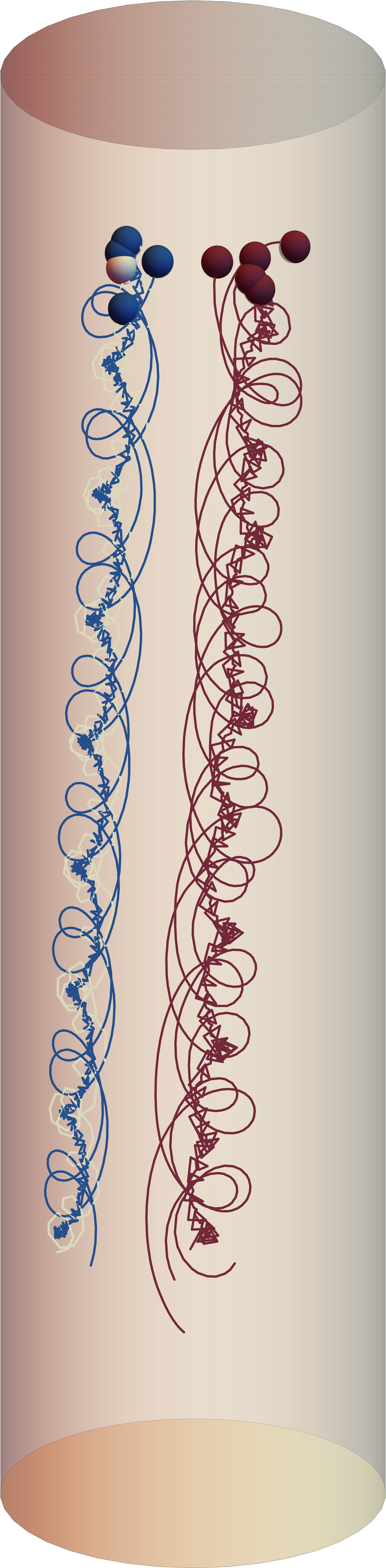}\\
\includegraphics[height=0.13\textheight]{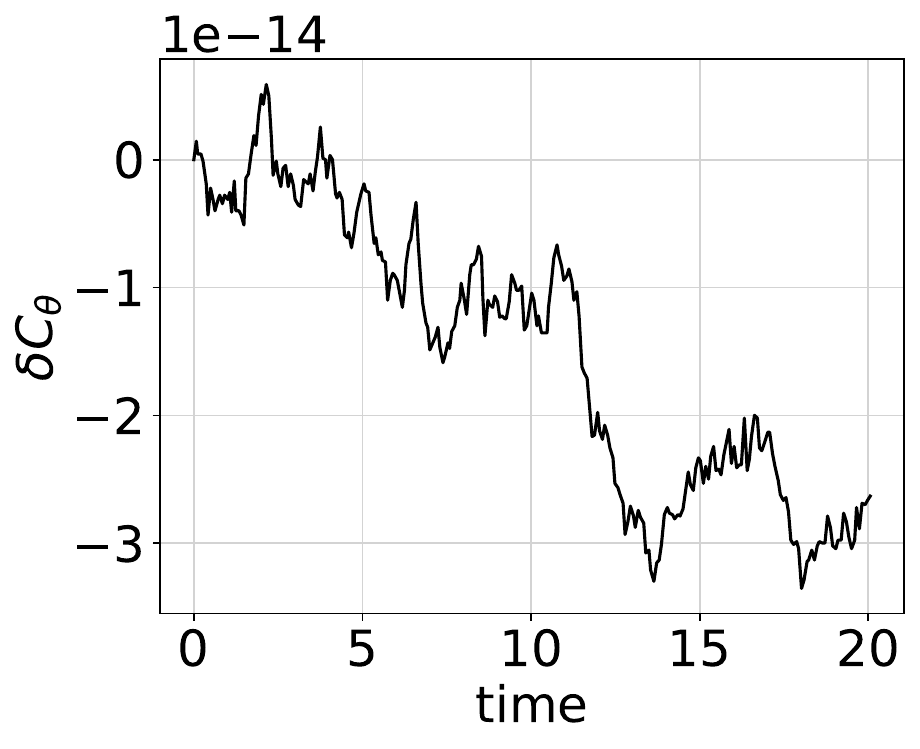}&&
\includegraphics[height=0.13\textheight]{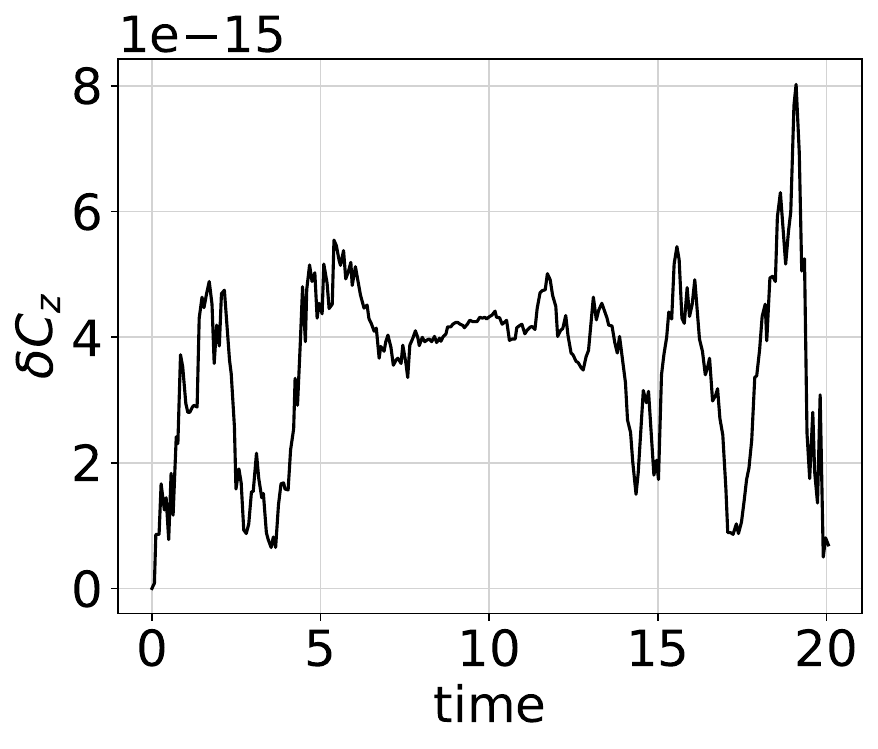}&&
\end{tabular}
 \caption{ Vortex leapfrogging for $\frac{\lambda}{R} =100$ in a 10 vortex system. Five vortices with random splittings of vortex strengths adding up to a net positive circulation $\sum_i\tau_i=+1$  are marked red,  while  remain vortices with random splittings of vortex strengths adding up to a net negative circulation $\sum_i\tau_i=-1$ are shown in blue, with the same color code for trajectories. Numerical errors in $L^2, C_\theta, C_z$ are also shown. }
 \label{lpf2} 
\end{figure}

In this section, we specialize to membranes of cylindrical geometry which have zero Gaussian curvature i.e. $K(x)=0 $ in Eq.~(\ref{membeq2}). This greatly simplifies the coupled system of equations and one can arrive at an analytic vortex solution in this setup. Note that even in the absence of Gaussian curvature, the extrinsic geometry of the cylinder modifies the momentum exchange between the membrane fluid and the external embedding fluids and has significant impact on the multi-vortex dynamics we are going to study. We refer to Appendix Sec.~\ref{avcons} for a detailed computation of the single vortex solution and here summarize the key results. The incompressible nature of the membrane fluid can be utilized to express the membrane velocity field confined to the cylinder surface in terms of a scalar stream function $\bm{\psi}$ as follows (here $\theta$ and $z$ are the usual angular and axial coordinates along the cylinder surface of fixed radius R):
\beqa
v_\theta =\partial_z \bm {\psi} \hspace{1cm}
v_z = -\frac{1}{R} \partial_\theta \bm {\psi} 
\eeqa
As shown in detail in Appendix Sec.~\ref{avcons}, in the limit of thin tubular membranes, the stream function $\bm{\psi}$ for a point vortex situated at the origin $(\theta,z)=(0,0)$ takes the following simple form
\beqa
\bm{\psi}(\theta, z)= \frac{\tau}{4 \pi \eta_{2D}}  \left(- \log \left[1-2 e^{-\frac{|z|}{R}} \cos \theta+e^{-\frac{2 |z|}{R}}\right]+ \sqrt{\frac{\lambda}{2R}}~ e^{-\frac{\sqrt{2} |z|}{\sqrt{\lambda R}}}\right).\
\label{strmhc}
\eeqa
where $\tau$ is the strength of the circulation. The logarithmic term in the stream function is expected from 2D vortex dynamics Ref.~\cite{newton}, while the
additional exponential correction arises from the exchange of momentum between the 2D membrane fluid and external 3D fluids, illustrating the quasi-2D nature of the system. The exponential correction term originates from the zero mode in the compact $\theta$ direction of the tube (and features in the $\theta$ component of the vortex velocity field computed afterwards).  The associated decay length in the exponential term is given by $\sqrt{\lambda R}$ which is larger than the tube radius R.  For a vortex located at $(\theta_0,z_0)$ on the membrane tube, the velocity field at any location $(\theta,z)$ is given by
\beqa
&v_\theta [\theta,z,\theta_0,z_0]=\frac{\tau (z-z_0)}{4 \pi R ~\eta_{2D} |z-z_0|}\left(\frac{2-2 \cos(\theta-\theta_0)~ e^{\frac{|z-z_0|}{R}}}{1+ e^{\frac{2|z-z_0|}{R}} - 2 \cos(\theta-\theta_0)~ e^{\frac{|z-z_0|}{R}}}-e^{-\sqrt{\frac{2}{R \lambda}}|z-z_0|}\right)\nn \\
&v_z [\theta,z,\theta_0,z_0] = -\frac{\tau}{4 \pi \eta_{2D} R} ~ \frac{\sin(\theta- \theta_0)}{ \cos(\theta- \theta_0) -\cosh\frac{z-z_0}{R}}
\label{vhc}
\eeqa
As explained in Appendix A, the above velocity field is constructed out of the stream function $\bm{\psi}$ via $v_\alpha = \epsilon_{\alpha \beta} D^\beta \bm{\psi}$, such that the incompressibility constraint $D^\alpha v_\alpha=0$ is automatically satisfied. Moreover, the velocity fields vanish at large separation in the z direction.
The vortex solution is presented in Fig.~\ref{kernel}. As evident from the streamlines, the cylinder topology enforces the creation of an additional saddle defect in the flow such that the net index is zero, consistent with Poincar\'e Index theorem. We now cast the dynamics of the multi-vortex system such that a vortex is simply advected by the local fluid flow generated
by the remaining vortices. Following standard treatment in vortex literature, such dynamics of the many-vortex system can be encapsulated in the form of a Hamiltonian on the cylinder, utilizing the cylinder stream function Eq.~(\ref{strmhc})
\beqa
H= \sum_{i \neq j} \frac{ \tau_i \tau_j }{4 \pi \eta_{2D}}  \left(- \log \left[1-2 e^{-\frac{|z_i -z_j|}{R}} \cos (\theta_i -\theta_j) +e^{-\frac{2 |z_i -z_j|}{R}}\right]+ \sqrt{\frac{\lambda}{2R}}~ e^{-\frac{\sqrt{2} |z_i -z_j|}{\sqrt{\lambda R}}}\right)
\label{hm}
\eeqa
where $Q_i =\sqrt{|\tau_i|} R ~ \theta_i$ and its associated conjugate momentum is $P_i=\sqrt{|\tau_i|} z_i$, in terms of which the dynamics is governed by the usual Hamilton's equations $\dot{Q}_i= \partial_{P_i} H,~ \dot{P}_i= -\partial_{Q_i} H$. Just like the vortex stream function, the Hamiltonian features logarithmic interactions along with an extra interaction term arising from the zero mode in the compact angular direction of the cylinder. For spherical membranes, the Hamiltonian also receives a dominant contribution from a similar zero mode that gives rise to global rotation Ref.~\cite{rs21}. In the usual coordinates $(\theta,z)$,
the dynamical equations for the multi-vortex system thus reads
\beqa
&R ~ \dot{\theta}_i=  \sum_{j \neq i}^N  v_{\theta}[\theta_i,z_i,\theta_j,z_j] \nn\\
&\dot{z}_i=  \sum_{j \neq i}^N   v_{z}[\theta_i,z_i,\theta_j,z_j] 
\label{dyneq}
\eeqa
where the velocity field $(v_\theta,v_z)$ is given by Eq.~(\ref{vhc}). Written explicitly, the dynamical equations take the following form
\beqa
& R\dot{\theta}_i=\sum_{j \neq i}^N\frac{\tau_j (z_i-z_j)}{4 \pi R ~\eta_{2D} |z_i-z_j|}\left(\frac{2-2 \cos\left(\theta_i-\theta_j\right)~ e^{\frac{|z_i-z_j|}{R}}}{1+ e^{\frac{2|z_i-z_j|}{R}} - 2 \cos\left(\theta_i-\theta_j\right)~ e^{\frac{|z_i-z_j|}{R}}}-e^{-\sqrt{\frac{2}{R \lambda}}~|z_i-z_j|}\right)\nn \\
&\dot{z}_i = -\sum_{j \neq i}^N\frac{\tau_j}{4 \pi \eta_{2D} R} ~ \frac{\sin\left(\theta_i-\theta_j\right)}{ \cos\left(\theta_i-\theta_j\right) -\cosh\frac{z_i-z_j}{R}}.
\label{vdneq}
\eeqa The translational symmetry of the Hamiltonian leads to the conservation of the following quantities
\beqa
C_\theta \equiv =\sum_i \tau_i \theta_i  \hspace {1cm} C_z = \sum_i \tau_i z_i
\label{ctrans}
\eeqa
which is ensured in our multi-vortex simulations to a good accuracy (Numerical errors of the order of $10^{-14}$). Note that there are fewer conserved quantities arising from the symmetries of the Hamiltonian and thus in general, one expects non-integrable dynamics for multi-vortex systems, as will be evident in the simulations carried out in the next section.   We further define a quantity
\beqa
L^2 \equiv \frac{1}{2} \big{|}\sum_{i \neq j} \tau_i \tau_j l_{ij}^2 \big{|}
\label{Ldef}
\eeqa
where $l_{ij}^2 = R^2 ~\text{Min} \left(\theta_i -\theta_j, 2\pi -(\theta_i -\theta_j) \right)^2 +(z_i-z_j)^2 $ is the distance function appropriate for the cylinder geometry. For the two vortex systems, we will denote the quantity in Eq.~(\ref{Ldef}) by $L_{12}^2$ and reserve the notation $L^2$ for multi-vortex systems. The quantity $L^2$  is conserved in flat and spherical membranes but violated in membrane tubes due to loss of in-plane rotational symmetry, as will be explicitly demonstrated in the simulations presented in Sec.~\ref{cylsim}. 
%
\section{Vortex dynamics in membrane tubes}
\label{cylsim}
In this section, we numerically simulate the vortex dynamics on the surface of the membrane tube using the dynamical equations Eq.~(\ref{dyneq}). The vortex trajectories are computed using a fourth-order Runge–Kutta (RK4) integration scheme with an adaptive time step. At each iteration, the time step is adjusted to ensure that the relative decrease in distance between any two approaching vortices does not exceed a specified threshold (0.05 or 0.1 in the present simulations). Additionally, the time step is constrained by a fixed upper bound, ranging from 0.01 to 0.001, depending on the simulation. This upper bound in timesteps maintains the maximum local integration error per time step in the order of $1.0\times e^{-10}$ to $1.0 \times e^{-15}$. The simulations are implemented in Python 3, with minor performance optimizations carried out using the Numba library. In all cases, we include a 3D plot showing the vortex trajectories on the surface of the membrane tube. Trajectories of vortices with positive circulation (anti-clockwise) are colored blue, while vortices of negative circulation (clockwise) are colored red. The initial locations of vortices are marked by a colored dot, following the same color code as the trajectories. We first discuss the two vortex case for $\frac{\lambda}{R}=100$ in Figs.~(\ref{ca}-\ref{cf}) for equal (or opposite) vortex strength. This typically results in closed orbits for vortex pairs of same circulation and helical geodesics for vortex pairs of opposite circulation. Next, in Fig.~(\ref{figuneq}) we show the trajectories for two vortices of unequal strength but same sign, giving rise to concentric orbits.  Fig.~(\ref{figlc}) illustrates interesting effects resulting from tuning $\frac{\lambda}{R}$ . Multi-vortex systems are explored in Figs.~(\ref{mvca}-\ref{mvcd}). Vortex leapfrogging and relative periodic orbits are illustrated in various situations Fig.~(\ref{lpf0}-\ref{lpf2}) along with the associated numerical errors in conservation of $ C_\theta$ and $C_z$. Finally, we study the stability of vortex rings in Fig.~(\ref{vring}).\\\\
\subsection{The two-vortex system}  
\begin{figure}[htbp!]
\begin{tabular}{lcc}
\includegraphics[height=0.2\textheight]{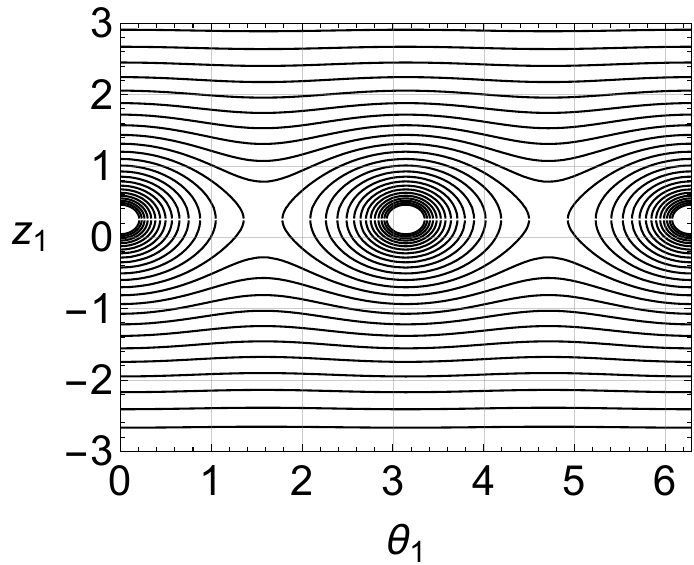}&&\hspace{0.1\textwidth}
\includegraphics[height=0.2\textheight]{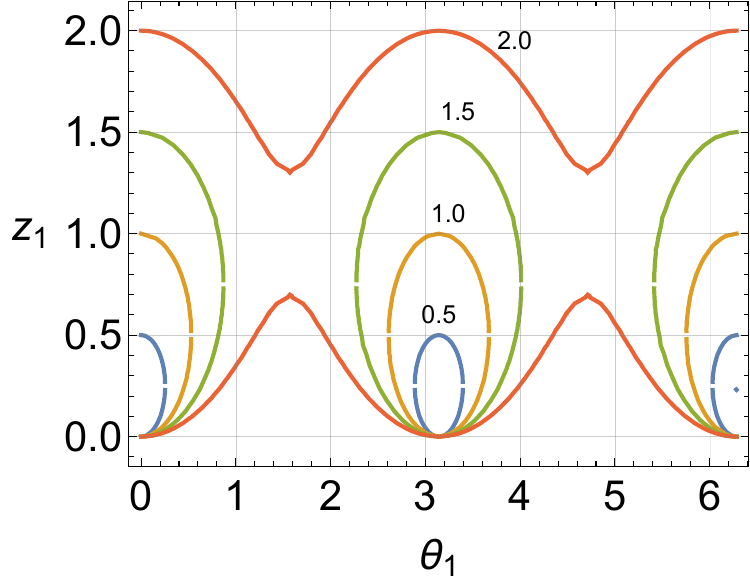}
\end{tabular}
 \caption{ Left: Level curves of the Hamiltonian  in the $(\theta_1, z_1)$ plane  for $\lambda/R=100$  and generic values of $C_\theta$ and $C_z$, Right: Level curves of the Hamiltonian in the $(\theta_1, z_1)$ plane for  $C_z$= 0.5 (blue), 1 (orange), 1.5 (green) and  2 (red)  at a fixed value of  $C_\theta=0$. This is achieved by tuning the initial separation along z-direction for the two vortex system, with $\theta_1=\theta_2=0$. Notice that the red curve for $C_z$ =2  traverses the entire angular direction of the cylinder.   }
 \label{levelcz} 
\end{figure}

We first consider a two-vortex system of the same strength $\tau$ with $\lambda/R=100$. In Fig.~(\ref{ca}), we show the variation of vortex locations $(\theta,z)$ and $L_{12}^2$ defined in Eq.(\ref{Ldef}) with time and  the rightmost plot shows a trace of the vortex trajectories. From the plots, it is clear that a two vortex system which is initially closely spaced will orbit each other in a closed orbit, with the inter-vortex distance oscillating in time.  The closed orbits indicate a near-equilibrium
configuration modulated by the delicate balance provided by the Hamiltonian. This is in contrast to flat and spherical fluid membranes, where $L_{12}^2$ is conserved and arises due to the breaking of in-plane rotational symmetry in tubular membranes. However, the orbit shape gets remarkably distorted for vortices, which are initially well separated. This is illustrated in Fig.~(\ref{cb}) where the vortices are initially well separated along the angular $\hat{\theta}$ direction. This can also lead to pinching of a single closed orbit into two separate orbits. For example, two vortices initially well separated along the longitudinal $\hat{z}$ direction of the tube prefer to go in separate orbits, encircling the tube, see Fig.~(\ref{cc}).\\\\
\textbf{Comments on pinching of orbits:} A mathematical understanding of these results emerges easily from the associated conserved quantities Eq.~(\ref{ctrans}) and the conservation of the Hamiltonian Eq.~(\ref{hm}). Using Eq.~(\ref{ctrans}) one can eliminate say $(\theta_2, z_2)$ in favor of $(\theta_1,z_1)$, thus the Hamiltonian Eq.~(\ref{hm}) for two vortices becomes a function of only one of the vortex co-ordinates i.e., $(\theta_1,z_1)$ and the constants of motion i.e. $C_\theta, C_z$. Analyzing the level curves of this Hamiltonian in the $(\theta_1,z_1)$ plane, one readily observes that for the situations described in Fig.~\ref{ca} and Fig.~\ref{cb} the contour plots are closed loops, while for Fig.~\ref{cc}, they traverse the entire angular direction of the cylinder, The level curves of the Hamiltonian in the $(\theta_1,z_1)$ plane is displayed in the left panel of Fig.~\ref{levelcz} for generic choice of constants of motion $C_\theta$ and $C_z$. In the right panel, we show the level curves of the Hamiltonian for 4 different values $C_z=0.5, 1, 1.5 , 2$  at a fixed value of $C_\theta=0$. This is achieved by tuning the initial separation along z-direction for the two vortex system, with $\theta_1=\theta_2=0$. We observe that for lower values of $C_z$, i.e. less initial separation in z, the level curves are closed loops, while for higher values of z-separation, the level curves traverse the entire angular direction of the cylinder leading to the pinching of orbits.

\begin{figure}[htbp!]
\begin{tabular}{lcc}
\includegraphics[height=0.2\textheight]{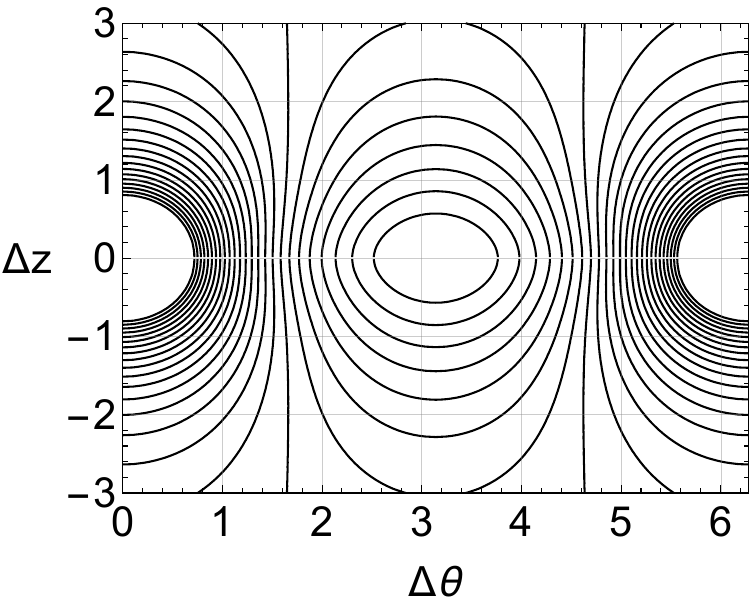}&&\hspace{0.1\textwidth}
\includegraphics[height=0.2\textheight]{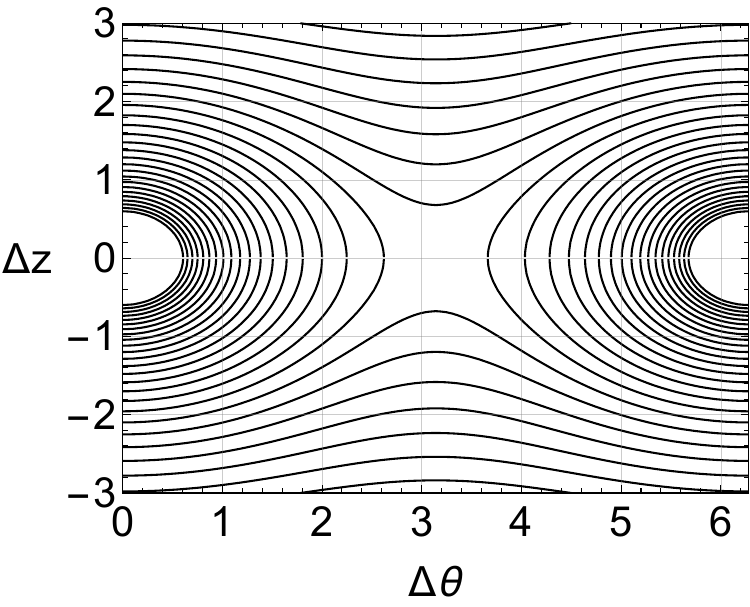}
\end{tabular}
 \caption{  Level curves of the Hamiltonian  in the $(\Delta \theta, \Delta z)$ plane  for Left: $\lambda/R=100$ and Right:  for $\lambda/R=1$,   }
 \label{levelsaff} 
\end{figure}

For the two-vortex system of opposite strength (also known as vortex pair), we observe vastly different interactions, see Fig.~(\ref{cd}-\ref{cf}). A vortex pair separated along the angular (transverse) direction of the tube move in vertical parallel trajectories (Fig.~\ref{cd}) while vortex pairs separated along the longitudinal direction of the tube move together along the transverse direction in separate closed orbits (Fig.~\ref{ce}). Arbitrarily situated vortex pairs move together along helical geodesics, in accordance with a conjecture by Kimura Ref.~\cite{km}. Note that in all the situations with vortex pairs, the net circulation is zero and $L^2$ is conserved in time.  These results are evident from Eq.~(\ref{vdneq}) where for equal and opposite strength $\tau_1= -\tau_2$ we find that $\dot{\theta}_1 =\dot{\theta}_2$ and $\dot{z}_1 =\dot{z}_2$. Using this fact, it is straightforward to compute the pitch of the helix $z_P$. The time period T for one complete rotation around the membrane tube of radius R for the helical motion in Fig.~(\ref{cf}) for a vortex pair with $\tau_1 =-\tau_2 \equiv \tau$ can be easily constructed from the fact that in this situation $\dot{\theta_1} =\dot{\theta_2}$  and $\dot{z}_1= \dot{z}_2$, which implies that the $\theta$ and $z$ separation between the vortices remain constant in time. Thus, we can utilize the velocity expression in Eq.~(\ref{vdneq})  to find the time period T (note that the velocities depend on the separation in $\theta$ and $z$ and hence are constants as well)
\beqa
& \frac{2 \pi R}{T}= \bigg{|} \frac{\tau (z^i_1-z^i_2)}{4 \pi R ~\eta_{2D} |z^i_1-z^i_2|}\left(\frac{2-2 \cos\left(\theta^i_1-\theta^i_2\right)~ e^{\frac{|z^i_1-z^i_2|}{R}}}{1+ e^{\frac{2|z^i_1-z^i_2|}{R}} - 2 \cos\left(\theta^i_1-\theta^i_2\right)~ e^{\frac{|z^i_1-z^i_2|}{R}}}-e^{-\sqrt{\frac{2}{R \lambda}}~|z^i_1-z^i_2|}\right)\bigg{|}\nn
\eeqa
where  $(\theta^i_1,  z^i_1)$ and $(\theta^i_2, z^i_2)$  are the initial vortex locations. Thus, the pitch of the helix $z_P$ can be computed from the $\dot{z}_1= \dot{z}_2$ expression in Eq.~(\ref{vdneq}) and  given by 
\beqa 
z_P=\frac{ \frac{ 2 \pi R \sin\left(\theta^i_1-\theta^i_2\right)}{ \cos\left(\theta^i_1-\theta^i_2\right) -\cosh\frac{z^i_1-z^i_2}{R}} }{\frac{2-2 \cos\left(\theta^i_1-\theta^i_2\right)~ e^{\frac{|z^i_1-z^i_2|}{R}}}{1+ e^{\frac{2|z^i_1-z^i_2|}{R}} - 2 \cos\left(\theta^i_1-\theta^i_2\right)~ e^{\frac{|z^i_1-z^i_2|}{R}}}-e^{-\sqrt{\frac{2}{R \lambda}}~|z^i_1-z^i_2|}} 
\eeqa
where  $(\theta^i_1,  z^i_1)$ and $(\theta^i_2, z^i_2)$  are the initial locations of the vortex pair.\\\\
\textbf{Comments on confined and unconfined dynamics of vortices}: The confined dynamics of vortices of same circulation (or equivalently, the unconfined dynamics of vortices of opposite circulation) can be easily understood from the conservation of the vortex Hamiltonian Eq.~\ref{hm} and the conservation of $C_z$, Eq.~\ref{ctrans}. For vortices of same circulation $\tau_1= \tau_2 \equiv \tau$, $z_1 = \frac{C_z}{\tau}- z_2$ which implies $|z_1-z_2| =\big{|}\frac{C_z} {\tau} - 2 z_2 \big{|}$. Thus, the Hamiltonian decays exponentially as $z_2 \rightarrow \infty$, which is forbidden by energy conservation. This in turn explains the bounded orbits for vortices of same strength. On the other hand, for vortices of equal and opposite strength, then $|z_1-z_2| =\big{|}\frac{C_z} {\tau} \big{|}$ which is a constant. Thus, the Hamiltonian conservation allows the vortices to drift together to large $z$ leading to unconfined dynamics in this situation.\\\\
Additionally, we note that the ratio of the vortex strengths $\tau_1/\tau_2$ and $\lambda/R$ provide additional tuning parameters that can be used to control the vortex trajectories, as illustrated in Fig.~(\ref{figuneq}) and Fig.~(\ref{figlc}) respectively. Fig.~(\ref{figuneq}) shows two concentric orbits arising from unequal vortex strengths,  whereas in Fig.~(\ref{figlc}) we simulate the same initial configuration as Fig.~(\ref{cb}) but with  $\lambda/R =5$. In this situation, the single closed orbit of Fig.~(\ref{cb}) breaks into two separate local closed orbits, as seen in  Fig.~(\ref{figlc}). This curvature mediated pinching of orbits can also be understood in terms of the level curves of the Hamiltonian Eq.~(\ref{hm}). In this situation, translational symmetry can be used to express the Hamiltonian of the two-vortex system in terms of separation $\Delta \theta$ and $\Delta z$. We investigate the level curves of the Hamiltonian in the $(\Delta \theta, \Delta z)$ plane as shown in Fig.~\ref{levelsaff} for large and small values of $\lambda/R$. The variations in $(\Delta \theta, \Delta z)$  gets localized upon reducing $\lambda/R$, consistent with Fig.~(\ref{figlc}).
\subsection{Essential features of the multi-vortex system} 
\begin{figure}[htbp!]
\begin{tabular}{lcc}
\includegraphics[height=0.13\textheight]{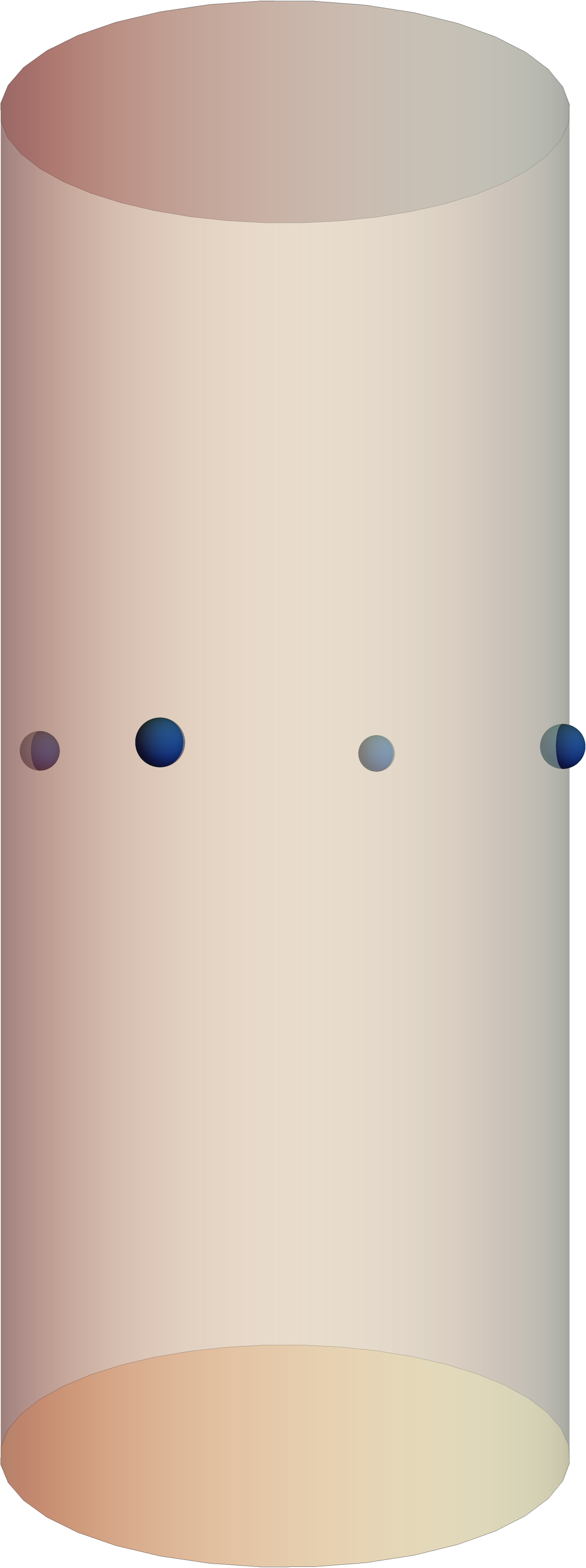}&&\hspace{0.2\textwidth}
\includegraphics[height=0.13\textheight]{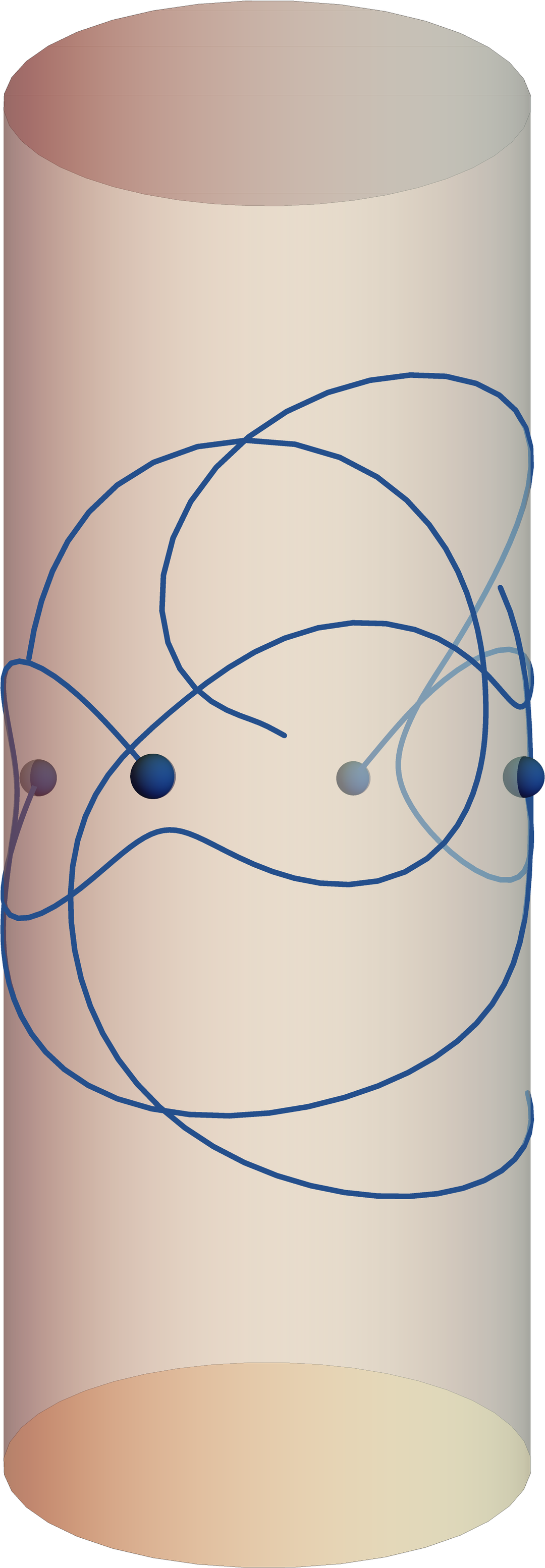}
\end{tabular}
 \caption{ Left: An equilibria of 4 vortices arranged in a ring, with intervortex separation $\frac{ \pi R}{2}$ , Right: Mild perturbation along $\theta$ or $z$ or both  renders the system unstable. }
 \label{vring} 
\end{figure}

Moving on to multi-vortex assemblies Fig.~(\ref{mvca}-\ref{mvcd}), we first simulate a set of 4 vortices arranged along the transverse ($\hat{\theta}$) direction of the membrane tube, of the same (Fig.~\ref{mvca}) and alternating (Fig.~\ref{mvcb}) circulations. All simulations are performed with $\lambda/R=100$. A distinct feature of multi-vortex assemblies of the same strength (Fig.~\ref{mvca}) is that their dynamics remains confined in a region determined by the initial distribution, with $L^2$ oscillating in time. This can be again explained from the conservation of the Hamiltonian Eq.~\ref{hm} like the two vortex system. On the other hand, the multi-vortex systems of alternating circulation (with zero net circulation) show unconfined dynamics, yet $L^2$ remains conserved in time, see Fig.~\ref{mvcb}. Vortices initially situated along the longitudinal direction of the cylinder Fig.~(\ref{mvcc}) and Fig.~(\ref{mvcd}) share the same features as their transverse counterparts, ie.  Fig.~(\ref{mvca}) and Fig.~(\ref{mvcb}), except the fact that the compact cylinder topology enforces the vortices of alternating circulations to exhibit closed orbits wrapping the entire membrane tube, see Fig.~(\ref{mvcd}). In Fig.~(\ref{lpf0}-\ref{lpf2}) we demonstrate the interesting phenomenon of vortex leapfrogging in the tubular fluid membrane. In such situations, the vortices move in periodic or quasi-periodic orbits along with translation along the tube surface. They are fairly easy to design using the results from two vortex systems. As observed in Fig.~(\ref{cd}-\ref{cf}), vortex pairs (of opposite circulation) translate together while vortices of same circulation perform periodic orbits with no net translation. In order to achieve relative periodic orbits (leapfrogging), one can take a vortex pair (of circulation $\tau_1=+1$ and $\tau_2=-1$) and divide them into two groups A and B, such that the circulations in group A add up to +1 while that of Group B add up to -1 (this is also known as ``Vortex splitting"). The vortices in Group A and Group B will translate together, while within each individual group we will have periodic or quasi periodic orbits of the split vortices. This is illustrated in several leapfrogging scenarios in Fig.~(\ref{lpf0}-\ref{lpf2}). Finally, we note that equilibria of vortex rings can be constructed for N vortices arranged along the transverse direction of the tube with intervortex separation $\frac{2 \pi R}{N}$. Mild perturbation along $\theta$ or $z$ (or both) in any one of the vortices renders the system unstable. This is illustrated in Fig.~(\ref{vring}) for $N=4$. A more systematic perturbative analysis of vortex rings and crystals in our setup is left for upcoming works. 

\section{Conclusion}
\label{cncl}
In this paper, we focus on the hydrodynamic interactions of vortex assemblies embedded in tubular fluid membranes.  The tubular membrane geometry is distinct from flat and spherical membranes, due to loss of in-plane rotational symmetry. We find that the confined cylinder topology (in the angular direction) leads to several interesting features in the dynamics of vortices in membrane tubes. We first construct an analytic vortex solution in this tubular fluid membrane setup and show that the cylinder topology enforces the creation of an additional saddle defect in the flow, consistent with Poincar\'e Index theorem. Next, the incompressibility condition of the membrane fluid flow is utilized to construct a Hamiltonian on the cylinder that governs the dynamics of the multi-vortex system. The Hamiltonian  depends on the system parameters, the Saffman length $\lambda$, the tube radius R and the circulation strength $\tau$ of the vortices. This offers experimentally accessible tuning parameters to control the dynamics of the multi-vortex system.  Focusing on a simple system of two vortices of same circulation strength $\tau$, we show that the breaking of rotational symmetry leads to non-conservation of the inter-vortex separation, unlike flat and spherical fluid membranes. The vortices generically move in closed orbits, which can be severely distorted by tuning the initial inter-vortex separation as well as the parameter $\frac{\lambda}{R}$. In particular, we observe several situations where a single closed orbit breaks into two upon tuning system parameters. We also illustrate this pinching from a mathematical point of view by considering the level sets of the associated vortex Hamiltonian. Two vortices with opposite circulation generically move together along helical trajectories, thus confirming a conjecture by Kimura on the dynamics of vortex pairs Ref.~\cite{km}. We also explore relative periodic orbits and relative equilibria including vortex leapfrogging. \\\\
The work needs extension in several directions: first, although we performed the analysis of vortex dynamics in the limit of thin membranes for ease of numerical implementation, the full  vortex solution constructed in Appendix \ref{avcons} can be utilized to perform vortex simulations in situations where the Saffman length is comparable to membrane radius as well as asymmetric viscosities of external fluids coupled to the membrane. Addition of short range repulsive interactions and thermal noise will be interesting and important for studying diffusion of motor proteins in membrane tubes, since the diffusion constant can be related to the hydrodynamic mobility via the Stokes-Einstein relation. Secondly, although the current model deals with membranes of fixed geometry, it will be worth pursuing similar studies of vortex dynamics in membranes of dynamical geometry, together with corrections from inertia terms in the Navier-Stokes equation.  Besides being of experimental relevance (Ref.~\cite{bss11}), we believe our results will be interesting for understanding the collective dynamics of vortices and rotors in tubular fluid interfaces and membranes. We believe many of our results will be applicable to more general setups involving rotating motors (both living and non-living) in tubular fluid interfaces, analogous to the studies in flat membranes Ref.~\cite{lenz2003,lenz2004, nmsh19, nmsh22} and vortex simulations in superconductors, Ref.~\cite{cyn17}. Dynamics of rotating matter in fluid interfaces is currently an active area of research Ref.\cite{rt0,rt1,rt2,rt3,rt4,rt5,rt6,rt7,rt8,rt9, lushiv2015,ylv2015,mzc}, both from an experimental and theoretical perspective. It will also be interesting to connect our results to vortex and mass dynamics in inviscid cylinder geometries,  Ref.\cite{mstcyl, boattocyl2020}. Let us note that the vortex dynamics studied in this work  bears many resemblances to the study of vortex dynamics  in inviscid fluids, Ref.~ \cite{mstcyl}, as expected from the symmetry considerations.  We leave a more detailed comparison of our results with existing numerical simulations and experiments for upcoming works. An analysis of thermal fluctuations in this setup, along the lines of the work of Sokolov and Diamant Ref.~\cite{sokolov_haim2018} is also left for the future.
\section{Acknowledgments}
We are very thankful to Naomi Oppenheimer, Haim Diamant, Mark Henle, Sarthak Bagaria, Samyak Jain, Prasad Perlekar and Mustansir Barma.  R.S is supported by DST INSPIRE Faculty fellowship, India (Grant No.IFA19-PH231), NFSG and OPERA Research Grant from Birla Institute of Technology and Science, Pilani (Hyderabad Campus). The work is dedicated to the memory of Professor Alex J. Levine, UCLA.
\appendix
\section{Construction of the Vortex solution in the Tubular Fluid Membrane}
\label{avcons}
In this Appendix, we will sketch the construction of the flow sourced by a point vortex embedded in a tubular fluid membrane, extending Ref.\cite{henlev2010, rs21,sarthak22}. We will first provide the vortex solution for cylindrical fluid membranes of arbitrary radius and finally consider the limit of thin tubular membrane, which is of relevance to our study in this paper.  The cylinder surface metric in the usual $(\theta,z)$ coordinates is 
\beqa
ds^2 = R^2 d\theta^2 + d z^2
\eeqa
and 
\beqa
 \bm{\epsilon}_{\alpha \beta}= \begin{pmatrix} 
 0    & R\\ 
  -R & 0
\end{pmatrix}
\label{levicv}
\eeqa
is the corresponding Levi Civita tensor.  The incompressibility condition of the membrane fluid (Eq.(\ref{membeq1}) of main text) allows us to write the 2D velocity field on the cylinder surface in terms of a scalar stream function $\bm{\psi}$ as follows:
\beqa
v_\alpha = \epsilon_{\alpha \beta} D^\beta \bm{\psi}.\nn
\eeqa
Since covariant derivatives on a scalar function commute, this automatically ensures that the incompressibility condition is met.
Decomposing $\bm{\psi}$ in terms of eigenmodes of the Laplace Beltrami operator for the cylinder surface, we get  
\beqa
v_\alpha= \epsilon_{\alpha \beta} D^\beta\left(\int d\Lambda ~A(\Lambda)~ e^{i \Lambda_\mu x^\mu} \right)
\label{avmain}
\eeqa
where $\Lambda_\alpha \equiv (n,q \equiv \frac{k}{R})$ and ${x}^\alpha \equiv (\theta,z)$  such that $ \Lambda_\mu x^\mu = n \theta+ \frac{k}{R} z$. The discrete index ``n" appears in the mode decomposition due to the periodicity along the compact $\theta$ direction of the membrane geometry. Note that $\int d\Lambda$  is shorthand for $\sum_{n=-\infty}^{\infty}\int_{-\infty}^{\infty} dq$. The unknown coefficients $ A(\Lambda)$ will be solved below using the stress-balance condition Eq.(\ref{membeq2}) of main text. In the stress balance condition, we insert the external point torque $\sigma_\alpha^{ext} =\tau~ \epsilon_{\alpha \gamma} D^\gamma{\delta^2(\theta- \theta_0, z-z_0)}$ and also take an anti-symmetric derivative on both sides of the equation to eliminate the membrane pressure p. The expression for the traction vector $T_\alpha$ appearing on the RHS of the stress balance condition in Eq.(\ref{membeq2}) requires a straightforward but somewhat long computation, which we just summarize below.\\

 \textbf{ Computation of $T_\alpha$ :} In order to compute the in-plane components of the traction vector $T_\alpha$ we utilize the known solution for Stokes flow in 3D (Eq.(\ref{membeq3}) of main text) in cylindrical coordinates following Happel and Brenner Ref.\cite{hb}:
\beqa
&\vec{v}^{\pm}(r, \theta, z)= \vec{\nabla} f^{\pm}(r, \theta, z)+\vec{\nabla} \times[g^{\pm}(r, \theta, z) \hat{z}] 
+r \partial_{r}\left[\vec{\nabla} h^{\pm}(r, \theta, z)\right]+\partial_{z} h^{\pm}(r, \theta, z) \hat{z}, \nn\\
&p^{\pm}(r, \theta, z)=-2 \eta_{\pm} \partial_{z}^{2} h^{\pm}(r, \theta, z),\nn
\eeqa

where $f^{\pm}, g^{\pm},h^{\pm}$ are harmonic functions of the Laplacian operator in 3D in cylindrical coordinates. Such harmonic functions on the cylinder admit a decomposition.

$$
f^{\pm}(r, \theta, z)=\sum_{n=-\infty}^{\infty}\int_{-\infty}^{\infty} dq~  e^{i \Lambda_\mu x^{\mu}} F^{\pm}(\Lambda)~ \xi_{n}^{\pm}(q r)
$$
and likewise for $ g^{\pm},h^{\pm}$. In the above expression,
$$
\xi_{n}^{+}(q r) \equiv K_{n}(|q| r), \quad \xi_{n}^{-}(q r) \equiv I_{n}(|q| r)
$$
where $K_n$ and $I_n$ are modified Bessel functions of order $n$ of first and second kind, respectively. The covariant components of the external fluid velocities thus read as follows: \\[20pt]
$v_{r}^{\pm}(\Lambda, r)=|q| \widetilde{\xi}_{n}^{\pm}(q r) F^{\pm}(\Lambda)+\frac{i n}{r} \xi_{n}^{\pm}(q r) G^{\pm}(\Lambda)$ $+H^{\pm}(\Lambda)\left[-|q| \widetilde{\xi}_{n}^{\pm}(q r)+r\left(q^{2}+\frac{n^{2}}{r^{2}}\right) \xi_{n}^{\pm}(q r)\right]$\\
$v_{\theta}^{\pm}(\Lambda, r)=i m \xi_{n}^{\pm}(q r) F^{\pm}(\Lambda)-|q| r \widetilde{\xi}_{n}^{\pm}(q r) G^{\pm}(\Lambda)$ $+i n H^{\pm}(\Lambda)\left[|q| r \widetilde{\xi}_{n}^{\pm}(q r)-\xi_{n}^{\pm}(q r)\right]$\\
$v_{z}^{\pm}(\Lambda, r)=i q \xi_{n}^{\pm}(q r) F^{\pm}(\Lambda)+i q H^{\pm}(\Lambda)\left[|q| r \tilde{\xi}_{n}^{\pm}(q r)+\xi_{n}^{\pm}(q r)\right] .$\\[5pt]

where $\left.\widetilde{\xi}^{+}(q r) \equiv \frac{d K_{n}(u)}{d u}\right|_{u=|q| r},\left.\quad \widetilde{\xi}(q r) \equiv \frac{d I_{n}(u)}{d u}\right|_{u=|q| r}$\\[5pt]
The 6 unknown coefficients $F^{\pm}, G^{\pm},H^{\pm}$ can be solved in terms of membrane mode coefficient $A$ using the following 6 equations derived from the no-slip boundary condition (Eq.(\ref{membeq5}) of main text):
\beqa
&v_{r}^{\pm}(\Lambda, R)=0, \quad \Lambda^{\alpha} v_{\alpha}^{\pm}(\Lambda, R)=0 \nn\\
&i \epsilon^{\alpha \gamma} \Lambda_{\gamma} v_{\alpha}^{\pm}(\Lambda, R)=-\Lambda_{\beta} \Lambda^{\beta} A(\Lambda)\nn
\eeqa
The solution for the 6 coefficients $F^{\pm}, G^{\pm},H^{\pm}$ is next obtained in \textit{Mathematica}. Plugging in these solutions, the required antisymmetric derivative of the traction vector $T_\alpha$ is given as follows:\\[5pt]
\beqa
\left.\epsilon^{\alpha \gamma} D_{\gamma} \sigma_{\alpha r}^{\pm}\right|_{r=R}=\eta_{\pm} \int D \Lambda e^{i \Lambda_{\mu} x^{\mu}} \frac{A(\Lambda)}{R^{3}} C^{\pm}(\Lambda)
\eeqa
where 
\beqa
C^{\pm}(n,k)=\frac{2 n^{2}\left[\rho^{\pm}(n,k)\right]^{3}+\left(n^{2}+k^{2}\right)^{2}\left[\rho^{\pm}(n,k)\right]^{2}+2 \rho^{\pm}(n,k)\left(k^{4}-n^{4}\right)-\left(k^{2}+n^{2}\right)^{3}}{\rho^{\pm}(n,k)~ k^{2}-\left[\rho^{\pm}(n,k)-n\right]\left[\rho^{\pm}(n,k)+n\right]\left[\rho^{\pm}(n,k)+2\right]}
\eeqa
and
\begin{equation}
k \equiv q R . \quad \rho^{\pm}(n,k) \equiv \frac{|k| \tilde{\xi}_{n}^{\pm}(k)}{\xi_{n}^{\pm}(k)}.
\end{equation}
More explicitly,
\beqa
&\rho_+ [n,k] = \frac{|k| \frac{d  K_n[u]}{du} \big{|}_{u= |k|}}{K_n[|k|]}, \hspace{1cm} \rho_- [n,k] = \frac{|k| \frac{d  I_n[u]}{du} \big{|}_{u= |k|}}{I_n[|k|]}
\eeqa   \\
The fundamental equation of stress balance  on the cylindrical membrane (after taking anti-symmetric derivative of both sides of Eq.(\ref{membeq2}) of main text) thus reads as
\beqa
-\epsilon^{\alpha \beta}D_\beta \left[\tau~ \epsilon_{\alpha \gamma} D^\gamma\right]\underbrace{\left(\frac{1}{4 \pi ^2 R } \int d \Lambda~ e^{i \Lambda_\beta (x -x_0)^\beta} \right)}_{\delta^2(\theta- \theta_0, z-z_0)}=\int d\Lambda ~\frac{\eta_{2D}~  A(\Lambda)}{R^4}\underbrace{ c_n(k)}_{\text{ membrane stress + Traction}} e^{i \Lambda_\alpha x^\alpha}\nn
\eeqa
with 
\beqa
c_n(k)=(n^2 +k^2)^2 -\frac{R}{\lambda_-} C^- (n,k) +\frac{R}{\lambda_+} C^+ (n,k)\nn\\
\eeqa
We solve for $A(\Lambda)$ from the above equation as
\beqa
A(\Lambda) = \frac{\tau R}{4 \pi^2 \eta_{2D}} \frac{n^2+k^2}{c_n(k)} e^{-i\Lambda_\alpha x_0^\alpha }
\eeqa
Plugging this solution of $A(\Lambda)$ into Eq.(\ref{avmain}), we get
\beqa
v_\alpha= \epsilon_{\alpha \beta} D^\beta \underbrace{\left(\int d\Lambda ~ \frac{\tau R}{4 \pi^2 \eta_{2D}} \frac{n^2+k^2}{c_n(k)} e^{i\bm{\Lambda}\cdot(\bm{x} -\bm{x}_0) } \right)}_{\equiv \bm{\psi}}
\eeqa
where we can now identify the stream function of the flow field at $(\theta,z)$ sourced by a vortex of strength $\tau$ situated at the origin 
\beqa
\bm{\psi} (\theta, z) = \frac{\tau}{4 \pi^2 \eta_{2d}}  \sum\limits_{n=-\infty}^{\infty}~\int\limits_{-\infty}^{\infty} dk~ \frac{n^2+ k^2}{c_n(k)} e^{i \left( n \theta + \frac{k}{R}z\right)}.
\label{avortex_strm}
\eeqa
The vortex flow velocity is given in terms of $\bm{\psi}$ as
\beqa
v_\theta =\partial_z \bm {\psi}\nn \\
v_z = -\frac{1}{R} \partial_\theta \bm {\psi} 
\label{avortex_gensol}
\eeqa

In the high curvature limit and assuming $\eta_-=\eta_+\equiv \eta$, 
\begin{equation}
\lim _{R / \lambda \rightarrow 0} c_{n}(k)= \begin{cases}k^{4}+2 k^{2} \frac{R } {\lambda} & n=0 \\[5pt]
 \left(k^{2}+n^{2}\right)^{2} & n \geq 1\end{cases}
\end{equation}

In this limit, we can perform the integrals and sums analytically, leading to the following stream function for a vortex of strength $\tau$ located at the origin.

\beqa
\bm{\psi}(\theta, z)= \frac{\tau}{4 \pi \eta_{2D}}  \left(- \log \left[1-2 e^{-\frac{|z|}{R}} \cos \theta+e^{-\frac{2 |z|}{R}}\right]+ \sqrt{\frac{\lambda}{2R}}~ e^{-\frac{\sqrt{2} |z|}{\sqrt{\lambda R}}}\right)\
\label{avortex_hcstrm}
\eeqa
which is the Eq.(\ref{strmhc}) of main text.\\
The corresponding vortex flow  at the location $(\theta,z)$ sourced by a vortex of strength $\tau$,  situated at $(\theta_0,z_0)$ is given by
\beqa
&v_\theta =\frac{\tau}{4 \pi \eta_{2D} |z-z_0|}\left(\frac{2-2 \cos(\theta-\theta_0)~ e^{\frac{|z-z_0|}{R}}}{1+2 e^{\frac{|z-z_0|}{R}} - 2 \cos(\theta-\theta_0)~ e^{\frac{|z-z_0|}{R}}}-\frac{e^{-\sqrt{\frac{2}{R \lambda}}|z-z_0|}(z-z_0)}{R}\right)\nn \\
&v_z = -\frac{\tau}{4 \pi \eta_{2D} R} ~ \frac{\sin(\theta- \theta_0)}{ \cos(\theta- \theta_0) -\cosh\frac{z-z_0}{R}}
\label{avortex_hc}
\eeqa
which is Eq.(\ref{vhc}) of main text. Note that in the limit $z\rightarrow z_0$, the vortex flow Eq.(\ref{avortex_hc}) simplifies to
\beqa
&\lim _{z\rightarrow z_0}v_\theta= 0 \nn\\
&\lim _{z\rightarrow z_0} v_z= \frac{\tau ~\cot\left(\frac{\theta- \theta_0}{2}\right)}{4 \pi R~ \eta_{2d}}
\label{azlim}
\eeqa

and in the limit  $\theta\rightarrow \theta_0$, the vortex flow Eq.(\ref{avortex_hc}) simplifies to
\beqa
&\lim _{\theta\rightarrow \theta_0}v_\theta= -\frac{\tau (z-z_0)}{4 \pi R~ \eta_{2D}~ |z-z_0|}\left(e^{-\sqrt{\frac{2}{R \lambda}}|z-z_0|} + \frac{2}{-1+ e^{\frac{|z-z_0|}{R}} }\right)\nn\\
&\lim _{\theta\rightarrow \theta_0} v_z= 0
\label{athetalim}
\eeqa

\textbf{Data Availability} The data that supports the findings of this study are available within the article and its supplementary material.

\bibliographystyle{unsrt}
\bibliography{main}

\end{document}